\documentclass[a4paper,12pt]{article}
\pdfoutput=1
\usepackage{epsfig,graphicx,xcolor,amsbsy,amssymb,latexsym,amsfonts,amsmath,setspace}
\usepackage{pstricks}
\usepackage{color}
\usepackage[numbers,square,comma, compress]{natbib}
\usepackage{placeins}

\usepackage{eurosym}
\usepackage[a4paper]{geometry}
\usepackage{graphicx}
\usepackage{tikz}
\usepackage{xcolor}
\usepackage{amsmath,amssymb,amsfonts,pstricks,setspace}

\numberwithin{equation}{section}

\newcommand{\bea}{\begin{eqnarray}\displaystyle}
\newcommand{\eea}{\end{eqnarray}}

\newcommand{\figref}[1]{Fig.~\protect\ref{#1}}

\title{
\vspace{-1.8cm}
\bf{Dual Little Strings from F-Theory \\
and Flop Transitions }\\[15pt]}
\author{\large \textsc{Stefan~Hohenegger\footnote{\tt s.hohenegger@ipnl.in2p3.fr}~,~Amer Iqbal\footnote{\tt  amer@alum.mit.edu}~,~ Soo-Jong Rey\,\footnote{\tt rey.soojong@gmail.com}}}
\date{}

\begin{document}

\maketitle

\begin{center}
\renewcommand{\thefootnote}{\fnsymbol{footnote}}\vspace{-0.5cm}
${}^{\footnotemark[1]\footnotemark[2]\footnotemark[3]}$ Fields, Gravity \& Strings, CTPU\\
Institute for Basic Sciences, Daejeon 34047 \rm KOREA\\[0.4cm]
${}^{\footnotemark[1]}$ Universit\'e de Lyon\\
UMR 5822, CNRS/IN2P3, Institut de Physique Nucl\'eaire de Lyon\\ 4 rue Enrico Fermi, 69622 Villeurbanne Cedex, \rm FRANCE\\[0.4cm]
${}^{\footnotemark[2]}$ Abdus Salam School of Mathematical Sciences \\ Government College University, Lahore, PAKISTAN\\[0.4cm]
${}^{\footnotemark[2]}$ Center for Theoretical Physics, Lahore, PAKISTAN\\[0.4cm]
${}^{\footnotemark[3]}$ School of Physics and Astronomy \& Center for Theoretical Physics\\
Seoul National University, Seoul 08826 \rm KOREA\\[1cm]
\end{center}

\begin{abstract}
\noindent
A particular two-parameter class of little string theories can be described by $M$ parallel M5-branes probing a transverse affine $A_{N-1}$ singularity.
We previously discussed the duality between the theories labelled by $(N,M)$ and $(M,N)$. In this work, we propose that these two are in fact only part of a larger web of dual theories. We provide evidence that the theories labelled by $(N,M)$ and $(\tfrac{NM}{k},k)$ are dual to each other, where $k=\text{gcd}(N,M)$. To argue for this duality, we use a geometric realization of these little string theories in terms of F-theory compactifications on toric, non-compact Calabi-Yau threefolds $X_{N,M}$ which have a double elliptic fibration structure. We show explicitly for a number of examples that $X_{NM/k,k}$ is part of the extended moduli space of $X_{N,M}$, \emph{i.e.} the two are related through symmetry transformations and flop transitions. By working out the full duality map, we provide a simple check at the level of the free energy of the little string theories. \end{abstract}

\newpage

\tableofcontents

\onehalfspacing

\vskip1cm

\section{Introduction}
\noindent
Little string theories are a class of interacting, non-local, ultraviolet complete quantum theories in six dimensions (or lower), which nevertheless have a local energy-momentum tensor. These theories have recently attracted a lot of renewed attention from various viewpoints \cite{Harvey:2014cva,Aharony:2015zea,Lin:2015zea,Kim:2015gha,DelZotto:2015rca,Aganagic:2015cta,Giveon:2015raa,Hohenegger:2015btj,Bhardwaj:2015oru,Hohenegger:2016eqy,Haouzi:2016ohr}. They can be obtained from string theory through a particular decoupling limit that preserves numerous 'stringy' properties, but suppresses gravitational interactions. The limit in type II string theory requires to take the string coupling constant to zero ($g_{\text{st}}\to 0$), while keeping the string length $\ell_{\text{st}}$ fixed. Depending on the details, we can construct various types of little string theories \cite{LSTR1,LSTR2,LSTR3,LSTR4,LSTR4.5,LSTR5,LSTR6,LSTR7}.
 
Maximally supersymmetric little string theories have sixteen supercharges (for a review, see \cite{Aharony:1998ub}) and come in two different incarnations that are related by T-duality upon circle compactification:
\begin{itemize}
\item type IIa little string theory of $A_{N-1}$ type with $\mathcal{N}=(1,1)$ supersymmetry:\\
This non-chiral LST can be obtained through the decoupling limit of a stack of $N$ NS5-branes in type IIB string theory, or through the decoupling limit of an $\mathbb{R}^4/\mathbb{Z}_N$ orbifold background in type IIA string theory,
\item type IIb little string theory of $A_{N-1}$ type with $\mathcal{N}=(2,0)$ supersymmetry:\\
This chiral LST can be obtained through the decoupling limit of a stack of $N$ NS5-branes in type IIA string theory, or through the decoupling limit of an $\mathbb{R}^4/\mathbb{Z}_N$ orbifold background in type IIB string theory. 
\end{itemize}
One way to make T-duality between these two theories tangible is through a nonperturbative construction of little string theories within the framework of F-theory compactification \cite{Morrison:1996na,Morrison:1996pp}. Indeed, the theories mentioned above can be described in terms of toric, non-compact Calabi-Yau threefolds that feature a double elliptic fibration structure. We studied these Calabi-Yau threefolds in previous papers \cite{Hohenegger:2015btj,Hohenegger:2016eqy} and denoted them by $X_{N,1}$. In this F-theory framework, the little string T-duality corresponds to the exchange of the two elliptic fibrations \cite{Bhardwaj:2015oru,Hohenegger:2015btj}, relating type IIa and type IIb little string theories of type $A_{N-1}$.

More recently, little string theories with eight supercharges have been studied \cite{Bhardwaj:2015oru, Hohenegger:2016eqy,Morrison:2016djb,Johnson:2016qar}. A two-parameter class of such theories can be obtained from type IIa or IIb little string theories of type $A_{N-1}$ through a particular orbifolding procedure \cite{Haghighat:2013tka,Bhardwaj:2015oru,Hohenegger:2016eqy}. Specifically, we consider the decoupling limit of a stack of NS5-branes that probe an orbifold singularity (rather than flat $\mathbb{R}^4$ as above), \emph{i.e.} for generic $N,M\in\mathbb{N}$, we have 
\begin{itemize}
\item $\mathbb{Z}_N$ orbifold of IIa little string theory of type $A_{M-1}$:\\
This little string theory generically preserves $\mathcal{N}=(1,0)$ supersymmetry and is described through the decoupling limit of $M$ NS5-branes in type IIB string theory probing an $\mathbb{R}^4/\mathbb{Z}_N$ orbifold background.  
\item $\mathbb{Z}_M$ orbifold of IIb little string theory of type $A_{N-1}$: \\
This little string theory generically preserves $\mathcal{N}=(1,0)$ supersymmetry and is described through the decoupling limit of $N$ NS5-branes in type IIA string theory probing an $\mathbb{R}^4/\mathbb{Z}_M$ orbifold background.
\end{itemize}
Extending the discussion in \cite{Ooguri:1995wj}, the type IIb orbifolded little string theory can also be described as a decoupling limit of $N$ parallel NS5-branes that probe an affine $A_{M-1}$ ALE space. By lifting to M-theory, this realisation allows a description of the BPS excitations of the little string theories through systems of parallel M5-branes with M2-branes stretched between them. For the latter, the partition functions can be computed in a very efficient manner, using either the world-sheet theory of the M-string (which is the one-dimensional intersection between the M5- and M2-branes) \cite{Haghighat:2013tka,Haghighat:2013gba} or a dual realisation in terms of $(p,q)$-brane webs in type II string theory \cite{Hohenegger:2013ala,Hohenegger:2015cba}. 

The description we will mostly focus on in this paper, however, is in terms of F-theory on a two-parameter class of toric, non-compact Calabi-Yau threefolds $X_{N,M}$, generalising the geometric approach to the little string theories with sixteen supercharges mentioned above. Indeed, denoting by $Z_{\text{IIa}}^{(N,M)}$ and $Z_{\text{IIb}}^{(M,N)}$ the little string partition functions and by $\mathcal{Z}_{X_{N,M}}(\mathbf{t},\mathbf{T},m,\epsilon_{1,2})$ the refined topological string partition function\footnote{Here $\mathbf{t}$ and $\mathbf{T}$ are two sets of K\"ahler parameters (associated with a double elliptic fibration of $X_{N,M}$) and $\epsilon_{1,2}$ as well as $m$ are deformation parameters.} on $X_{N,M}$, we proposed in the previous work \cite{Hohenegger:2016eqy}
\begin{align}
&Z^{(N,M)}_{\text{IIa}}(\mathbf{T},\mathbf{t},m,\epsilon_{1,2})={\cal Z}_{X_{N,M}}(\mathbf{t},\mathbf{T},m,\epsilon_{1,2})\,,&&
Z^{(M,N)}_{\text{IIb}}(\mathbf{T},\mathbf{t},m,\epsilon_{1,2})={\cal Z}_{X_{M,N}}(\mathbf{T},\mathbf{t},m,\epsilon_{1,2})\, , \nonumber
\end{align}
which makes T-duality manifest in the sense that $Z^{(N,M)}_{\text{IIa}}(\mathbf{T},\mathbf{t},m,\epsilon_{1,2})=Z^{(N,M)}_{\text{IIb}}(\mathbf{t},\mathbf{T},m,\epsilon_{1,2})$.
This connection between little string theories and the Calabi-Yau threefolds $X_{N,M}$ also reveals a number of other additional dualities. For example, since $X_{N,M}$ can be represented as a particular resolution of a $\mathbb{Z}_N\times \mathbb{Z}_M$ orbifold of $X_{1,1}$ (which at the boundary of the moduli space resembles the resolved conifold), it follows that $X_{N,M}$ and $X_{M,N}$ are dual to each other. This duality also implies that the $\mathbb{Z}_N$ orbifolds of IIa LST of type $A_{M-1}$ (and respectively the $\mathbb{Z}_M$ orbifolds of IIb LST of type $A_{N-1}$) are self-dual under T-duality \cite{Hohenegger:2016eqy} in certain regions of the $(\mathbf{t},\mathbf{T})$ moduli space. This has been checked explicitly at the level of the free energies associated with the partition functions $Z^{(N,M)}_{\text{IIa}}$ and $Z^{(M,N)}_{\text{IIb}}$\cite{Hohenegger:2016eqy}.

In this paper, we consider further duality relations between theories that are characterised by different $(N,M)$. Based on a number of explicit examples, we propose that theories characterised by $(N,M)$ (with generic $M,N\in\mathbb{N}$ and $\text{gcd}(M,N)=k$) and $(\tfrac{NM}{k},k)$ are dual to each other. The corresponding Calabi-Yau threefolds are related through a chain of transformations involving flop transitions \footnote{By $\sim$, we will denote Calabi-Yau threefolds related by symmetry transformations and flop transitions.}:
\begin{align}
&X_{N,M} \quad \sim \quad X_{NM/k,k}\,,&&\text{for} &&k=\text{gcd}(M,N)\,.
\nonumber
\end{align}
Moreover, we make a proposal for the full duality map, \emph{i.e.} we provide the K\"ahler parameters of $X_{NM/k,k}$ completely in terms of the K\"ahler parameters of $X_{N,M}$. Finally, based on results of our previous work \cite{Hohenegger:2016eqy}, we perform a simple check at the level of the free energies $\Sigma_{N,M}$ associated with the partition functions $\mathcal{Z}_{X_{N,M}}$. Indeed, in \cite{Hohenegger:2016eqy}, we proposed that the free energies $\Sigma_{N,M}$ enjoy a special property which we called \emph{self-similarity}: in a particular region of the parameter space and in the Nekrasov-Shatashvili limit \cite{Nekrasov:2009rc,Mironov:2009uv}, we have $\lim_{\epsilon_2\to 0}\epsilon_2 \Sigma_{N,M}=NM\lim_{\epsilon_2\to 0}\epsilon_2 \Sigma_{1,1}$. We show for generic $M,N\in\mathbb{N}$ that the proposed duality map, when restricted to the above mentioned region in parameter space, corresponds to an $Sp(2,\mathbb{Z})$ transformation, under which ${\cal Z}_{X_{1,1}}$ is invariant. This is a strong indication that indeed $\mathcal{Z}_{X_{N,M}}=\mathcal{Z}_{X_{NM/k,k}}$ after application of the duality map.

This proposed duality has other consequences and in fact gives rise to a whole web of dual theories. Indeed, for any pair of integers $(M,N)$ and $(M',N')$ with $MN=M'N'$ and $\text{gcd}(M,N)=k=\text{gcd}(M',N')$, the proposed duality also implies
\begin{align}
X_{N,M}\quad \sim \quad X_{N',M'}\,.
\nonumber
\end{align}
In this way, we have for example that the Calabi-Yau threefolds parametrised by $(30,1)\sim (15,2)\sim (10,3)\sim (6,5)\sim (5,6)\sim (3,10)\sim (2,15)\sim (1,30)$ are all related to one another. Reformulating this relation in a more abstract fashion, we propose that the theories parametrised by $(M,N)$ and $(M',N')$ as above are part of the extended moduli space \cite{Reid,CT1,CT2,CT3,CT4} of the Calabi-Yau threefold $X_{NM/k,k}$. Moreover, little string theories described by $(N,M)$ with $\text{gcd}(N,M)=k=1$ actually preserve sixteen (rather than eight) supercharges (even if $N\neq 1$ and  $M\neq 1$): indeed, in this case $X_{N,M}\sim X_{NM,1}$, which can also be realised as an elliptic fibration over an affine $A_{NM-1}$ space. From the point of view of the little string theories, this means that $\mathbb{Z}_{M}$ orbifolds of type II little string theories of type $A_{N-1}$ with $k=1$ are dual to IIa or IIb LSTs of type $A_{NM-1}$ preserving $\mathcal{N}=(1,1)$ or $\mathcal{N}=(2,0)$ supersymmetry, respectively. However, as we shall discuss in detail, the little string theories labelled by $(N,M)$ and $(NM,1)$ live in different points of the parameter space: the radii of the two transverse circles on which the little strings are compactified are different in the two cases. 

This paper is organised as follows: In section~\ref{Sect:LSandCYs}, we review in more detail the relation between little string theories and toric Calabi-Yau threefolds $X_{N,M}$. We will also discuss flop transitions that define the extended moduli space of the latter. In section~\ref{Sect:PartitionFunctions}, we review the topological string partition functions of $X_{N,M}$ as well as its relation to ${\cal Z}_{X_{1,1}}$ in a particular region in the K\"ahler moduli space. In section~\ref{Sect:GenericMapMN}, we present numerous explicit examples for the duality $X_{N,M}\sim X_{NM/k,k}$ (with $k=\text{gcd}(N,M)$) and present checks for the duality at the level of the free energies. In section~\ref{Sect:GenericNM}, we make a proposal of the duality map for generic $(N,M)$ and show that it satisfies the same checks as all the examples of section~\ref{Sect:GenericMapMN}. 
Finally, section~\ref{Sect:Conclusions} contains our conclusions. 

\section{Little Strings and Calabi-Yau Threefolds}\label{Sect:LSandCYs}

Little string theories can be described in several different but U-dual equivalent ways. For example, they can be realised as ADE orbifold compactifications of type II string theory (in the presence of NS5-branes) subject to a particular limit in which gravity decouples \cite{LSTR1,LSTR2,LSTR3,LSTR4,LSTR4.5,LSTR5,LSTR6,LSTR7}. Alternatively, LSTs can also be described by systems of M5-branes in M-theory that probe a transverse ADE orbifold. However, the description that is most useful for our current purposes uses the F-theory: indeed, by compactifying F-theory on appropriate elliptically fibered Calabi-Yau threefolds,  little string theories with both $\mathcal{N}=(2,0)$ and $\mathcal{N}=(1,0)$ supersymmetry can be obtained. In fact, a classification of little string theories using this F-theory approach was carried out recently in \cite{Bhardwaj:2015oru}. 

These Calabi-Yau threefolds are elliptically fibered over a two complex dimensional base, which in the decoupling limit becomes non-compact. In general, the charge lattice of the compactified theory is given by the lattice of two-cycles in the non-compact base, and BPS states are given by D3-branes wrapping holomorphic two curves in the base. The little strings in particular arise from D3-branes wrapping a curve of self intersection zero in the base. Calabi-Yau threefolds that have such a curve with vanishing self-intersection in the base actually have a double elliptic fibration, since the base itself is an elliptic fibration over the complex plane \cite{schoen}. As was first discovered in \cite{Bhardwaj:2015oru,Hohenegger:2015btj}, this double fibration structure is crucially related to T-duality of the little string theory.
 
 \subsection{Calabi-Yau Threefolds $X_{N,M}$}
Calabi-Yau threefolds $X_{N,M}$ which realise type IIa little strings in an orbifold background of type $A_N$ are toric and are given by the resolution of a $\mathbb{Z}_{N}\times \mathbb{Z}_{M}$ orbifold of $X_{1,1}$ \cite{Hohenegger:2015btj}. Here $X_{1,1}$ is itself a Calabi-Yau threefold with a double elliptic fibration over the complex plane \cite{Hollowood:2003cv}. To understand this, we begin with a geometry that is similar to $X_{1,1}$, namely the (deformed) conifold. The latter can be described as a hypersurface in $\mathbb{C}^4$
\begin{align}
&xy-zw=\epsilon\,,&&\text{for} &&(x,y,z,w)\in \mathbb{C}^4\,.\label{DeformedConi}
\end{align}
Here $\epsilon>0$ corresponds to the deformation that regularises an otherwise singular geometry.  To make the fibration structure visible, we re-write Eq.(\ref{DeformedConi}) in the following manner
\bea\label{uplane1}
xy=u\,,\,\,\,zw=u-\epsilon\,,
\eea
which corresponds to a $\mathbb{C}^{\times}\times\mathbb{C}^{\times}$ fibration over the u-plane: the first $\mathbb{C}^{\times}$ fibration degenerates at $u=0$ and the second one at $u=\epsilon$. The $\mathbb{S}^1$'s of these two $\mathbb{C}^{\times}$ fibrations together with the path in the $u$-plane connecting the points $u=0$ and $u=\epsilon$ give an $\mathbb{S}^3$. 
We can compactify the $\mathbb{C}^{\times}$ fibers to obtain a double elliptic fibration over the u-plane,
\bea\label{uplane2}
y^2=x^3+f_{1}(u)\,x+g_{1}(u)\,,\,\,\,\,\,\,\,w^2=z^3+f_{2}(u)\,z+g_{2}(u)\,.
\eea
Here $f_{1}(u),g_{1}(u),f_{2}(u),g_{2}(u)$ are holomorphic functions of $u$ such that $\Delta_{1}(u)=4f_{1}^3+27g_{1}^2$ vanishes only at $u=0$ and $\Delta_{2}(u)=4f_{2}^3+27g_{2}^2$ vanishes only at $u=\epsilon$ so that the two elliptic fibrations have fibers of type $I_{1}$ at $u=0$ and $u=\epsilon$, respectively. The geometry of this Calabi-Yau threefold and its mirror was studied in detail in a recent paper \cite{atsushi}.

The geometry described above is singular in the limit $\epsilon\to 0$. Besides the deformation Eq.(\ref{DeformedConi}), we have an alternative way of dealing with this singularity: we can also resolve the conifold to obtain a $\mathbb{P}^1$. Indeed, instead of Eq.(\ref{DeformedConi}), we describe the resolution of the conifold through the equation

\bea
\left(\begin{array}{cc}x & z \\ w & y\end{array}\right)\left(\begin{array}{c}\lambda_1 \\ \lambda_2\end{array}\right)=0\,,\label{ResolvedConifold}
\eea
where $\lambda=\frac{\lambda_1}{\lambda_2}$ is the coordinate on the blowup $\mathbb{P}^1$ mentioned above. The resulting geometry can also be described as a $\mathbb{C}^4$ quotient by $\mathbb{C}^{\times}$,
\bea
(A_{1},A_{2},B_{1},B_{2})\mapsto (\lambda\,A_{1},\lambda\,A_{2},\lambda^{-1}\,B_{1},\lambda^{-1}\,B_{2})\,,\label{DefA1A2}
\eea
where $(A_1,A_2,B_1,B_2)$ are coordinates of $\mathbb{C}^4$ such that\footnote{Notice that the relation $\text{det}\left[\left(\begin{array}{c}A_{1} \\ A_{2}\end{array}\right)\left(\begin{array}{cc}B_{1} & B_{2}\end{array}\right)\right]=xy-zw=0$ for the unresolved conifold is manifestly realised in this parametrisation.}
 \bea
 \left(\begin{array}{cc}x & z \\ w & y\end{array}\right)=\left(\begin{array}{c}A_{1} \\ A_{2}\end{array}\right)\left(\begin{array}{cc}B_{1} & B_{2}\end{array}\right)\,.
 \eea
The resolution is particularly useful for discussing orbifolds of the conifold:  
let $\Gamma_{1}\times \Gamma_{2}\subset SU(2)\times SU(2)$  be a pair of ADE subgroups of $SU(2)$ which act on $(A_{1},A_{2},B_{1},B_{2})\in \mathbb{C}^4$ by embedding in $SU(4)$:
\begin{align}
\Gamma_{1}:\,\left(\begin{array}{c}A_{1} \\ B_{1}\end{array}\right)&\mapsto \gamma_{1}\,\left(\begin{array}{c}A_{1} \\ B_{1}\end{array}\right)\,,&&\text{for}&&\gamma_{1}\in \Gamma_1\nonumber\\
\left(\begin{array}{c}A_{2} \\ B_{2}\end{array}\right) &\mapsto \left(\begin{array}{c}A_{2} \\ B_{2}\end{array}\right)\nonumber\\
\Gamma_{2}:\,\left(\begin{array}{c}A_{1} \\ B_{2}\end{array}\right)&\mapsto \gamma_{2}\,\left(\begin{array}{c}A_{1} \\ B_{2}\end{array}\right)\,,&&\text{for} &&\gamma_{2}\in \Gamma_2\nonumber\\
\left(\begin{array}{c}A_{2} \\ B_{1}\end{array}\right) &\mapsto \left(\begin{array}{c}A_{2} \\ B_{1}\end{array}\right)\,.\label{OrbifoldAction}
\end{align}
Modding the geometry (\ref{ResolvedConifold}) by this action we get the $\Gamma_{1}\times \Gamma_{2}$ orbifold of the resolved conifold. In the particular case  $\Gamma_1\times\Gamma_2=\mathbb{Z}_{N}\times \mathbb{Z}_{M}$ for $M,N\in\mathbb{Z}$ the action Eq.(\ref{OrbifoldAction}) takes the form \cite{Aganagic:1999fe}:
\bea
(A_{1},A_{2},B_{1},B_{2})\mapsto (\omega^{a}_{N}\omega^{b}_{M}\,A_{1},A_{2},\omega^{-a}_{N}B_{1},\omega^{-b}_{M}\,B_{2})\,,
\eea
where $(\omega_{N},\omega_{M})=(e^{\frac{2\pi i}{N}},e^{\frac{2\pi i}{M}})$ and $a,b\in \mathbb{Z}$.\footnote{For completeness we mention that we can also describe $\mathbb{Z}_{N}\times \mathbb{Z}_{M}$ orbifolds of the deformed conifold. Indeed, when the $\mathbb{C}^{\times}\times \mathbb{C}^{\times}$ fibration of the deformed conifold is compactified to a double elliptic fibration, we can introduce the the orbifold geometry 
\bea\label{uplane2}
y^2=x^3+f_{1}(u^N)\,x+g_{1}(u^N)\,,\,\,\,\,\,\,\,w^2=z^3+f_{2}(u^M)\,z+g_{2}(u^M)\,,
\eea
so that there is a $I_{N}$ and $I_{M}$ fiber in the two elliptic fibrations.} 

In order to describe little string theories, we have to replace the deformed conifold by the doubly elliptically fibered Calabi-Yau manifold $X_{1,1}$ which can be understood as a compactification of $\mathbb{C}^{\times}\times \mathbb{C}^{\times}$ fibers of the resolved conifold to double elliptic fibers, as we shall discuss in the following section. In the general case of a $\Gamma_{1}\times \Gamma_{2}$ orbifold of $X_{1,1}$ we obtain a geometry with two fibers of the corresponding $ADE$ type in the two elliptic fibrations, which describes $(\Gamma_1,\Gamma_2)$ little string theories \footnote{For a recent discussions on multiple fibrations, see \cite{Anderson:2016cdu}.}. 

\subsubsection{The dual brane webs}
\begin{figure}[h]
  \centering
  \includegraphics[width=5in]{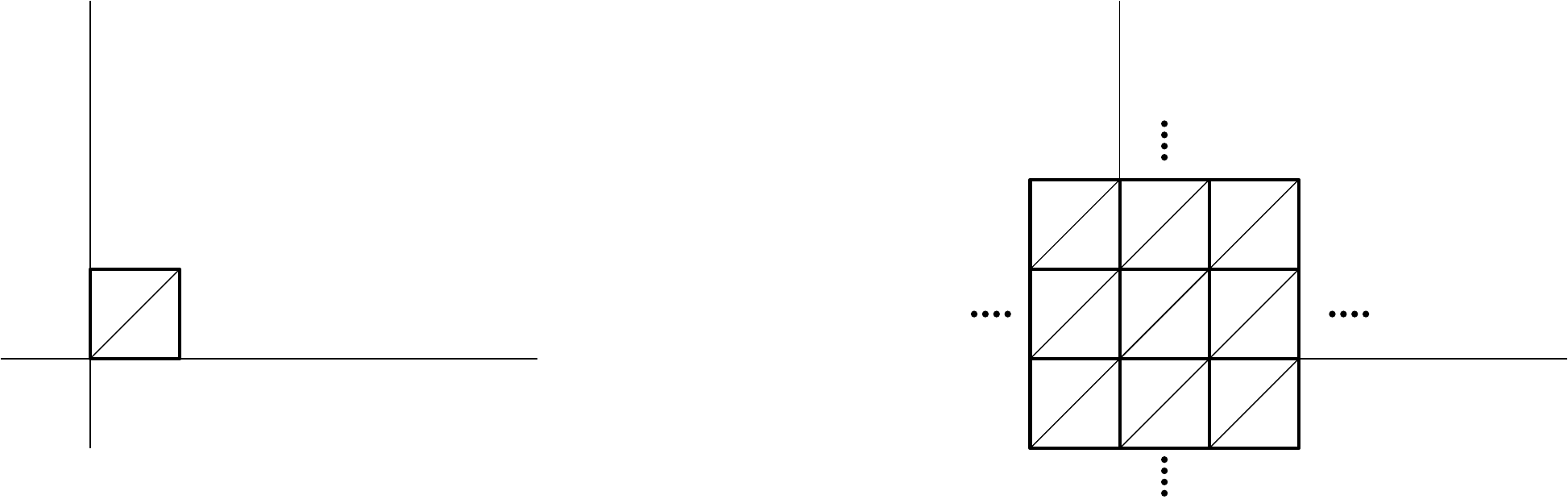}
  \caption{\emph{The Newton polygon of the resolved conifold (left) and the Calabi-Yau threefold $X_{1,1}$ (right). The polygon of the latter is obtained by tiling the plane with the Newton polygon of the resolved conifold. Equivalently, this tiling can be though of drawing the Newton polygon of the resolved conifold on a (doubly periodic) torus.}}
  \label{tile1}
\end{figure}

\begin{figure}[h]
  \centering
\begin{tikzpicture}
\draw[-](-0.5,0) -- (5,0);
\draw[-] (0,-0.5) -- (0,4);
\draw[ultra thick] (0,0) -- (1.5,0) -- (1.5,1) -- (0,1) -- (0,0);
\draw[-] (0,0.5) -- (1.5,0.5);
\draw[-] (0,1) -- (1.5,1);
\draw[-] (0.5,0) -- (0.5,1);
\draw[-] (1,0) -- (1,1);
\draw[-] (1.5,0) -- (1.5,1);
\draw[-] (0,0.5) -- (0.5,1);
\draw[-] (0,0) -- (1,1);
\draw[-] (0.5,0) -- (1.5,1);
\draw[-] (1,0) -- (1.5,0.5);
\draw[<->] (0,1.5) -- (1.5,1.5);
\node at (0.75,1.8) {$N$};
\draw[<->] (2,0) -- (2,1);
\node at (2.3,0.5) {$M$};
\node at (2,-1.5) {(a)};
\end{tikzpicture}  
\hspace{2cm}
\begin{tikzpicture}
\draw[-](-0.5,1) -- (5,1);
\draw[-] (1.5,-0.5) -- (1.5,4);
\draw[ultra thick] (0,0) -- (4.5,0) -- (4.5,3) -- (0,3) -- (0,0);
\draw[-] (0,0.5) -- (4.5,0.5);
\draw[ultra thick,-] (0,1) -- (4.5,1);
\draw[-] (0,1.5) -- (4.5,1.5);
\draw[ultra thick,-] (0,2) -- (4.5,2);
\draw[-] (0,2.5) -- (4.5,2.5);
\draw[-] (0.5,0) -- (0.5,3);
\draw[-] (1,0) -- (1,3);
\draw[ultra thick,-] (1.5,0) -- (1.5,3);
\draw[-] (2,0) -- (2,3);
\draw[-] (2.5,0) -- (2.5,3);
\draw[ultra thick,-] (3,0) -- (3,3);
\draw[-] (3.5,0) -- (3.5,3);
\draw[-] (4,0) -- (4,3);
\draw[-] (0,2.5) -- (0.5,3);
\draw[-] (0,2) -- (1,3);
\draw[-] (0,1.5) -- (1.5,3);
\draw[-] (0,1) -- (2,3);
\draw[-] (0,0.5) -- (2.5,3);
\draw[-] (0,0) -- (3,3);
\draw[-] (0.5,0) -- (3.5,3);
\draw[-] (1,0) -- (4,3);
\draw[-] (1.5,0) -- (4.5,3);
\draw[-] (2,0) -- (4.5,2.5);
\draw[-] (2.5,0) -- (4.5,2);
\draw[-] (3,0) -- (4.5,1.5);
\draw[-] (3.5,0) -- (4.5,1);
\draw[-] (4,0) -- (4.5,0.5);
\node at (-0.7,1.5) {$\Large \cdots$};
\node at (5.2,1.5) {$\Large \cdots$};
\node at (2.25,4) {$\Large \vdots$};
\node at (2.25,-0.5) {$\Large \vdots$};
\node at (2,-1.5) {(b)};
\end{tikzpicture}  

 \caption{\emph{(a) The Newton polygon of the $\mathbb{Z}_{N}\times \mathbb{Z}_{M}$ orbifold of the resolved conifold. (b) The Newton polygon of $X_{N,M}$ is obtained through a tiling of the plane with the Newton polygon of the orbifold of the resolved conifold.}  }
 \label{tile2}
\end{figure}
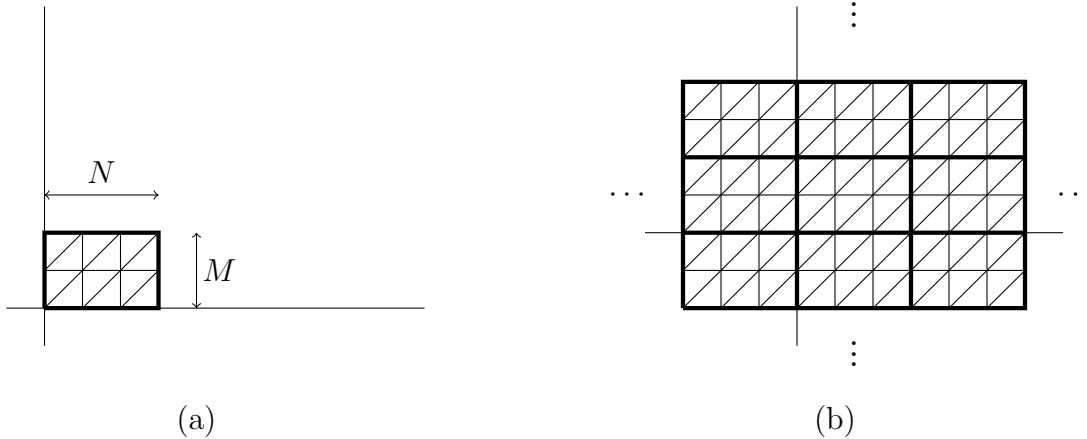

\begin{figure}[htb]
\begin{center}
\begin{tikzpicture}[scale = 0.60]
\draw[ultra thick,green!50!black] (-6,0) -- (-5,0);
\draw[ultra thick,red] (-5,-1) -- (-5,0);
\draw[ultra thick,blue] (-5,0) -- (-4,1);
\draw[ultra thick,green!50!black] (-4,1) -- (-3,1);
\draw[ultra thick,red] (-4,1) -- (-4,2);
\draw[ultra thick,red] (-3,1) -- (-3,0);
\draw[ultra thick,blue] (-3,1) -- (-2,2);
\draw[ultra thick,blue] (-4,2) -- (-3,3);
\draw[ultra thick,green!50!black] (-5,2) -- (-4,2);
\draw[ultra thick,red] (-3,3) -- (-3,4);
\draw[ultra thick,green!50!black] (-3,3) -- (-2,3);
\draw[ultra thick,red] (-2,2) -- (-2,3);
\draw[ultra thick,green!50!black] (-2,2) -- (-1,2);
\draw[ultra thick,blue] (-2,3) -- (-1,4);
\draw[ultra thick,blue] (-1,2) -- (0,3);
\draw[ultra thick,red] (-1,2) -- (-1,1);
\draw[ultra thick,green!50!black] (-4,4) -- (-3,4);
\draw[ultra thick,blue] (-3,4) -- (-2,5);
\draw[ultra thick,green!50!black] (-2,5) -- (-1,5);
\draw[ultra thick,green!50!black] (-1,4) -- (0,4);
\draw[ultra thick,green!50!black] (0,3) -- (1,3);
\draw[ultra thick,red] (-2,5) -- (-2,6);
\draw[ultra thick,red] (-1,4) -- (-1,5);
\draw[ultra thick,red] (0,3) -- (0,4);
\draw[ultra thick,blue] (1,3) -- (2,4);
\draw[ultra thick,blue] (0,4) -- (1,5);
\draw[ultra thick,blue] (-1,5) -- (0,6);
%
\draw[ultra thick,green!50!black] (2,4) -- (3,4);
\draw[ultra thick,green!50!black] (1,5) -- (2,5);
\draw[ultra thick,green!50!black] (0,6) -- (1,6);
%
\draw[ultra thick,red] (1,3) -- (1,2);
\draw[ultra thick,red] (2,4) -- (2,5);
\draw[ultra thick,red] (1,5) -- (1,6);
\draw[ultra thick,red] (0,6) -- (0,7);
\draw[ultra thick,green!50!black] (-1,10) -- (0,10);
\node[rotate=90] at (0,8) {{\Huge$\ldots$}};
\node[rotate=90] at (2,9) {{\Huge$\ldots$}};
\node[rotate=90] at (4,10) {{\Huge$\ldots$}};
\node[rotate=90] at (6,11) {{\Huge$\ldots$}};
\node[rotate=90] at (-2,7) {{\Huge$\ldots$}};
\draw[ultra thick,red] (0,9) -- (0,10);
\draw[ultra thick,blue] (0,10) -- (1,11);
\draw[ultra thick,red] (1,11) -- (1,12);
\draw[ultra thick,red] (2,11) -- (2,10);
\draw[ultra thick,green!50!black] (1,11) -- (2,11);
\draw[ultra thick,blue] (2,11) -- (3,12);
\draw[ultra thick,red] (3,12) -- (3,13);
\draw[ultra thick,red] (4,12) -- (4,11);
\draw[ultra thick,green!50!black] (3,12) -- (4,12);
\draw[ultra thick,blue] (4,12) -- (5,13);
\draw[ultra thick,green!50!black] (5,13) -- (6,13);
\draw[ultra thick,red] (5,13) -- (5,14);
\draw[ultra thick,blue] (6,13) -- (7,14);
\draw[ultra thick,red] (6,13) -- (6,12);
\draw[ultra thick,green!50!black] (7,14) -- (8,14);
\draw[ultra thick,red] (7,14) -- (7,15);
\draw[ultra thick,blue] (1,6) -- (2,7);
\draw[ultra thick,red] (2,7) -- (2,8);
\draw[ultra thick,red] (3,7) -- (3,6);
\draw[ultra thick,green!50!black] (2,7) -- (3,7);
\draw[ultra thick,blue] (3,7) -- (4,8);
\draw[ultra thick,red] (4,8) -- (4,9);
\draw[ultra thick,red] (5,8) -- (5,7);
\draw[ultra thick,green!50!black] (4,8) -- (5,8);
\draw[ultra thick,green!50!black] (3,6) -- (4,6);
\draw[ultra thick,green!50!black] (5,7) -- (6,7);
\draw[ultra thick,blue] (5,8) -- (6,9);
\draw[ultra thick,red] (6,9) -- (6,10);
\draw[ultra thick,blue] (2,5) -- (3,6) ;
\draw[ultra thick,blue] (4,6) -- (5,7) ;
\draw[ultra thick,green!50!black] (6,9) -- (7,9);
\node at (8,9) {{\Huge $\ldots$}};
\draw[ultra thick,blue] (3,4) -- (4,5);
\draw[ultra thick,red] (3,4) -- (3,3);
\draw[ultra thick,red] (4,5) -- (4,6);
\draw[ultra thick,green!50!black] (4,5) -- (5,5);
\node at (7,7) {{\Huge $\ldots$}};
\draw[ultra thick,green!50!black] (10,14) -- (11,14);
\draw[ultra thick,red] (11,14) -- (11,13);
\draw[ultra thick,blue] (11,14) -- (12,15);
\node at (9,14) {{\Huge $\ldots$}};
\draw[ultra thick,red] (12,15) -- (12,16);
\draw[ultra thick,green!50!black] (12,15) -- (13,15);

\node[rotate=90] at (11,12) {{\Huge $\ldots$}};
\node at (6,5) {{\Huge $\ldots$}};
\draw[ultra thick,blue] (10,9) -- (11,10);
\draw[ultra thick,green!50!black] (9,9) -- (10,9);
\draw[ultra thick,red] (11,10) -- (11,11);
\draw[ultra thick,green!50!black] (11,10) -- (12,10);
\draw[ultra thick,red] (10,9) -- (10,8);
\draw[ultra thick,blue] (9,7) -- (10,8);
\draw[ultra thick,green!50!black] (8,7) -- (9,7);
\draw[ultra thick,red] (9,6) -- (9,7);
\draw[ultra thick,green!50!black] (10,8) -- (11,8);
\draw[ultra thick,blue] (8,5) -- (9,6);
\draw[ultra thick,green!50!black] (9,6) -- (10,6);
\draw[ultra thick,green!50!black] (7,5) -- (8,5);
\draw[ultra thick,red] (8,4) -- (8,5);
%
%
%
\draw[ultra thick,green!50!black] (-3,9) -- (-2,9);
\draw[ultra thick,blue] (-2,9) -- (-1,10);
\draw[ultra thick,red] (-1,10) -- (-1,11);
\draw[ultra thick,red] (-2,8) -- (-2,9);
\node[rotate=90] at (-5.5,0) {$=$};
\node at (-5.3,-0.2) {{\tiny$1$}};
\node[rotate=90] at (-4.5,2) {$=$};
\node at (-4.3,1.8) {{\tiny$2$}};
\node[rotate=90] at (-3.5,4) {$=$};
\node at (-3.3,3.8) {{\tiny$3$}};
\node[rotate=90] at (-2.5,9) {$=$};
\node at (-2.3,8.75) {{\tiny$M$}};
\node[rotate=90] at (9.5,6) {$=$};
\node at (9.7,5.8) {{\tiny$1$}};
\node[rotate=90] at (10.5,8) {$=$};
\node at (10.7,7.8) {{\tiny$2$}};
\node[rotate=90] at (11.5,10) {$=$};
\node at (11.7,9.8) {{\tiny$3$}};
\node[rotate=90] at (12.7,15) {$=$};
\node at (12.9,14.75) {{\tiny$M$}};
\node at (-5,-0.5) {$-$};
\node at (-4.8,-0.7) {{\tiny $1$}};
\node at (-3,0.5) {$-$};
\node at (-2.8,0.3) {{\tiny $2$}};
\node at (-1,1.5) {$-$};
\node at (-0.8,1.3) {{\tiny $3$}};
\node at (1,2.5) {$-$};
\node at (1.2,2.3) {{\tiny $4$}};
\node at (3,3.5) {$-$};
\node at (3.2,3.3) {{\tiny $5$}};
\node at (8,4.5) {$-$};
\node at (8.25,4.3) {{\tiny $N$}};
\node at (-1,10.5) {$-$};
\node at (-0.8,10.3) {{\tiny $1$}};
\node at (1,11.5) {$-$};
\node at (1.2,11.3) {{\tiny $2$}};
\node at (3,12.5) {$-$};
\node at (3.2,12.3) {{\tiny $3$}};
\node at (5,13.5) {$-$};
\node at (5.2,13.3) {{\tiny $4$}};
\node at (7,14.5) {$-$};
\node at (7.2,14.3) {{\tiny $5$}};
\node at (12,15.5) {$-$};
\node at (12.25,15.3) {{\tiny $N$}};
\draw[thick, <->] (-5,-2) -- (-3.05,-2);
\node at (-4,-2.5) {$t_1$};
\draw[thick, <->] (-2.95,-2) -- (-1.05,-2);
\node at (-2,-2.5) {$t_2$};
\draw[thick, <->] (-0.95,-2) -- (0.95,-2);
\node at (0,-2.5) {$t_3$};
\draw[thick, <->] (1.05,-2) -- (2.95,-2);
\node at (2,-2.5) {$t_4$};
\draw[thick, <->] (3.05,-2) -- (4.95,-2);
\node at (4,-2.5) {$t_5$};
\node at (6.5,-2) {{\Huge $\ldots$}};
\draw[thick, <->] (8.05,-2) -- (9.95,-2);
\node at (9,-2.5) {$t_N$};
\draw[thick, <->] (-7,0) -- (-7,1.95);
\node at (-7.5,1) {$T_1$};
\draw[thick, <->] (-7,2.05) -- (-7,3.95);
\node at (-7.5,3) {$T_2$};
\node[rotate=90] at (-7,6.5) {{\Huge $\ldots$ }};
\draw[thick, <->] (-7,9.05) -- (-7,10.95);
\node at (-7.5,10) {$T_M$};
\end{tikzpicture}

\end{center}
\caption{\sl The 5-brane web dual to the Calabi-Yau threefold $X_{N,M}$.}
\label{Fig:WebToric}
\end{figure}
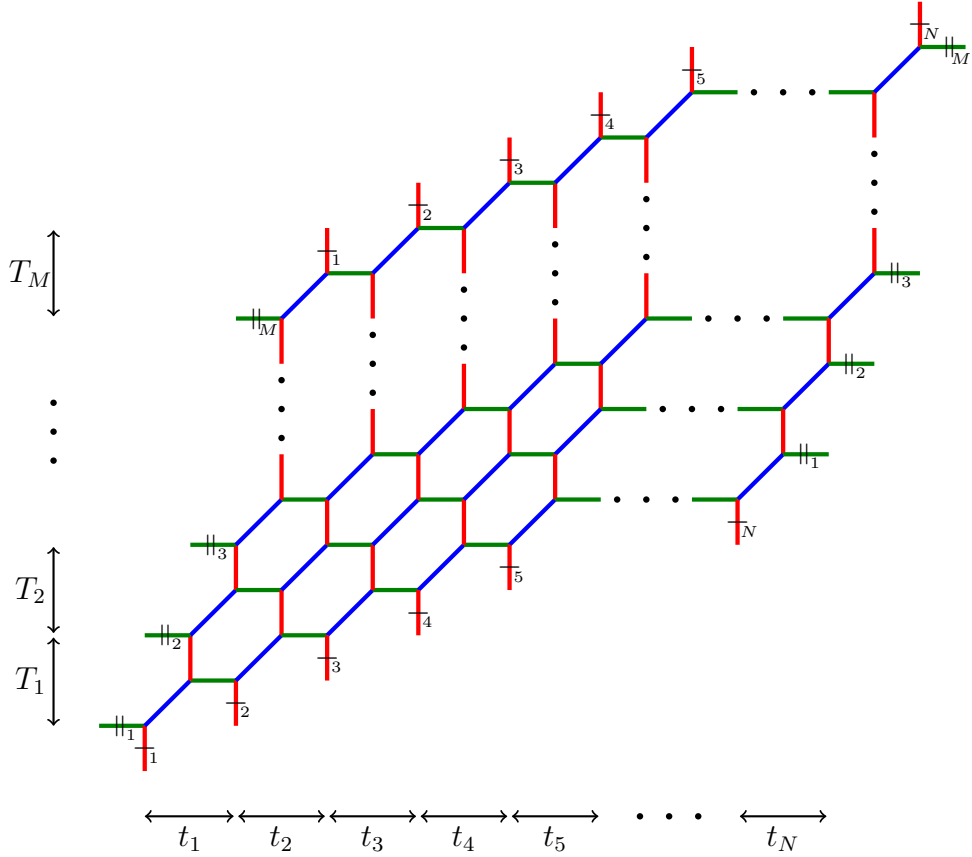

The toric Calabi-Yau threefolds $X_{N,M}$ obtained from $\mathbb{Z}_M\times \mathbb{Z}_N$ orbifolds of $X_{1,1}$, as described in the previous subsubsection, can be dualised to $(p,q)$ 5-brane webs \cite{Aharony:1997bh,Leung:1997tw} which in turn realise certain five- and six-dimensional gauge theories \cite{Hollowood:2003cv, Aharony:1997bh}. This dual 5-brane web is in fact identical to the toric web of $X_{N,M}$, which is dual to the Newton polygon \cite{Aharony:1997bh}. The Newton polygon of $X_{1,1}$ (as well as the one of the resolved conifold) are shown in \figref{tile1}. 
The $\mathbb{Z}_{N}\times \mathbb{Z}_{M}$ orbifold of the resolved conifold is also toric and the corresponding Newton polygon is given by a rectangle of size $N\times M$ \cite{Aganagic:1999fe}. Similar to $X_{1,1}$, the Newton polygon of $X_{N,M}$ is described by a tiling of the plane by the Newton polygon of the orbifolded resolved conifold, \emph{i.e.} the rectangle of size $N\times M$ as shown in \figref{tile2}.
The periodic tiling in \figref{tile2}(b) then implies that these webs are living on $\mathbb{T}^2$ rather than on $\mathbb{R}^2$, and the two elliptic fibrations of $X_{N,M}$ are dual to the two circles on which the 5-branes are wrapped.
The web diagram of $X_{N,M}$ can be obtained as the dual graph of the Newton polygon and is shown in \figref{Fig:WebToric}. The diagonal (blue) lines along the direction $(1,1)$ in \figref{Fig:WebToric} are due to the triangulation of the Newton polygon. 

Furthermore, the lengths of various lines in the 5-brane web correspond to K\"ahler parameters of $X_{N,M}$. From \figref{Fig:WebToric}, it can be seen that there are $MN+2$ independent parameters \cite{Haghighat:2013tka}. This counting arises as follows: the web is parametrised by $MN+M+N$ parameters, corresponding to the vertical and horizontal distances as well as the blue diagonal distances. However, there are also $M+N-2$ constraints which arise from the gluing-conditions on $\mathbb{T}^2$, which impose that the vertical lines at the top and the bottom of the diagram and the horizontal lines at the left- and right end of the diagram can be glued together in a consistent fashion.

Specifically, we denote by $T_{i}$ the separation between the horizontal lines (corresponding to D5-branes from the $(p,q)$-brane web perspectives) and by $t_{a}$ the separation between the vertical lines (corresponding to NS5-branes). Furthermore, we will collectively denote
\begin{align}
&\mathbf{T}=(T_1,T_2,\ldots,T_M)\,,&&\text{and} &&\mathbf{t}=(t_1,t_2,\ldots,t_N)\,,
\end{align}
and we denote the following quantities
\begin{align}
&\tau=\frac{i}{2\pi}(T_1+T_2+\ldots+T_M)\,,&&\text{and}&&\rho=\frac{i}{2\pi}(t_1+t_2+\ldots+t_N)\,.
\end{align}
Notice, from the point of view of the Calabi-Yau threefold, $\tau$ and $\rho$ are just a particular combination of the K\"ahler moduli $(\mathbf{t},\mathbf{T})$. In fact \cite{Hohenegger:2016eqy},  one may choose to parametrise the moduli space by $(\tau,T_1,\ldots,T_{M-1},\rho,t_1,\ldots,t_{N-1})$. However, from the point of view of the dual little string theories, $\tau$ and $\rho$ are the radii of two circles of the little string theory geometry. In this regard, they are treated as fixed (but generic) external parameters rather than moduli.  

\subsection{Flop Transitions in $X_{N,M}$}
In this section, we discuss flop transitions which connect topologically distinct Calabi-Yau threefolds. We also discuss the flop transition from the brane web perspectives.

\subsubsection{Resolved Conifold and Flop Transitions}\label{Sect:Resolution}
A Calabi-Yau threefold can degenerate in many different ways including a divisor collapsing to a point or a curve collapsing to a point. The flop transition is of the latter type in the sense that a rational curve shrinks to a zero size. In the neighborhood of such a singularity, the Calabi-Yau threefold is described by 
\bea\label{sing2}
xy-zw=0\,,~~~~~~~~x=z_{1}+iz_{2}\,,y=z_{1}-iz_{2}\,,z=-z_{3}-iz_{4}\,,w=z_{3}-iz_{4}\,.
\eea
which resembles the conifold as discussed above. As before, we can resolve the singularity 
\bea\label{tf}
\frac{x}{z}=\frac{w}{y}\qquad 
\mbox{such that} \qquad x=\lambda\, z\,~~~~~\mbox{and}~~~~~~w=\lambda\,y\,.
\eea
Here, $\lambda$ is the local coordinate on the blowup $\mathbb{P}^1$ and $(x,w,\lambda)$ are the coordinates on the Calabi-Yau threefold in a patch over $\mathbb{P}^1$. In the second patch, the coordinates are $(z,y,\lambda^{-1})$. The two equations relating $(x,w)$ to $(z,y)$ in Eq.(\ref{tf}) are the transition functions for the bundle ${\cal O}(-1)\oplus {\cal O}(-1)$ over $\mathbb{P}^1$.

However, there are two possible choices:
\begin{align}
&\frac{x}{z}=\frac{w}{y} \qquad 
\mbox{such that} \qquad x=\lambda y\,&&\text{and}&&w=\lambda y\,,\\\nonumber
&\frac{x}{w}=\frac{z}{y}\qquad 
\mbox{such that} \qquad x=\lambda' w\,&&\text{and}&&z=\lambda' y\,.
\end{align}
The difference between the two choices corresponds to interchanging $z$ and $w$, and this is known as a flop transition. Another way of describing this transition is through the quotient construction of the conifold in terms of $(A_1,A_2,B_1,B_2)\in \mathbb{C}^4$ (see Eq.(\ref{DefA1A2})),
\bea
|A_1|^2+|A_{2}|^2-|B_1|^2-|B_2|^2=r\,,
\eea
where the parameter $r$ determines the size of the resolved $\mathbb{P}^1$. The flop transition is given by $r\mapsto -r$ so that the roles of $A_{1,2}$ and $B_{1,2}$ are interchanged. Thus, in terms of the original K\"ahler class, if the $\mathbb{P}^1$ has area $r$, then, after the flop transition, it has area $-r$. However, in the flopped geometry, we can define a different K\"ahler class with respect to which the flopped curve has a positive area. Thus, at $r=0$ we are at the boundary of the K\"ahler cone and the flop $r\mapsto -r$ takes us out of the K\"ahler cone of the Calabi-Yau we started with and into the K\"ahler cone of another Calabi-Yau threefold. This extended moduli space of the K\"ahler class is discussed in more detail in section~\ref{Sect:ExtendedModuliSpace}.
\subsubsection{Flop Transitions in $X_{1,1}$ and $Sp(2,\mathbb{Z})$}
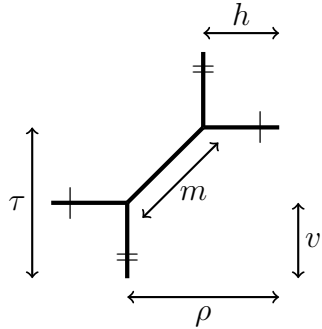
\begin{figure}[htbp]
\begin{center}
\begin{tikzpicture}[scale=0.5]
\draw[ultra thick] (-3,-1) -- (-1,-1);
\draw[ultra thick] (-1,-3) -- (-1,-1);
\draw[ultra thick] (-1,-1) -- (1,1);
\draw[ultra thick] (1,1) -- (3,1);
\draw[ultra thick] (1,1) -- (1,3);
\node at (-2.5,-1) {$|$};
\node at (2.5,1) {$|$};
\node at (1,2.5) {$=$};
\node at (-1,-2.5) {$=$};
\draw[thick, <->] (-1,-3.5) -- (3,-3.5);
\node at (1,-4) {$\rho$};
\draw[thick, <->] (1,3.5) -- (3,3.5);
\node at (2,4) {$h$};
\draw[thick, <->] (-3.5,-3) -- (-3.5,1);
\node at (-3.9,-1) {$\tau$};
\draw[thick, <->] (3.5,-3) -- (3.5,-1);
\node at (3.9,-2) {$v$};
\draw[thick,<->] (-0.6,-1.4) -- (1.4,0.6);
\node at (0.75,-0.75) {$m$};
\end{tikzpicture}
\end{center}
\caption{\emph{Toric diagram of the Calabi-Yau manifold $X_{1,1}$. The parameters $(\rho,\tau,m)$ are those appearing in the partition function with $\tau=v+m$ and $\rho=m+h$.}}
\label{Fig:WebToricCase1}
\end{figure}

In order to facilitate the computations in later part of this paper, we discuss the flop transition from the perspective of the brane web. The web diagram of $X_{1,1}$ is shown in \figref{Fig:WebToricCase1} with a choice of the $NM+2=1+2=3$ independent parameters $(h,v,m)$ (which correspond to the K\"ahler parameters of three $\mathbb{P}^1$'s) or, equivalently, $(\rho,\tau,m)$. The local geometry of all three $\mathbb{P}^1$'s is that of the resolved conifold and hence these can undergo flop transitions. 
Indeed, the second figure in Eq.(\ref{Flop11}) shows the 5-brane web after a flop of the curve corresponding to the parameter $m$. The last two figures of (\ref{Flop11}) give different ways of representing the web dual to the flopped geometry~\footnote{In the last step, we have performed an $SL(2,\mathbb{Z})$ transformation that acts in the following manner on the various lines: $
(1,0)\longrightarrow (1,1)$, $(0,1)\longrightarrow (0,1)$ and $(1,-1)\longrightarrow (1,0)$.}. Eq.(\ref{Flop11}) also indicates the change in the K\"ahler parameters caused by the flop transition: the K\"ahler parameters of all curves intersecting the curve that undergoes a flop transition change depending on the intersection number \cite{thomasyau}~\footnote{Notice, since the vertical and horizontal lines are identified, both $h$ and $v$ touch at both ends the line $m$ that is flopped (\emph{i.e.} they have intersection number $2$). This explains the shifts by $2m$ rather than simply $m$.}.
\begin{align}
\scalebox{0.8}{\parbox{3.1cm}{
\begin{tikzpicture}
\draw[ultra thick] (-1,0) -- (0,0);
\node at (-0.9,0.3) {{\footnotesize $h$}};
\draw[ultra thick] (0,-1) -- (0,0);
\node at (0.2,-0.9) {{\footnotesize $v$}};
\draw[ultra thick] (0,0) -- (1,1);
\node at (0.7,0.3) {{\footnotesize $m$}};
\draw[ultra thick] (1,1) -- (2,1);
\node at (1.9,1.3) {{\footnotesize $h$}};
\draw[ultra thick] (1,1) -- (1,2);
\node at (0.8,1.9) {{\footnotesize $v$}};
\node at (0,-0.5) {$=$};
\node at (1,1.5) {$=$};
\node at (-0.5,0) {$|$};
\node at (1.5,1) {$|$};
\end{tikzpicture}}
}
&&\Longrightarrow&&
\scalebox{0.87}{\parbox{3.8cm}{
\begin{tikzpicture}
\draw[ultra thick] (-1,0) -- (0,0);
\node at (-0.8,0.4) {{\footnotesize$h+2m$}};
\draw[ultra thick] (0,1) -- (0,0);
\node at (0.7,0.9) {{\footnotesize$v+2m$}};
\draw[ultra thick] (0,0) -- (1,-1);
\node at (0.1,-0.7) {{\footnotesize$-m$}};
\draw[ultra thick] (1,-1) -- (2,-1);
\node at (1.8,-0.6) {{\footnotesize $h+2m$}};
\draw[ultra thick] (1,-1) -- (1,-2);
\node at (0.3,-1.9) {{\footnotesize $v+2m$}};
\node at (1,-1.5) {$=$};
\node at (0,0.5) {$=$};
\node at (-0.5,0) {$|$};
\node at (1.5,-1) {$|$};
\end{tikzpicture}}
}
&&=&&
\scalebox{0.87}{\parbox{3.8cm}{
\begin{tikzpicture}
\draw[ultra thick] (-1,1) -- (0,0);
\node at (-0.9,0.3) {{\footnotesize $-m$}};
\draw[ultra thick] (0,0) -- (0,-1);
\node at (-0.7,-0.9) {{\footnotesize $v+2m$}};
\draw[ultra thick] (0,0) -- (1,0);
\node[rotate=90] at (0.5,0.7) {{\footnotesize $h+2m$}};
\draw[ultra thick] (1,0) -- (1,1);
\node at (1.7,0.9) {{\footnotesize $v+2m$}};
\draw[ultra thick] (1,0) -- (2,-1);
\node at (1.8,-0.3) {{\footnotesize $-m$}};
\node at (0,-0.5) {$=$};
\node at (1,0.5) {$=$};
\node[rotate=-45] at (-0.75,0.75) {$|$};
\node[rotate=-45] at (1.75,-0.75) {$|$};
\end{tikzpicture}}}
&&\Longrightarrow&&
\scalebox{0.87}{\parbox{3.3cm}{
\begin{tikzpicture}
\draw[ultra thick] (-1,0) -- (0,0);
\node at (-0.9,0.3) {{\footnotesize $-m$}};
\draw[ultra thick] (0,-1) -- (0,0);
\node at (0.7,-0.9) {{\footnotesize $v+2m$}};
\draw[ultra thick] (0,0) -- (1,1);
\node at (1.1,0.3) {{\footnotesize $h+2m$}};
\draw[ultra thick] (1,1) -- (2,1);
\node at (1.9,1.3) {{\footnotesize $-m$}};
\draw[ultra thick] (1,1) -- (1,2);
\node at (0.3,1.9) {{\footnotesize $v+2m$}};
\node at (0,-0.5) {$=$};
\node at (1,1.5) {$=$};
\node at (-0.5,0) {$|$};
\node at (1.5,1) {$|$};
\end{tikzpicture}
}}\label{Flop11}
\end{align}
Furthermore, in the first diagram on the left in Eq.(\ref{Flop11}), we have 
\begin{align}
&\rho=h+m\,,&&\tau=v+m\,,
\end{align}
while in the final diagram on the right of (\ref{Flop11}) we have
\begin{align}\label{FT1}
&\rho'=h+2m-m=\rho\,,\nonumber\\
&\tau'=v+h+4m=\rho+\tau+2m\,,\nonumber\\
&m'=h+2m=\rho+m\,.
\end{align}
The transformation $(\tau,\rho,m)\mapsto (\tau',\rho',m')$ corresponds to an automorphism of $X_{1,1}$ acting as an $Sp(2,\mathbb{Z})$ transformation. To understand this, we recall that the geometry mirror to $X_{1,1}$ is a genus-two curve with period matrix given by 
\begin{align}
\Omega=\left(\begin{array}{cc}\tau & m \\ m &\rho\end{array}\right)\,.
\end{align}
The automorphism group of this curve is $Sp(2,\mathbb{Z})$, which acts on the period matrix by the generalisation of the fractional linear transformations:
\begin{align}
&\Omega\mapsto (A\Omega+B)(C\Omega+D)^{-1}\,,&&\text{with} &&\left(\begin{array}{cc}A & B \\ C &D\end{array}\right)\in Sp(2,\mathbb{Z})\,.\label{DefOmega11}
\end{align}
Specifically, the matrices $A,B,C,D$ (with integer entries) satisfy
\begin{align}
&A^TD-C^TB=1\!\!1_{2\times 2}=DA^T-CB^T\,,&&A^TC=C^TA\,,&&B^TD=D^TB\,.\label{Sp2ZTrafos}
\end{align}
Using the transformation of the parameters under a flop transition given by Eq.(\ref{FT1}), we can arrange them into a transformed period matrices
\begin{align}
\Omega'=\left(\begin{array}{cc}\tau' & m' \\ m' &\rho'\end{array}\right)\,,
\end{align}
which is indeed related to Eq.(\ref{DefOmega11}) by
\begin{align}
\Omega'=(A\Omega+B)(C\Omega+D)^{-1}\,,
\end{align}
with
\begin{align}
&A=\left(\begin{array}{cc}1 & 1 \\ 0 &1\end{array}\right)\,,&&B=\left(\begin{array}{cc}0 & 0 \\ 0 &0\end{array}\right)\,,&&C=\left(\begin{array}{cc}0 & 0 \\ 0 &0\end{array}\right)\,,&&D=\left(\begin{array}{cc}1 & 0 \\ -1 &1\end{array}\right)\,.
\end{align}
These matrices satisfy Eq.(\ref{Sp2ZTrafos}),
as is required for a $Sp(2,\mathbb{Z})$ transformation~\footnote{Notice also $\text{det}\Omega=\rho\tau-m^2=\text{det}\Omega'$.} .This indicates that the combined flop transition Eq.(\ref{Flop11}) corresponds to an $Sp(2,\mathbb{Z})$ transformation. 

\subsubsection{Flop Transitions and Constraints}\label{Sect:FlopsConstraints}
\begin{figure}[h]
  \centering
\scalebox{0.89}{\parbox{19cm}{\begin{tikzpicture}[scale = 1]
\draw[ultra thick] (0,0) -- (1,1);
\draw[ultra thick] (1,1) -- (3,1) -- (3,3) -- (1,3) -- (1,1);
\draw[ultra thick] (3,1) -- (4,0);
\draw[ultra thick] (1,3) -- (0,4);
\draw[ultra thick] (3,3) -- (4,4);
\node at (2,0.5) {$L_1$};
\node at (2,3.5) {$L_1$};
\node at (0.5,2) {$L_2$};
\node at (3.5,2) {$L_2$};
\node at (2,-1) {(a)};
\draw[ultra thick] (8,2) -- (9,2) -- (9,3) -- (8,3) -- (8,2);
\draw[ultra thick] (6,1) -- (7,1);
\draw[ultra thick] (7,0) -- (7,1);
\draw[ultra thick] (7,1) -- (8,2);
\node at (7.1,1.7) {$E_1$};
\draw[ultra thick] (9,2) -- (10,1);
\draw[ultra thick] (10,1) -- (11,1);
\draw[ultra thick] (10,1) -- (10,0);
\draw[ultra thick] (7,1) -- (8,2);
\node at (9.9,1.7) {$-E_4$};
\draw[ultra thick] (9,3) -- (10,4);
\draw[ultra thick] (10,4) -- (11,4);
\draw[ultra thick] (10,4) -- (10,5);
\node at (7.1,3.3) {$-E_3$};
\draw[ultra thick] (8,3) -- (7,4);
\draw[ultra thick] (7,4) -- (6,4);
\draw[ultra thick] (7,4) -- (7,5);
\node at (9.9,3.3) {$E_2$};
\node at (8.5,1.7) {$L_1$};
\node at (8.5,3.3) {$L_1$};
\node at (7.7,2.5) {$L_2$};
\node at (9.3,2.5) {$L_2$};
\node at (8.5,-1) {(b)};
\draw[ultra thick] (15,2) -- (16,2) -- (17,3) -- (17,4) -- (16,4) -- (15,3) -- (15,2);
\draw[ultra thick] (13,1) -- (14,1);
\draw[ultra thick] (14,0) -- (14,1);
\draw[ultra thick] (14,1) -- (15,2); 
\draw[ultra thick] (16,2) -- (16,1);
\draw[ultra thick] (17,3) -- (18,3);
\draw[ultra thick] (17,4) -- (18,5);
\draw[ultra thick] (18,5) -- (19,5);
\draw[ultra thick] (18,5) -- (18,6);
\draw[ultra thick] (16,4) -- (16,5);
\draw[ultra thick] (15,3) -- (14,3);
\node at (14.1,1.7) {$E_1$};
\node at (17.9,4.3) {$E_2$};
\node at (15.1,3.7) {$E_3$};
\node at (16.9,2.3) {$E_4$};
\node at (17.8,3.5) {$L_2-E_4$};
\node at (14.1,2.5) {$L_2-E_3$};
\node[rotate=270] at (15.5,1.1) {$L_1-E_4$};
\node[rotate=270] at (16.5,4.9) {$L_1-E_3$};
\node at (16.5,-1) {(c)};
\end{tikzpicture}}}
  \caption{\emph{(a) Four-cycles of the type $\mathbb{P}^1\times \mathbb{P}^1$ glued together. (b) The four-cycles are blown up with curves $E_{1,2}$ and $-E_{3,4}$. (c) The geometry after a flop of the lines $-E_3$ and $-E_4$.}}
  \label{curves}
\end{figure}
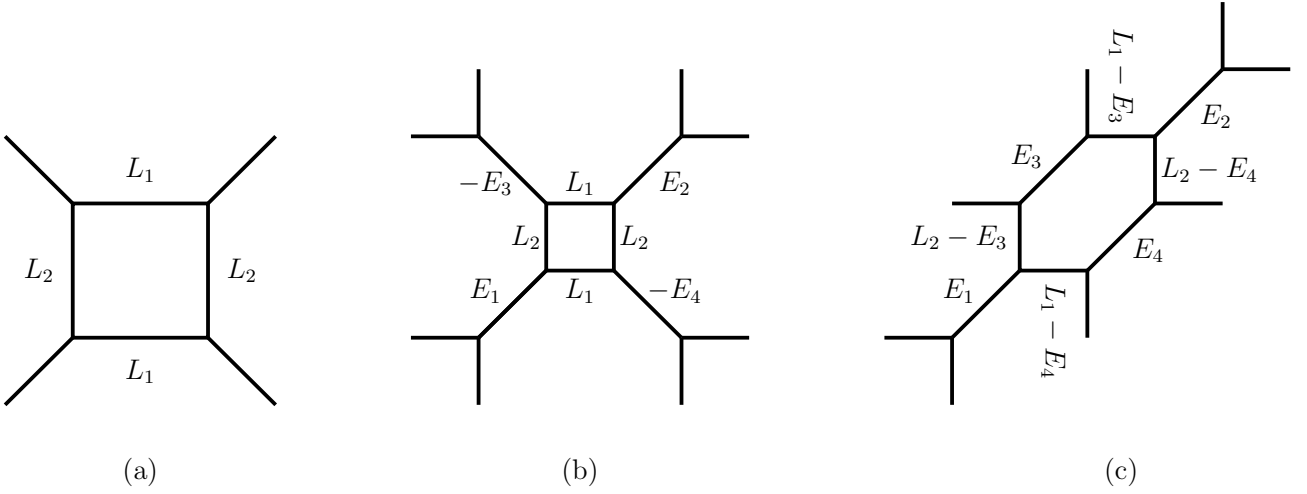

The orbifold $X_{N,M}$ has many curves which can undergo flop transitions. In Eq.(\ref{ExampleFlop}), we show the generic case, where a generic such curve in the web diagram undergoes a flop, and the way the K\"ahler parameters change: the transformation not only changes the sign of the K\"ahler parameter of the curve that undergoes a flop, but also modifies the K\"ahler parameter of any curve with non-trivial intersection with the former Eq.(\ref{ExampleFlop}).

\begin{align}
\scalebox{0.95}{\parbox{3cm}{
\begin{tikzpicture}
\draw[ultra thick] (-1,0) -- (0,0);
\node at (-0.8,0.3) {{\footnotesize $h_i$}};
\draw[ultra thick] (0,-1) -- (0,0);
\node at (0.3,-0.9) {{\footnotesize $v_i$}};
\draw[ultra thick] (0,0) -- (1,1);
\node at (0.7,0.3) {{\footnotesize $m$}};
\draw[ultra thick] (1,1) -- (2,1);
\node at (1.9,1.3) {{\footnotesize $h_j$}};
\draw[ultra thick] (1,1) -- (1,2);
\node at (0.7,1.9) {{\footnotesize $v_j$}};
\end{tikzpicture}}
}
&&\Longrightarrow&&
\scalebox{0.95}{\parbox{3.8cm}{
\begin{tikzpicture}
\draw[ultra thick] (-1,0) -- (0,0);
\node at (-0.8,0.3) {{\footnotesize$h_i+m$}};
\draw[ultra thick] (0,1) -- (0,0);
\node at (0.7,0.9) {{\footnotesize$v_j+m$}};
\draw[ultra thick] (0,0) -- (1,-1);
\node at (0.1,-0.7) {{\footnotesize$-m$}};
\draw[ultra thick] (1,-1) -- (2,-1);
\node at (1.8,-0.7) {{\footnotesize $h_j+m$}};
\draw[ultra thick] (1,-1) -- (1,-2);
\node at (0.3,-1.9) {{\footnotesize $v_i+m$}};
\end{tikzpicture}}
}\label{ExampleFlop}
\end{align}
We can apply the result of this generic transformation to more complex cases. For example, the Calabi-Yau threefolds $X_{N,M}$ are non-compact with a collection of four-cycles forming the compact part. These four-cycles $S_{a}$ are all $\mathbb{P}^1\times \mathbb{P}^1$ blown up at four points glued to each other, as depicted in \figref{curves}.
The six curves $(E_1,E_2,E_3,E_4,L_1,L_2)$ shown in \figref{curves} are all exceptional curves which are locally ${\cal O}(-1) \oplus { \cal O}(-1) \mapsto  \mathbb{P}^1$. They have the following intersection form 
\begin{align}
&L_{i}\cdot L_{j}=1-\delta_{ij}\,,&&L_{i}\cdot E_{a}=0\,, &&E_{a}\cdot E_{b}=-\delta_{ab}\,,&&\text{for} &&\begin{array}{l} i=1,2 \\  a,b=1,2,3\,.\end{array}
\end{align}
After the blow-ups, the areas of the curves can be represented as follows
 \begin{align}
&A(E_{a})=m_{a} \,,&&A(L_{1}-E_{4})=h_{1}\,,&&A(L_{1}-E_{3})=h_{2}\,,\nonumber\\
& &&A(L_{2}-E_{3})=v_{1}\,,&&A(L_{2}-E_{4})=v_{2}\,, \label{areas}
 \end{align}
for which the corresponding K\"ahler form is given by
\bea
\omega=(v_{1}+m_{3})L_{1}+(h_{1}+m_{4})L_{2}-m_{1}E_{1}-m_{2}E_{2}-m_{3} E_{3}-m_{4} E_{4}\,.
\eea
We can check that the above K\"ahler form gives areas for the curves $E_{1,2,3,4}$ and $L_{1}-E_{4}$ and $L_{2}-E_{3}$ that are consistent with Eq.(\ref{areas}). However, the areas of $L_{1}-E_{3}$ and $L_{2}-E_{4}$ under this K\"ahler form give two constraints:
\begin{align}
&h_{2}=h_{1}+m_{4}-m_{3}\,,&&\text{and} &&v_{2}=v_{1}+m_{3}-m_{4}\,. \label{c1}
\end{align}
The presence of two constraints is consistent with the fact that there are eight curves in \figref{curves}, of which only six are independent. The canonical class of $S_a$ is given by
\bea
-K_{S_a}=2L_{1}+2L_{2}-E_{1}-E_{2}-E_{3}-E_{4}\,,
\eea
and its area is given by
\begin{align}
s_{a}&=\omega \cdot (-K_{S_a})=2(h_{1}+m_{4}+v_{1}+m_{3})-m_{1}-m_{2}-m_{3}-m_{4}\nonumber\\
&=2h_{1}+2v_{1}+m_{3}+m_{4}-m_{1}-m_{2}\nonumber\\
&=h_{1}+h_{2}+v_{1}+v_{2}+m_{3}+m_{4}-m_{1}-m_{2}\,.
\end{align}
Here, we have used the constraint Eq.(\ref{c1}) to simplify the expression in the second line. The area of the canonical class remains the same as the various curves in the geometry undergo flop transitions. 

\begin{figure}[hbtp]
  \centering
  \includegraphics[width=5in]{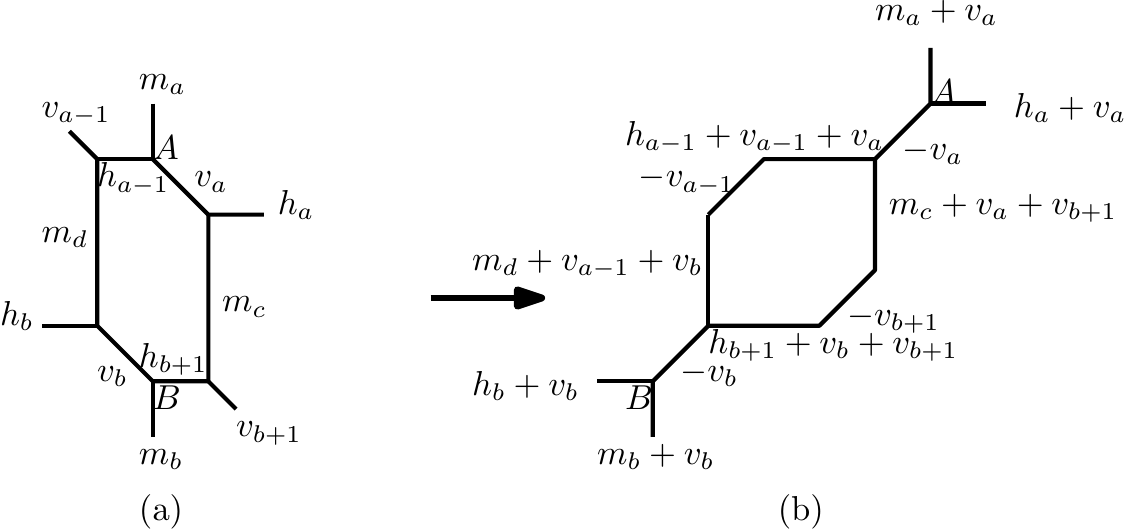}
 \caption{\sl Change of parameters and vertices $A, B$ after a flop transition. }\label{Fig:flop}
\end{figure}

Finally, another loop of cycles that is needed for the computations later on is shown in \figref{Fig:flop}, which corresponds to a four-cycle $S$ in the Calabi-Yau threefold, which is $\mathbb{P}^1\times \mathbb{P}^1$ blown up at four points.  
The loop corresponds to the curve class which is the anti-canonical class of $S$, whose area is given by
\bea
A:=v_{a}+m_{c}+v_{b+1}+h_{b+1}+v_{b}+m_{d}+v_{a-1}+h_{a-1}\,.
\eea
After a flop transition, the area of the curves in $S$ changes but the area of the canonical class remains the same
\begin{align}
A'&=-v_{a}+(m_{c}+v_{a}+v_{b+1})-v_{b+1}+(h_{b+1}+v_{b}+v_{b+1})-v_{b}+(m_{d}+v_{a-1}+v_{b})\\\nonumber
&-v_{a-1}+(h_{a-1}+v_{a-1}+v_{a})=A\,.
\end{align}

\subsection{Calabi-Yau Threefolds and Extended Moduli Space}\label{Sect:ExtendedModuliSpace}
In the previous section, we discussed flop transition from the point of view of Calabi-Yau manifolds and the associated brane webs. To understand the physical implications of such transitions, we recall that propagating strings see the underlying geometry in a very different way than point particles. This extended nature of strings is responsible for physically smooth processes which lead to topology changes of the geometry. In \cite{Reid}, Reid conjectured (this conjecture is known as \emph{Reid's fantasy}) that all Calabi-Yau threefolds are connected to each other through singular, perhaps non-K\"ahler, manifolds. Many families of Calabi-Yau threefolds are now known to be connected by conifold and flop transitions, as discussed above \cite{CT1,CT2,CT3,CT4}. The case of the conifold transition is well understood from string theory point of view that it requires non-perturbative effects to make the physics non-singular \cite{strominger,Greene:1995hu}. The conifold transition also leads to a change in the Hodge numbers of the Calabi-Yau threefold \cite{Greene:1995hu}.

\begin{figure}[h]
  \centering
  \includegraphics[width=7cm]{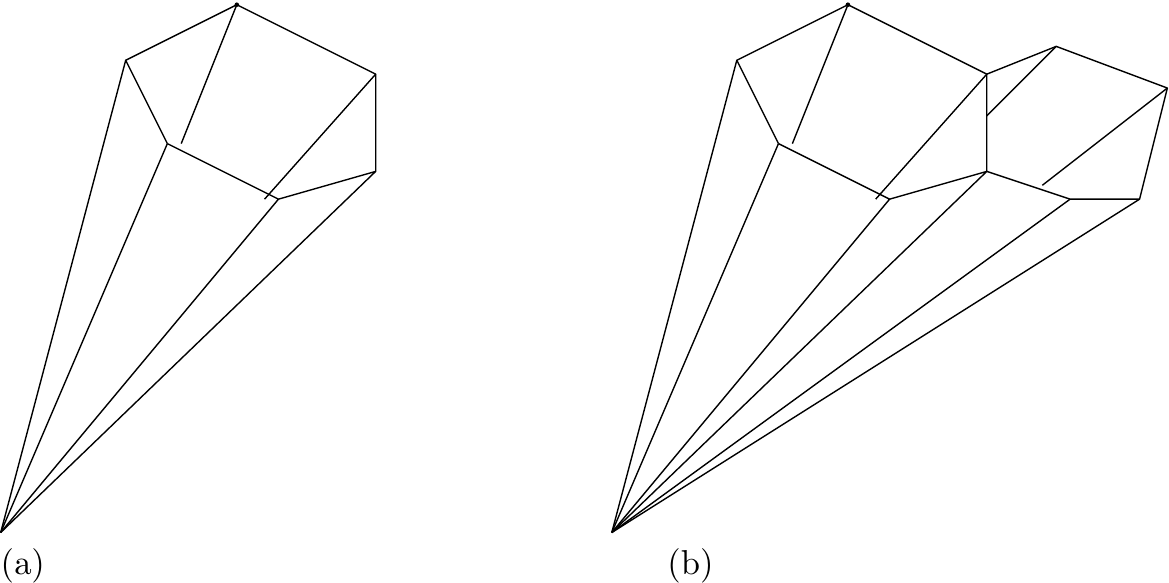}\\
  \caption{\sl (a) K\"ahler cone of a CY3fold. (b) Extended K\"ahler cone connecting two distinct CY3folds.}\label{KMS}
\end{figure}

To understand these transitions from the point of view of the K\"ahler moduli space of K\"ahler threefolds, we recall that the latter has the structure of a cone given by:
\begin{align}
&\int_{X} \omega \wedge \omega \wedge\omega>0\,,&&\int_{P_a} \omega \wedge \omega >0\,,&&\int_{\Sigma_i}\omega>0\,,
\end{align}
where $\omega$ is the K\"ahler form on $X$, $P_a$ are two-dimensional complex submanifolds and $\Sigma_i$ represent holomorphic curves in $X$.  As we discussed above, the walls of the K\"ahler cone are given by the shrinking of a rational curve with normal bundle ${\cal O}(-1)\oplus {\cal O}(-1)$ to zero size. Resolving this degeneration as described above leads again to a rational curve but in a Calabi-Yau threefold which is outside the K\"ahler cone we started with. In terms of the original K\"ahler form the new rational curve has negative area. Therefore, such flop transitions connects two Calabi-Yau threefolds with same hodge numbers  but different triple intersection numbers. By putting together the K\"ahler cones of the two Calabi-Yau threefolds, we get an extended moduli space, as shown in \figref{KMS}. In was shown recently that Calabi-Yau threefold related by flop transitions are connected by a path in the extended K\"ahler moduli space which consists of Ricci flat Calabi-Yau threefolds \cite{Rong:2010sf}.

There are many known examples of families of Calabi-Yau threefolds which connect to each other via flop transitions (and conifold transitions) and therefore fit to an extended K\"ahler moduli space \cite{CT1, Hubsch,Avram:1997rs, Candelas:2007ac,Candelas:2010ve}

\section{Partition Function: Relation between $(N,M)$ and $(1,1)$}\label{Sect:PartitionFunctions}
In this section, we review some results in the literature, which will turn out to be important in the following.
\subsection{Partition Functions}
In our previous works \cite{Hohenegger:2015btj,Hohenegger:2016eqy}, the topological string partition functions $\mathcal{Z}_{X_{N,M}}$ of the toric Calabi-Yau manifolds $X_{M,N}$ have been studied. The former can either be computed directly using the (refined) topological vertex formalism or alternatively as the M-string partition function: 

The partition function $\mathcal{Z}_{X_{N,M}}$ depends on the $MN+2$ independent  K\"ahler parameters of $X_{N,M}$ as well as two deformation parameters $\epsilon_{1,2}$. In \cite{Haghighat:2013tka,Haghighat:2013gba,Hohenegger:2013ala}, it has been computed for the particular case that the resolution of each intersection (called $m$) is the same. For the remaining parameters, the notation in Figure~\ref{Fig:WebToric} was used which appear in the following fashion
\begin{align}
&\bar{Q}_i=e^{-T_i}\,,&&\forall i=1,\ldots,M\,,\nonumber\\
&Q_a=e^{-t_a}\,,&&\forall a=1,\ldots,N\,,\nonumber\\
&Q_{ab}=Q_aQ_{a+1}\ldots Q_{b-1}\,,&&\text{for }1\leq a<b\leq N\, , 
\end{align}
as well as the two deformation parameters of the $\Omega$-background $\epsilon_{1,2}$ which appear through
\begin{align}
&q=e^{2\pi i\epsilon_1}\,&&\text{and} &&t=e^{-2\pi i\epsilon_2}\,.
\end{align}
With this notation, the partition function takes the form \cite{Hohenegger:2013ala}
\begin{align}
{\cal Z}_{X_{N,M}}&(\mathbf{t},\mathbf{T},m,\epsilon_{1,2})=W_N(\emptyset)^M\,\sum_{\alpha_a^{(i)}}Q_\tau ^{\sum_a|\alpha^{(M)}_a|}\,\left(\prod_{i=1}^M \overline{Q}_i^{\sum_a\left(\left|\alpha_a^{(i)}\right|-\left|\alpha_a^{(M)}\right|\right)}\right)\nonumber\\
&\times \prod_{i=1}^M\prod_{a=1}^N\frac{\vartheta_{\alpha_a^{(i+1)}\alpha_a^{(i)}}(Q_m;\rho)}{\vartheta_{\alpha_a^{(i)}\alpha_a^{(i)}}(\sqrt{t/q};\rho)}\,\prod_{1\leq a<b\leq N}\prod_{i=1}^M\frac{\vartheta_{\alpha_a^{(i)}\alpha_b^{(i+1)}}(Q_{ab}Q_m^{-1};\rho)\,\vartheta_{\alpha_a^{(i+1)}\alpha_b^{(i)}}(Q_{ab}Q_m;\rho)}{\vartheta_{\alpha_a^{(i)}\alpha_b^{(i)}}(Q_{ab}\sqrt{t/q};\rho)\,\vartheta_{\alpha_a^{(i)}\alpha_b^{(i)}}(Q_{ab}\sqrt{q/t};\rho)}\,,\label{FunctTopPartFctMN}
\end{align} 
with the prefactor
\begin{align}
&W_N(\emptyset;\mathbf{t},m,\epsilon_{1,2})=\lim_{\tau\to i\infty}{\cal Z}_{X_{N,1}}(\tau,\rho,t_{1},\cdots,t_{N-1},m,\epsilon_{1,2})\nonumber\\
&=\frac{1}{\prod_{n=1}^\infty(1-Q_\rho^n)}\prod_{i,j,k}\prod_{a,b=1}^N\frac{1-Q_\rho^{k-1}Q_{a,a+b}Q_m^{-1}t^{i-\frac{1}{2}}q^{j-\frac{1}{2}}}{1-Q_\rho^{k-1}Q_{a,a+b}\,t^{i-1}q^{j}}\,\frac{1-Q_\rho^{k-1}Q_{a,a+b-1}Q_m\, t^{i-\frac{1}{2}}q^{j-\frac{1}{2}}}{1-Q_\rho^{k-1}Q_{a,a+b}\,t^{i}q^{j-1}}\,.\nonumber\label{DefWNfact}
\end{align}
The $\alpha_a^{(i)}$ in (\ref{FunctTopPartFctMN}) are $N M$ independent integer partitions where, for a particular partition $\alpha_a^{(i)}=(\alpha_{a,1}^{(i)},\alpha_{a,2}^{(i)},\ldots,\alpha_{a,\ell}^{(i)})$ with total length $\ell$, 
\begin{align}
|\alpha_{a}^{(i)}|=\sum_{n=1}^\ell\alpha_{a,n}^{(i)}\,.
\end{align}
Finally, for any two integer partitions $\mu$ and $\nu$, the functions $\vartheta_{\mu\nu}$ in Eq.(\ref{FunctTopPartFctMN}) are defined as
\begin{align}
\vartheta_{\mu\nu}(x;\rho):=\prod_{(i,j)\in\mu}\vartheta\left(x^{-1}\,q^{-\mu_i+j-\tfrac{1}{2}}\,t^{-\nu_j^t+i-\tfrac{1}{2}};\rho\right)\,\prod_{(i,j)\in\nu}\vartheta\left(x^{-1}\,q^{\nu_i-j+\tfrac{1}{2}}\,t^{\mu_j^t-i+\tfrac{1}{2}};\rho\right)\,,
\end{align}
where (for $x=e^{2\pi iz}$)
\begin{align}
\vartheta(x;\rho):=\left(x^{\frac{1}{2}}-x^{-\frac{1}{2}}\right)\prod_{k=1}^{\infty}(1-x\,Q_\rho^k)(1-x^{-1}\,Q_\rho^k)=\frac{i\,\theta_1(\rho,x)}{Q_\rho^{1/8}\,\prod_{k=1}^\infty(1-Q_\rho^k)}\,.\label{DefVarthet}
\end{align}
Here, $\theta_1$ is the Jacobi theta function and we refer the reader to \cite{Eichler} for the proper definition along with several useful identities.

Finally, we can also define the BPS partition function that counts single particle BPS states of the little string theories associated with $X_{N,M}$
\begin{align}
\Sigma_{N,M}({\bf t},{\bf T},&m,\epsilon_{1,2})=\text{Plog}\,{\cal Z}_{X_{N,M}}(\tau,\rho,t_1,\ldots,t_{N-1},T_1,\ldots,T_{M-1},m,\epsilon_{1,2})\nonumber\\
&=\sum_{k=1}^{\infty}\frac{\mu(k)}{k}\,\text{log}{\cal Z}_{X_{N,M}}(k\tau,k\rho,k\,t_1,\ldots,k\,t_{N-1},k\,T_1,\ldots,k\,T_{M-1},k\,m,k\epsilon_{1,2})\,,\label{DefBPSCounting}
\end{align}
where $\text{Plog}$ is the plethystic logarithm and $\mu(k)$ is the M\"obius function. In \cite{Hohenegger:2016eqy}, we have studied the BPS counting functions (\ref{DefBPSCounting}) at the particular region in the full string moduli space
\begin{align}
Q_{1}=Q_{2}=\ldots=Q_{N}=Q_{\rho}^{\frac{1}{N}}&&\text{and} &&
\overline{Q}_{1}=\overline{Q}_{2}=\ldots =\overline{Q}_{M}=Q_{\tau}^{\frac{1}{M}}\,.\label{limit}
\end{align}
For a large class of examples, we found that, in the Nekrasov-Shatashvili limit $\epsilon_2\to 0$, 
\begin{align}
\lim_{\epsilon_2 \rightarrow 0}\,\epsilon_2\, \Sigma_{N, M}(\tfrac{\rho}{N}, \cdots, \tfrac{\rho}{N}, \tfrac{\tau}{M}, \cdots, \tfrac{\tau}{M}, m, \epsilon_{1,2}) = NM \, \lim_{\epsilon_{2 \rightarrow 0}} \,\epsilon_2\,\Sigma_{1,1} (\tfrac{\rho}{N}, \tfrac{\tau}{M}, m, \epsilon_{1,2})\,,\label{RelMNto11}
\end{align}
which we termed \emph{self-similarity} and which we conjectured to hold for generic $M,N\in\mathbb{N}$. As a consequence, in the Nekrasov-Shatashvili limit and for Eq.(\ref{limit}), the free energy of any configuration $(N,M)$ can be completely obtained from $\mathcal{Z}_{X_{1,1}}$. In Section 3.2, we will discuss ${\cal Z}_{X_{1,1}}$ in detail and show that it is invariant under $Sp(2,\mathbb{Z})$. It then follows from Eq.(\ref{RelMNto11}) that $Sp(2,\mathbb{Z})$ symmetry is inherited by the partition function of the $(N,M)$ case in the limit we discussed before. 
\subsection{The Case $X_{1,1}$}
The simplest configuration corresponds to the case $(N,M)=(1,1)$, which will play an important role throughout this work.
\subsubsection{The partition Function $\mathcal{Z}_{X_{1,1}}$}

As discussed in Section 2, the Calabi-Yau threefold $X_{1,1}$ has three K\"ahler parameters $(\tau,\rho,m)$. \figref{Fig:WebToricCase1} shows the three parameters in the dual 5-brane web diagram.  The topological string partition function $\mathcal{Z}_{X_{1,1}}$ depends on these three K\"ahler parameters and is given by \cite{Haghighat:2013gba,Hohenegger:2013ala,Hohenegger:2015btj}
\begin{align}
{\cal Z}_{X_{1,1}}&(\tau,\rho,m,\epsilon_{1,2})=W_1(\emptyset)\,\sum_{\alpha}Q_\tau ^{|\alpha|}\, \frac{\vartheta_{\alpha\alpha}(Q_m;\rho)}{\vartheta_{\alpha\alpha}(\sqrt{t/q};\rho)}\,,\label{FunctTopPartFctMN}
\end{align}
with the same notation as in (\ref{FunctTopPartFctMN}), \emph{i.e.} 
\begin{align}
&Q_\tau=e^{2\pi i\tau}\,,&&Q_\rho=e^{2\pi i\rho}\,,&&q=e^{2\pi i\epsilon_1}\,,&&t=e^{-2\pi i\epsilon_2}\,,&&Q_m=e^{2\pi i m}\,.
\end{align}
The prefactor (\ref{DefWNfact}) for $N=1$ takes the form 
\begin{align}
W_1(\emptyset;\rho,m,\epsilon_{1,2})&=\lim_{\tau\to i\infty}{\cal Z}_{X_{1,1}}(\tau,\rho,m,\epsilon_{1,2})\nonumber\\
&=\frac{1}{\prod_{n=1}^\infty(1-Q_\rho^n)}\prod_{i,j,k}\frac{1-Q_\rho^{k}Q_m^{-1}t^{i-\frac{1}{2}}q^{j-\frac{1}{2}}}{1-Q_\rho^{k}\,t^{i-1}q^{j}}\,\frac{1-Q_\rho^{k-1}Q_m\, t^{i-\frac{1}{2}}q^{j-\frac{1}{2}}}{1-Q_\rho^{k}\,t^{i}q^{j-1}}\,.\label{DefW1fact}
\end{align}
As was discussed in \cite{Hollowood:2003cv,Li:2004ef} (see also \cite{Dijkgraaf:1996xw}), the expansion of $\mathcal{Z}_{X_{1,1}}$ in a power series in $Q_\tau$ can be written in terms of the equivariant elliptic genus of the Hilbert scheme of points,
\begin{align}
\mathcal{Z}_{X_{1,1}}(\tau,\rho,m,\epsilon_{1,2})=W_1(\emptyset,\rho,m,\epsilon_{1,2})\sum_{k=0}^\infty Q_\tau^k\chi_{\text{ell}}(\text{Hilb}^k[\mathbb{C}^2])=\prod_{n,k,\ell,r,s}\left(1-Q_\rho^n Q_\tau^k Q_m^\ell q^r t^s\right)^{c(nk,\ell,r,s)}\,,\label{ExpansionZ11}
\end{align}
where the coefficients $c$ are obtained from a Fourier expansion of the equivariant elliptic genus of $\mathbb{C}^2$
\begin{align}
\chi_{\text{ell}}(\mathbb{C}^2)=\sum_{n,\ell,r,s}c(n,\ell,r,s)\,Q^n_\rho Q_m^\ell q^r t^s\,.
\end{align}
Notice that Eq.(\ref{ExpansionZ11}) is manifestly invariant under the exchange $\rho\longleftrightarrow \tau$ since $c$ only depends on the product $nk$. This is compatible with a particular Hecke structure \cite{Hohenegger:2015cba} found for the associated free energy $\Sigma_{1,1}$.

Furthermore, as was discussed in \cite{Hohenegger:2013ala}, the partition function $\mathcal{Z}_{X_{1,1}}$ can be written as a sum over the Narain lattice $\Gamma_{3,2}$, which is manifestly invariant under $SO(3,2,\mathbb{Z})$ \footnote{It was shown in \cite{Kawai:1997em} that the generating function of the elliptic genus of Hilbert scheme of points on K3 can be expressed as a sum over the Narain lattice $\Gamma_{3,2}$ and is given in terms of the weight ten Igusa cusp form. The two are not unrelated as the genus-one amplitude of ${\cal Z}_{X_{1,1}}$ is precisely the generating function studied in \cite{Dijkgraaf:1996xw,Kawai:1997em}. This is not a coincidence, as to be explained in \cite{toappear}.}. Since the latter is homeomorphic to $Sp(2,\mathbb{Z})$ as recalled below, the partition function $\mathcal{Z}_{X_{1,1}}$ is invariant under the latter. As a consequence of Eq.(\ref{RelMNto11}), this property is also inherited by $\mathcal{Z}_{X_{N,M}}$ in the NS-limit and Eq.(\ref{limit}).

Recall that an element of $\mathbb{R}^{3,2}$ that transform linearly under $SO(3,2, \mathbb{Z})$ can be mapped to a $(4 \times 4)$ matrix that transform  by conjugation under $Sp(2, \mathbb{Z})$. Denote 
\begin{equation}
\vec{a}=(a_{1},a_{2},a_{3},a_{4},a_{5})\in \mathbb{R}^{3,2}
\quad \mbox{and} \quad  M(\vec{a})=\frac{1}{2}\left(\begin{array}{cccc}a_{1} & a_{2}-a_{3}& 0& -a_{4}+a_{5} \\ a_{2}+a_{3} &-a_{1}&a_{4}-a_{5}&0\\
0&-a_{4}-a_{5}&a_{1}&a_{2}+a_{3}\\a_{4}+a_{5}&0&a_{2}-a_{3}&-a_{1}\end{array}\right).
\end{equation}
Then, the $SO(3,2,\mathbb{Z})$ action on $\vec{a}$ is realized as an $Sp(2,\mathbb{Z})$ action on $M$ given by $g\,M\,g^{-1}$, where $g\in Sp(2,\mathbb{Z})$ with invariant inner product given by Tr$(M(\vec{a})M(\vec{b}))=a_{1}b_{1}+a_{2}b_{2}-a_{3}b_{3}-a_{4}b_{4}+a_{5}b_{5}$.

\section{Mapping $(N,M)$ to $(\tfrac{MN}{k},k)$: Examples}\label{Sect:GenericMapMN}
In this section, we compile evidences that the Calabi-Yau threefolds $X_{N,M}$ and $X_{\frac{NM}{k},k}$ are related through a series of flop transitions.
\subsection{Relations between Newton Polygons}\label{Sect:RelNewtonPoly}
 Before presenting a series of examples, we first give a general argument to motivate that such a relation ought to exists. To this end, we consider the Newton polygon of $X_{N,M}$ tiling the plane $\mathbb{R}^2$. We start from the origin and follow a diagonal lines until we reach a point that is equivalent to the origin (see Figure~\ref{tile2Lines}).
\begin{figure}[h]
  \centering
\begin{tikzpicture}[scale=1.5]
\draw[-](-0.5,0) -- (5,0);
\draw[-] (0,-0.5) -- (0,4);
\draw[ultra thick] (0,0) -- (4.5,0) -- (4.5,3) -- (0,3) -- (0,0);
\draw[-] (0,0.5) -- (4.5,0.5);
\draw[ultra thick,-] (0,1) -- (4.5,1);
\draw[-] (0,1.5) -- (4.5,1.5);
\draw[ultra thick,-] (0,2) -- (4.5,2);
\draw[-] (0,2.5) -- (4.5,2.5);
\draw[-] (0.5,0) -- (0.5,3);
\draw[-] (1,0) -- (1,3);
\draw[ultra thick,-] (1.5,0) -- (1.5,3);
\draw[-] (2,0) -- (2,3);
\draw[-] (2.5,0) -- (2.5,3);
\draw[ultra thick,-] (3,0) -- (3,3);
\draw[-] (3.5,0) -- (3.5,3);
\draw[-] (4,0) -- (4,3);
\draw[-] (0,2.5) -- (0.5,3);
\draw[-] (0,2) -- (1,3);
\draw[-] (0,1.5) -- (1.5,3);
\draw[-] (0,1) -- (2,3);
\draw[-] (0,0.5) -- (2.5,3);
\draw[-] (0,0) -- (3,3);
\draw[-] (0.5,0) -- (3.5,3);
\draw[-] (1,0) -- (4,3);
\draw[-] (1.5,0) -- (4.5,3);
\draw[-] (2,0) -- (4.5,2.5);
\draw[-] (2.5,0) -- (4.5,2);
\draw[-] (3,0) -- (4.5,1.5);
\draw[-] (3.5,0) -- (4.5,1);
\draw[-] (4,0) -- (4.5,0.5);
\node at (-0.7,1.5) {$\Large \cdots$};
\node at (5.2,1.5) {$\Large \cdots$};
\node at (2.25,4) {$\Large \vdots$};
\node at (2.25,-0.5) {$\Large \vdots$};
%
\draw[red, ultra thick] (0,0) -- (3,3);
\node[red] at (0,0) {$\bullet$};
\node[red] at (0.5,0.5) {$\bullet$};
\node[red] at (1,1) {$\bullet$};
\node[red] at (1.5,1.5) {$\bullet$};
\node[red] at (2,2) {$\bullet$};
\node[red] at (2.5,2.5) {$\bullet$};
\node[red] at (3,3) {$\bullet$};
\node at (3,3.3) {$(\text{lcm}(M,N),\text{lcm}(M,N))$};
\draw[<->] (0.05,-0.3) -- (1.5,-0.3);
\node at (0.75,-0.5) {$N$};
\draw[<->] (-0.3,0.05) -- (-0.3,1);
\node at (-0.5,0.5) {$M$};
\end{tikzpicture}  
 \caption{\emph{Newton polygon of $X_{N,M}$ (with $\text{gcd}(M,N)=1$) and a path following only diagonal lines. The two points at $(0,0)$ and $(\text{lcm}(M,N),\text{lcm}(M,N))$ are equivalent. Furthermore, since $\text{gcd}(M,N)=1$, this path maps out all inequivalent points of the polygon.} }
 \label{tile2Lines}
\end{figure}

\begin{figure}[h]
  \centering
\begin{tikzpicture}[scale=1.5]
\draw[-](-0.5,0) -- (9.5,0);
\draw[-] (0,-0.5) -- (0,6.5);
\draw[ultra thick] (0,0) -- (6,0) -- (6,2) -- (0,2) -- (0,0);
\draw[ultra thick,-] (0,0) -- (9,0);
\draw[-] (0,0.5) -- (9,0.5);
\draw[-] (0,1) -- (9,1);
\draw[-] (0,1.5) -- (9,1.5);
\draw[ultra thick,-] (0,2) -- (9,2);
\draw[-] (0,2.5) -- (9,2.5);
\draw[-] (0,3) -- (9,3);
\draw[-] (0,3.5) -- (9,3.5);
\draw[ultra thick,-] (0,4) -- (9,4);
\draw[-] (0,4.5) -- (9,4.5);
\draw[-] (0,5) -- (9,5);
\draw[-] (0,5.5) -- (9,5.5);
\draw[ultra thick,-] (0,6) -- (9,6);
\draw[ultra thick,-] (0,0) -- (0,6);
\draw[-] (0.5,0) -- (0.5,6);
\draw[-] (1,0) -- (1,6);
\draw[-] (1.5,0) -- (1.5,6);
\draw[-] (2,0) -- (2,6);
\draw[-] (2.5,0) -- (2.5,6);
\draw[ultra thick,-] (3,0) -- (3,6);
\draw[-] (3.5,0) -- (3.5,6);
\draw[-] (4,0) -- (4,6);
\draw[-] (4.5,0) -- (4.5,6);
\draw[-] (5,0) -- (5,6);
\draw[-] (5.5,0) -- (5.5,6);
\draw[ultra thick,-] (6,0) -- (6,6);
\draw[-] (6.5,0) -- (6.5,6);
\draw[-] (7,0) -- (7,6);
\draw[-] (7.5,0) -- (7.5,6);
\draw[-] (8,0) -- (8,6);
\draw[-] (8.5,0) -- (8.5,6);
\draw[ultra thick,-] (9,0) -- (9,6);
\draw[-] (0,5.5) -- (0.5,6);
\draw[-] (0,5) -- (1,6);
\draw[-] (0,4.5) -- (1.5,6);
\draw[-] (0,4) -- (2,6);
\draw[-] (0,3.5) -- (2.5,6);
\draw[-] (0,3) -- (3,6);
\draw[-] (0,2.5) -- (3.5,6);
\draw[-] (0,2) -- (4,6);
\draw[-] (0,1.5) -- (4.5,6);
\draw[-] (0,1) -- (5,6);
\draw[-] (0,0.5) -- (5.5,6);
\draw[-] (0,0) -- (6,6);
\draw[-] (0.5,0) -- (6.5,6);
\draw[-] (1,0) -- (7,6);
\draw[-] (1.5,0) -- (7.5,6);
\draw[-] (2,0) -- (8,6);
\draw[-] (2.5,0) -- (8.5,6);
\draw[-] (3,0) -- (9,6);
\draw[-] (3.5,0) -- (9,5.5);
\draw[-] (4,0) -- (9,5);
\draw[-] (4.5,0) -- (9,4.5);
\draw[-] (5,0) -- (9,4);
\draw[-] (5.5,0) -- (9,3.5);
\draw[-] (6,0) -- (9,3);
\draw[-] (6.5,0) -- (9,2.5);
\draw[-] (7,0) -- (9,2);
\draw[-] (7.5,0) -- (9,1.5);
\draw[-] (8,0) -- (9,1);
\draw[-] (8.5,0) -- (9,0.5);
\node at (-0.7,3) {$\Large \cdots$};
\node at (9.5,3) {$\Large \cdots$};
\node at (4.5,6.5) {$\Large \vdots$};
\node at (4.5,-0.5) {$\Large \vdots$};
%
\draw[red, ultra thick] (0,0) -- (6,6);
\node[red] at (0.5,0.5) {$\bullet$};
\node[red] at (1,1) {$\bullet$};
\node[red] at (1.5,1.5) {$\bullet$};
\node[red] at (2,2) {$\bullet$};
\node[red] at (2.5,2.5) {$\bullet$};
\node[red] at (3,3) {$\bullet$};
\node[red] at (3.5,3.5) {$\bullet$};
\node[red] at (4,4) {$\bullet$};
\node[red] at (4.5,4.5) {$\bullet$};
\node[red] at (5,5) {$\bullet$};
\node[red] at (5.5,5.5) {$\bullet$};
\node[red] at (6,6) {$\bullet$};
\draw[ultra thick, blue] (0,0) -- (0.5,0) -- (6.5,6);
\node[red] at (0,0) {$\bullet$};
\node[blue] at (0.5,0) {$\bullet$};
\node[blue] at (1,0.5) {$\bullet$};
\node[blue] at (1.5,1) {$\bullet$};
\node[blue] at (2,1.5) {$\bullet$};
\node[blue] at (2.5,2) {$\bullet$};
\node[blue] at (3,2.5) {$\bullet$};
\node[blue] at (3.5,3) {$\bullet$};
\node[blue] at (4,3.5) {$\bullet$};
\node[blue] at (4.5,4) {$\bullet$};
\node[blue] at (5,4.5) {$\bullet$};
\node[blue] at (5.5,5) {$\bullet$};
\node[blue] at (6,5.5) {$\bullet$};
\node[blue] at (6.5,6) {$\bullet$};
%
\node[rotate=90] at (6,6.8) {$(\tfrac{NM}{k},\frac{NM}{k})$};
\node[rotate=90] at (6.5,7) {$(\tfrac{NM}{k}+1,\frac{NM}{k})$};
\draw[<->] (0.05,-0.3) -- (3,-0.3);
\node at (1.5,-0.5) {$N$};
\draw[<->] (-0.3,0.05) -- (-0.3,2);
\node at (-0.5,1) {$M$};
\end{tikzpicture}  
\caption{\emph{Newton polygon of $X_{N,M}$ (with $\text{gcd}(M,N)>2$) with two path following only diagonal lines. These two paths map out all inequivalent points of the Newton polygon.} }
 \label{tile3Lines}
\end{figure}
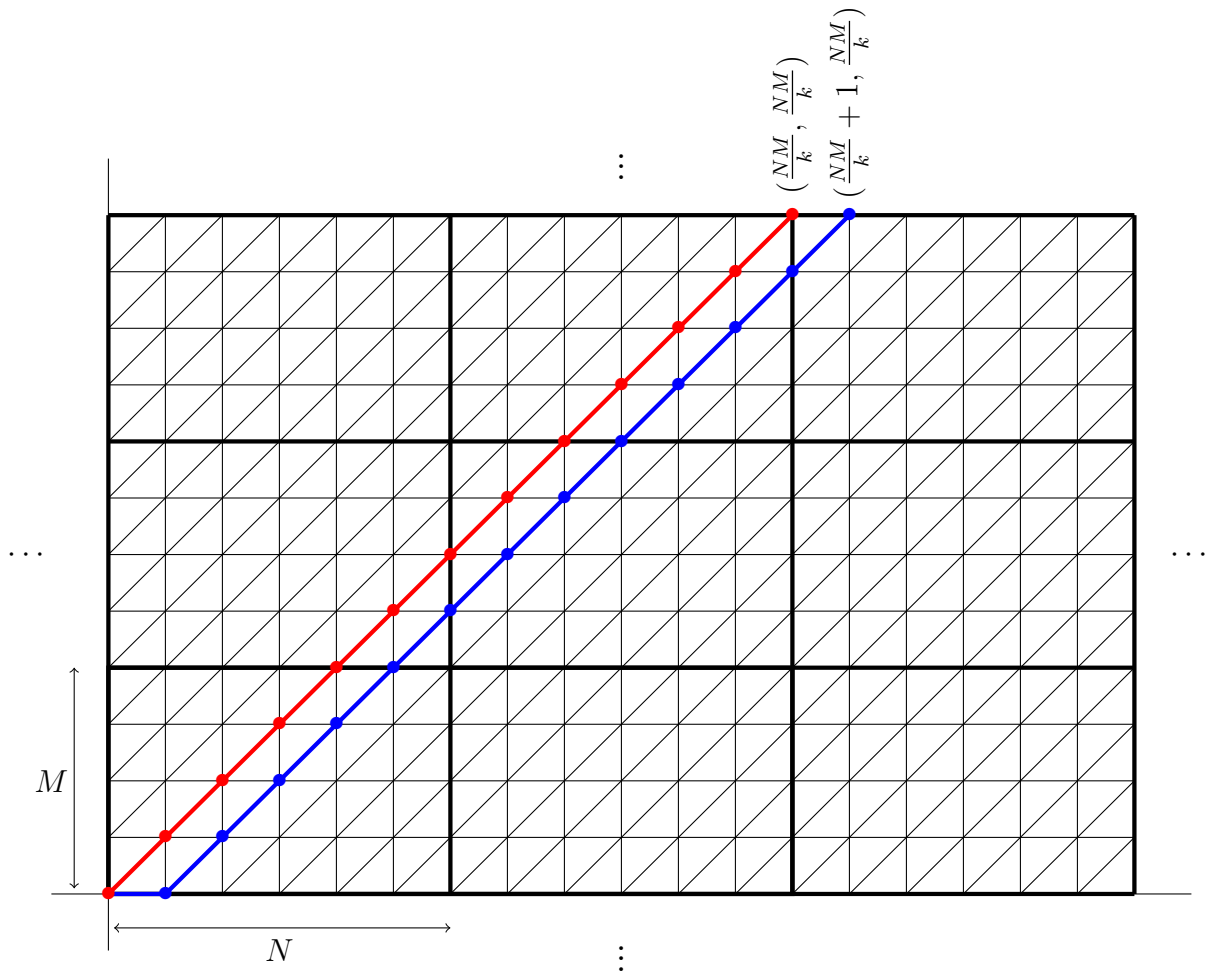

\noindent
This point has coordinates
\begin{align}
&(\text{lcm}(M,N),\text{lcm}(M,N))=\left(\tfrac{NM}{k},\tfrac{NM}{k}\right)\,,&&\text{with} &&k=\text{gcd}(M,N)\,.
\end{align}
If $k=1$ (as in the example in figure~\ref{tile2Lines}), this path has already mapped out all inequivalent points of the Newton polygon. If $k>1$, we can continue the procedure by following one horizontal line from the origin and tracing another diagonal path until we hit again an equivalent point. Following this procedure for $k$ steps maps out the entire Newton polygon (see Figure~\ref{tile3Lines}).
The fact that the Newton polygon can equivalently be mapped out by $k$ lines of length $\tfrac{NM}{k}$ suggests that it is equivalent to the Newton polygon of size $(\tfrac{NM}{k},k)$. In other words, it suggests that there exists a transformation which maps $X_{N,M}$ to $X_{\tfrac{NM}{k},k}$. In the following, we discuss numerous examples (using the toric webs of $X_{N,M}$) to show that such a transformation indeed exists.
\subsection{Relation between $(N,M)=(3,2)$ and $(6,1)$}
Our first example is the configuration $(N,M)=(3,2)$, which we show can be related to the case $(N,M)=(6,1)$. Since this is the simplest (non-trivial) example, we will work out this case in great detail.
\subsubsection{Starting Configuration $(N,M)=(3,2)$ and Consistency Conditions}
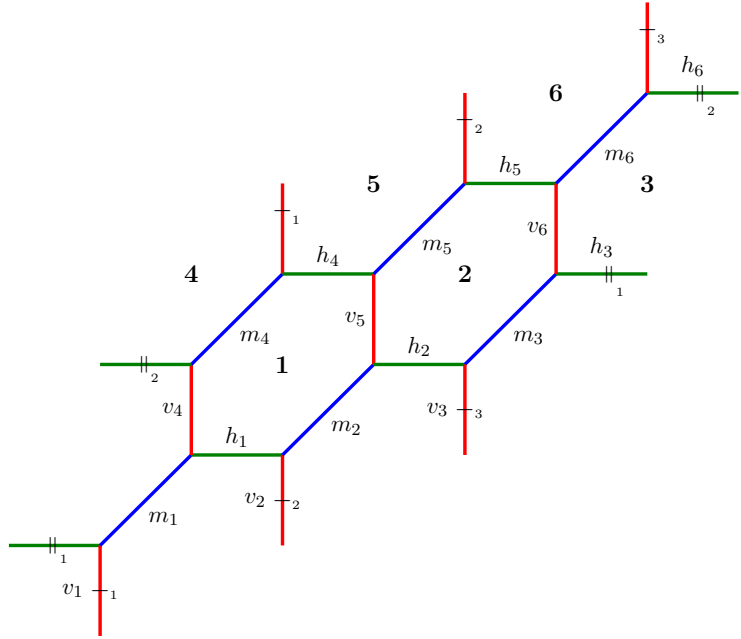
\begin{figure}[htb]
\begin{center}
\scalebox{0.8}{\parbox{12cm}{\begin{tikzpicture}[scale = 1.50]
\draw[ultra thick,green!50!black] (-6,0) -- (-5,0);
\draw[ultra thick,red] (-5,-1) -- (-5,0);
\draw[ultra thick,blue] (-5,0) -- (-4,1);
\draw[ultra thick,green!50!black] (-4,1) -- (-3,1);
\draw[ultra thick,red] (-4,1) -- (-4,2);
\draw[ultra thick,red] (-3,1) -- (-3,0);
\draw[ultra thick,blue] (-3,1) -- (-2,2);
\draw[ultra thick,blue] (-4,2) -- (-3,3);
\draw[ultra thick,green!50!black] (-5,2) -- (-4,2);
\draw[ultra thick,red] (-3,3) -- (-3,4);
\draw[ultra thick,green!50!black] (-3,3) -- (-2,3);
\draw[ultra thick,red] (-2,2) -- (-2,3);
\draw[ultra thick,green!50!black] (-2,2) -- (-1,2);
\draw[ultra thick,blue] (-2,3) -- (-1,4);
\draw[ultra thick,blue] (-1,2) -- (0,3);
\draw[ultra thick,red] (-1,2) -- (-1,1);
%
\draw[ultra thick,green!50!black] (-1,4) -- (0,4);
\draw[ultra thick,green!50!black] (0,3) -- (1,3);
\draw[ultra thick,red] (0,3) -- (0,4);
%
\draw[ultra thick,blue] (0,4) -- (1,5);
%
\draw[ultra thick,green!50!black] (1,5) -- (2,5);
%
\draw[ultra thick,red] (1,5) -- (1,6);
\draw[ultra thick,red] (-1,4) -- (-1,5);

%

%
%
%
%
\node[rotate=90] at (-5.5,0) {$=$};
\node at (-5.4,-0.15) {{\tiny$1$}};
\node[rotate=90] at (-4.5,2) {$=$};
\node at (-4.4,1.85) {{\tiny$2$}};
%
\node[rotate=90] at (0.6,3) {$=$};
\node at (0.7,2.8) {{\tiny$1$}};
\node[rotate=90] at (1.6,5) {$=$};
\node at (1.7,4.8) {{\tiny$2$}};
%
\node at (-5,-0.5) {$-$};
\node at (-4.85,-0.55) {{\tiny $1$}};
\node at (-3,0.5) {$-$};
\node at (-2.85,0.45) {{\tiny $2$}};
\node at (-1,1.5) {$-$};
\node at (-0.85,1.45) {{\tiny $3$}};
%
\node at (-3,3.7) {$-$};
\node at (-2.85,3.6) {{\tiny $1$}};
\node at (-1,4.7) {$-$};
\node at (-0.85,4.6) {{\tiny $2$}};
\node at (1,5.7) {$-$};
\node at (1.15,5.6) {{\tiny $3$}};
%
\node at (-4.3,0.3) {{\small $m_1$}};
\node at (-2.3,1.3) {{\small $m_2$}};
\node at (-0.3,2.3) {{\small $m_3$}};
\node at (-3.3,2.3) {{\small $m_4$}};
\node at (-1.3,3.3) {{\small $m_5$}};
\node at (0.7,4.3) {{\small $m_6$}};
\node at (-3.5,1.2) {{\small $h_1$}};
\node at (-1.5,2.2) {{\small $h_2$}};
\node at (0.5,3.3) {{\small $h_3$}};
\node at (-2.5,3.2) {{\small $h_4$}};
\node at (-0.5,4.2) {{\small $h_5$}};
\node at (1.5,5.3) {{\small $h_6$}};
\node at (-5.3,-0.5) {{\small $v_1$}};
\node at (-3.3,0.5) {{\small $v_2$}};
\node at (-1.3,1.5) {{\small $v_3$}};
\node at (-4.2,1.5) {{\small $v_4$}};
\node at (-2.2,2.5) {{\small $v_5$}};
\node at (-0.2,3.5) {{\small $v_6$}};
\node at (-3,2) {\bf 1};
\node at (-1,3) {\bf 2};
\node at (1,4) {\bf 3};
\node at (-4,3) {\bf 4};
\node at (-2,4) {\bf 5};
\node at (0,5) {\bf 6};
\end{tikzpicture}}}
\end{center}
\caption{\sl Parametrisation of the $(N,M)=(3,2)$ web diagram. The parameters $(h_i,v_i,m_i)$ with $i=1,\ldots,6$ are not all independent, but are subject to the conditions (\ref{InitC1}) -- (\ref{InitC6}).}
\label{Fig:WebToric23}
\end{figure}

The web diagram of the configuration $(N,M)=(3,2)$ is shown in Figure~\ref{Fig:WebToric23}
which is parametrised by 18 variables: $h_{i}$, $v_{i}$ and $m_i$ for $i=1,\ldots,6$. As discussed above, these parameters are not all independent. Indeed, for consistency, all lines of the same color in Figure~\ref{Fig:WebToric23} have to be parallel to each other. With the blue lines oriented along the direction $(1,1)$, these conditions can be formulated locally for each of the 6 hexagons (see also the discussion in section~\ref{Sect:FlopsConstraints})
\begin{itemize}
\item hexagon 1:
\begin{align}
&h_1+m_2=m_4+h_4\,,&&v_4+m_4=m_2+v_5\,,\label{InitC1}
\end{align}
\item hexagon 2:
\begin{align}
&h_2+m_3=m_5+h_5\,,&&v_5+m_5=m_3+v_6\,,\label{InitC2}
\end{align}
\item hexagon 3:
\begin{align}
&h_3+m_1=m_6+h_6\,,&&v_6+m_6=m_1+v_4\,,\label{InitC3}
\end{align}
\item hexagon 4:
\begin{align}
&h_6+m_4=m_3+h_3\,,&&v_3+m_3=m_4+v_1\,,\label{InitC4}
\end{align}
\item hexagon 5:
\begin{align}
&h_4+m_5=m_1+h_1\,,&&v_1+m_1=m_5+v_2\,,\label{InitC5}
\end{align}
\item hexagon 6:
\begin{align}
&h_5+m_6=m_2+h_2\,,&&v_2+m_2=m_6+v_3\,.\label{InitC6}
\end{align}
\end{itemize}
These equations leave 8 independent parameters. This agrees with the result found in \cite{Haghighat:2013gba} (see also \cite{Hohenegger:2016eqy}) that a generic $(M,N)$ web has $MN+2$ independent parameters.
\subsubsection{Duality Transformation}
\begin{figure}[htb]
\begin{center}
\scalebox{0.8}{\parbox{14cm}{\begin{tikzpicture}[scale = 1.5]
\draw[ultra thick] (-6,0) -- (-5,0);
\draw[ultra thick] (-5,-1) -- (-5,0);
\draw[ultra thick] (-5,0) -- (-4,1);
\draw[ultra thick] (-4,1) -- (-3,1);
\draw[ultra thick] (-4,1) -- (-4,2);
\draw[ultra thick] (-3,1) -- (-3,0);
\draw[ultra thick] (-3,1) -- (-2,2);
\draw[ultra thick] (-4,2) -- (-3,3);
\draw[ultra thick] (-5,2) -- (-4,2);
\draw[ultra thick] (-3,3) -- (-3,4);
\draw[ultra thick] (-3,3) -- (-2,3);
\draw[ultra thick] (-2,2) -- (-2,3);
\draw[ultra thick] (-2,2) -- (-1,2);
\draw[ultra thick] (-2,3) -- (-1,4);
\draw[ultra thick] (-1,2) -- (0,3);
\draw[ultra thick] (-1,2) -- (-1,1);

\draw[ultra thick] (-1,4) -- (0,4);
\draw[ultra thick] (0,3) -- (1,3);
\draw[ultra thick] (0,3) -- (0,4);
\draw[ultra thick] (0,4) -- (1,5);
\draw[ultra thick] (1,5) -- (2,5);
\draw[ultra thick] (1,5) -- (1,6);
\draw[ultra thick] (-1,4) -- (-1,5);
\node at (-6.2,0) {{\small \bf $a$}};
\node at (-5.2,2) {{\small \bf $b$}};
\node at (1.2,3) {{\small \bf $a$}};
\node at (2.2,5) {{\small \bf $b$}};
\node at (-5,-1.3) {{\small \bf $1$}};
\node at (-3,-0.3) {{\small \bf $2$}};
\node at (-1,0.7) {{\small \bf $3$}};
\node at (-3,4.3) {{\small \bf $1$}};
\node at (-1,5.3) {{\small \bf $2$}};
\node at (1,6.3) {{\small \bf $3$}};
\node at (-4.2,0.4) {{\small $m_1$}};
\node at (-2.2,1.4) {{\small $m_2$}};
\node at (-0.2,2.4) {{\small $m_3$}};
\node at (-3.2,2.4) {{\small $m_4$}};
\node at (-1.2,3.4) {{\small $m_5$}};
\node at (0.8,4.4) {{\small $m_6$}};
\node at (-3.5,1.2) {{\small $h_1$}};
\node at (-1.5,2.2) {{\small $h_2$}};
\node at (0.5,3.3) {{\small $h_3$}};
\node at (-2.5,3.2) {{\small $h_4$}};
\node at (-0.5,4.2) {{\small $h_5$}};
\node at (1.5,5.3) {{\small $h_6$}};
\node at (-5.3,-0.5) {{\small $v_1$}};
\node at (-3.3,0.5) {{\small $v_2$}};
\node at (-1.3,1.5) {{\small $v_3$}};
\node at (-4.2,1.5) {{\small $v_4$}};
\node at (-2.2,2.5) {{\small $v_5$}};
\node at (-0.2,3.5) {{\small $v_6$}};
\draw[dashed] (-6,2) -- (-3,-1);
\draw[dashed] (-5,4) -- (-1,0);
\draw[dashed] (-3,5) -- (1,1);
\draw[dashed] (-1,6) -- (2,3);
\end{tikzpicture}}
}
\end{center}
\caption{\sl Cutting the $(N,M)=(3,2)$ web diagram. The vertical lines $1,2,3$ and the horizontal lines $a,b$ are identified respectively.}
\label{Fig:Cutting23}
\end{figure}
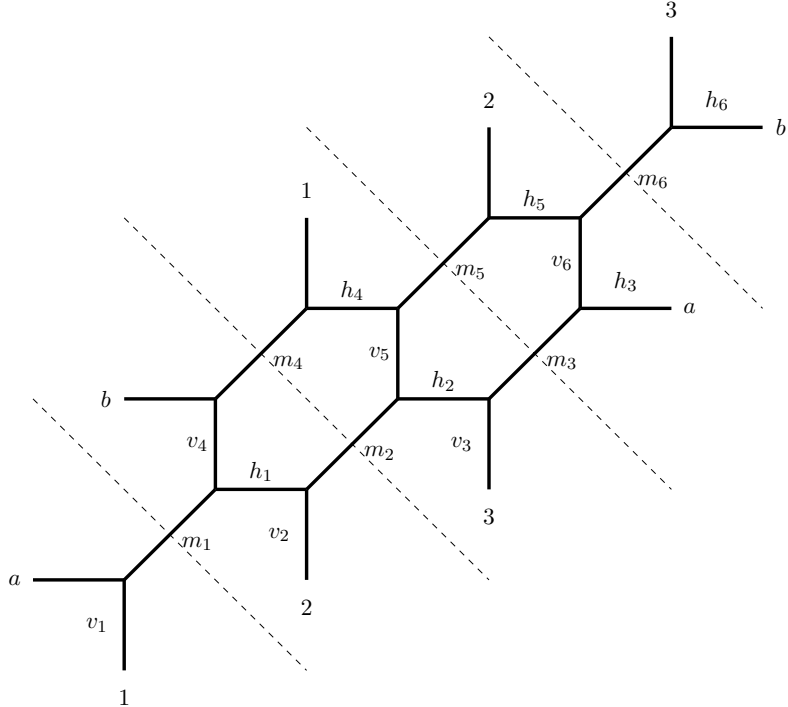

\begin{figure}[htbp]
\begin{center}
\scalebox{0.7}{\parbox{14cm}{\begin{tikzpicture}[scale = 1.75]
\draw[ultra thick] (0,0) -- (1,0);
\node at (0.5,0.2) {{\small$h_3$}};
\draw[ultra thick] (1,0) -- (1,-1);
\node at (0.8,-0.7) {{\small$v_1$}};
\draw[ultra thick] (1,-1) -- (2,-1);
\node at (1.5,-0.8) {{\small$h_4$}};
\draw[ultra thick] (2,-1) -- (2,-2);
\node at (1.8,-1.5) {{\small$v_5$}};
\draw[ultra thick] (2,-2) -- (3,-2);
\node at (2.5,-1.8) {{\small$h_2$}};
\draw[ultra thick] (3,-2) -- (3,-3);
\node at (2.8,-2.5) {{\small$v_3$}};
\draw[ultra thick] (3,-3) -- (4,-3);
\node at (3.5,-2.8) {{\small$h_6$}};
\draw[ultra thick] (4,-3) -- (4,-4);
\node at (3.8,-3.5) {{\small$v_4$}};
\draw[ultra thick] (4,-4) -- (5,-4);
\node at (4.5,-3.8) {{\small$h_1$}};
\draw[ultra thick] (5,-4) -- (5,-5);
\node at (4.8,-4.5) {{\small$v_2$}};
\draw[ultra thick] (5,-5) -- (6,-5);
\node at (5.5,-4.8) {{\small$h_5$}};
\draw[ultra thick] (6,-5) -- (6,-6);
\node at (5.8,-5.5) {{\small$v_6$}};
\draw[ultra thick] (6,-6) -- (7,-6);
\node at (6.5,-5.8) {{\small$h_3$}};
\draw[ultra thick] (1,0) -- (2,1);
\node at (1.4,0.7) {{\small $m_1$}};
\draw[ultra thick] (2,-1) -- (3,0);
\node at (2.4,-0.3) {{\small $m_5$}};
\draw[ultra thick] (3,-2) -- (4,-1);
\node at (3.4,-1.3) {{\small $m_3$}};
\draw[ultra thick] (4,-3) -- (5,-2);
\node at (4.4,-2.3) {{\small $m_4$}};
\draw[ultra thick] (5,-4) -- (6,-3);
\node at (5.4,-3.3) {{\small $m_2$}};
\draw[ultra thick] (6,-5) -- (7,-4);
\node at (6.4,-4.3) {{\small $m_6$}};
\draw[ultra thick] (1,-1) -- (0,-2);
\node at (0.3,-1.3) {{\small $m_4$}};
\draw[ultra thick] (2,-2) -- (1,-3);
\node at (1.3,-2.3) {{\small $m_2$}};
\draw[ultra thick] (3,-3) -- (2,-4);
\node at (2.3,-3.3) {{\small $m_6$}};
\draw[ultra thick] (4,-4) -- (3,-5);
\node at (3.3,-4.3) {{\small $m_1$}};
\draw[ultra thick] (5,-5) -- (4,-6);
\node at (4.3,-5.3) {{\small $m_5$}};
\draw[ultra thick] (6,-6) -- (5,-7);
\node at (5.3,-6.3) {{\small $m_3$}};
\draw[dashed] (0,-0.5) -- (2,-0.5);
\node at (-0.2,-2.2) {{\small \bf $1$}};
\node at (0.8,-3.2) {{\small \bf $2$}};
\node at (1.8,-4.2) {{\small \bf $3$}};
\node at (2.8,-5.2) {{\small \bf $4$}};
\node at (3.8,-6.2) {{\small \bf $5$}};
\node at (4.8,-7.2) {{\small \bf $6$}};
\node at (2.2,1.2) {{\small \bf $1$}};
\node at (3.2,0.2) {{\small \bf $2$}};
\node at (4.2,-0.8) {{\small \bf $3$}};
\node at (5.2,-1.8) {{\small \bf $4$}};
\node at (6.2,-2.8) {{\small \bf $5$}};
\node at (7.2,-3.8) {{\small \bf $6$}};
\node at (-0.2,0) {{\small \bf $a$}};
\node at (7.2,-6) {{\small \bf $a$}};
\end{tikzpicture}}
}
\end{center}
\caption{\sl Re-gluing of the $(2,3)$ web-diagram. The lines labelled $1,2,3,4,5,6$ and $a$ are identified, respectively. The dashed line signifies the cutting in order to obtain the web drawn in Figure~\ref{Fig:Diag61}.}
\label{Fig:Gluing23}
\end{figure}
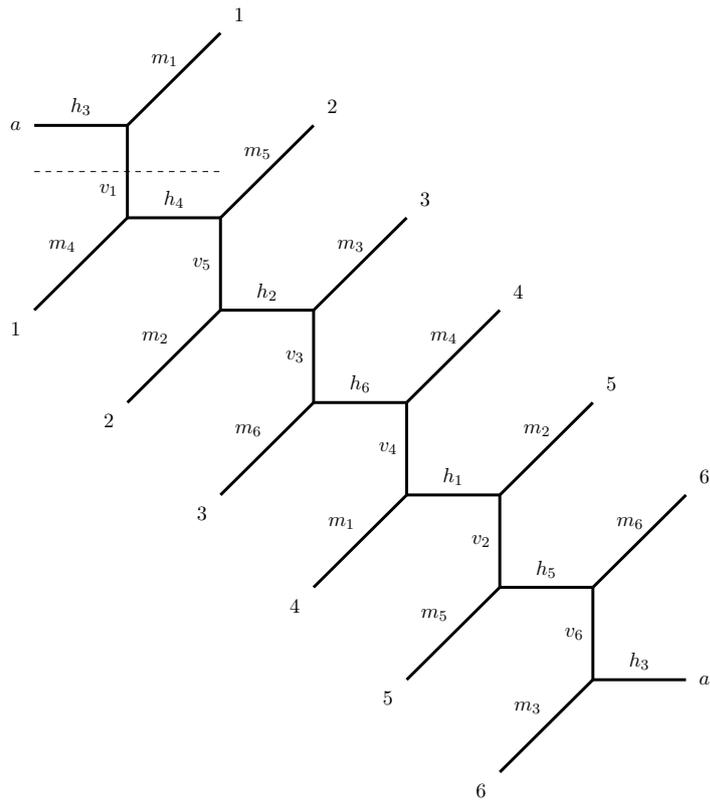

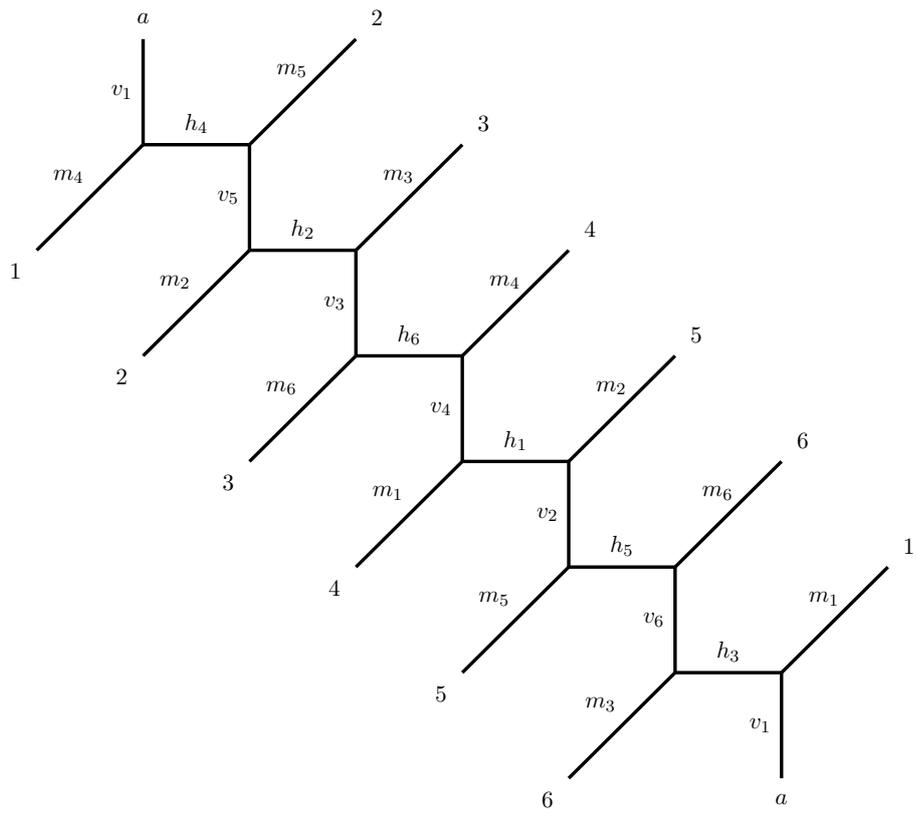
\begin{figure}[p]
\begin{center}
\scalebox{0.8}{\parbox{14cm}{\begin{tikzpicture}[scale = 1.75]
\draw[ultra thick] (1,0) -- (1,-1);
\node at (0.8,-0.5) {{\small$v_1$}};
\draw[ultra thick] (1,-1) -- (2,-1);
\node at (1.5,-0.8) {{\small$h_4$}};
\draw[ultra thick] (2,-1) -- (2,-2);
\node at (1.8,-1.5) {{\small$v_5$}};
\draw[ultra thick] (2,-2) -- (3,-2);
\node at (2.5,-1.8) {{\small$h_2$}};
\draw[ultra thick] (3,-2) -- (3,-3);
\node at (2.8,-2.5) {{\small$v_3$}};
\draw[ultra thick] (3,-3) -- (4,-3);
\node at (3.5,-2.8) {{\small$h_6$}};
\draw[ultra thick] (4,-3) -- (4,-4);
\node at (3.8,-3.5) {{\small$v_4$}};
\draw[ultra thick] (4,-4) -- (5,-4);
\node at (4.5,-3.8) {{\small$h_1$}};
\draw[ultra thick] (5,-4) -- (5,-5);
\node at (4.8,-4.5) {{\small$v_2$}};
\draw[ultra thick] (5,-5) -- (6,-5);
\node at (5.5,-4.8) {{\small$h_5$}};
\draw[ultra thick] (6,-5) -- (6,-6);
\node at (5.8,-5.5) {{\small$v_6$}};
\draw[ultra thick] (6,-6) -- (7,-6);
\node at (6.5,-5.8) {{\small$h_3$}};
\draw[ultra thick] (7,-6) -- (7,-7);
\node at (6.8,-6.5) {{\small$v_1$}};
\draw[ultra thick] (2,-1) -- (3,0);
\node at (2.4,-0.3) {{\small $m_5$}};
\draw[ultra thick] (3,-2) -- (4,-1);
\node at (3.4,-1.3) {{\small $m_3$}};
\draw[ultra thick] (4,-3) -- (5,-2);
\node at (4.4,-2.3) {{\small $m_4$}};
\draw[ultra thick] (5,-4) -- (6,-3);
\node at (5.4,-3.3) {{\small $m_2$}};
\draw[ultra thick] (6,-5) -- (7,-4);
\node at (6.4,-4.3) {{\small $m_6$}};
\draw[ultra thick] (7,-6) -- (8,-5);
\node at (7.4,-5.3) {{\small $m_1$}};
\draw[ultra thick] (1,-1) -- (0,-2);
\node at (0.3,-1.3) {{\small $m_4$}};
\draw[ultra thick] (2,-2) -- (1,-3);
\node at (1.3,-2.3) {{\small $m_2$}};
\draw[ultra thick] (3,-3) -- (2,-4);
\node at (2.3,-3.3) {{\small $m_6$}};
\draw[ultra thick] (4,-4) -- (3,-5);
\node at (3.3,-4.3) {{\small $m_1$}};
\draw[ultra thick] (5,-5) -- (4,-6);
\node at (4.3,-5.3) {{\small $m_5$}};
\draw[ultra thick] (6,-6) -- (5,-7);
\node at (5.3,-6.3) {{\small $m_3$}};
\node at (-0.2,-2.2) {{\small \bf $1$}};
\node at (0.8,-3.2) {{\small \bf $2$}};
\node at (1.8,-4.2) {{\small \bf $3$}};
\node at (2.8,-5.2) {{\small \bf $4$}};
\node at (3.8,-6.2) {{\small \bf $5$}};
\node at (4.8,-7.2) {{\small \bf $6$}};
\node at (3.2,0.2) {{\small \bf $2$}};
\node at (4.2,-0.8) {{\small \bf $3$}};
\node at (5.2,-1.8) {{\small \bf $4$}};
\node at (6.2,-2.8) {{\small \bf $5$}};
\node at (7.2,-3.8) {{\small \bf $6$}};
\node at (8.2,-4.8) {{\small \bf $1$}};
\node at (1,0.2) {{\small \bf $a$}};
\node at (7,-7.2) {{\small \bf $a$}};
\end{tikzpicture}}
}
\end{center}
\caption{\sl Re-gluing of the web-diagram in figure~\ref{Fig:Diag61}. The lines labelled $1,2,3,4,5,6$ and $a$ are identified respectively.}
\label{Fig:Diag61}
\end{figure}

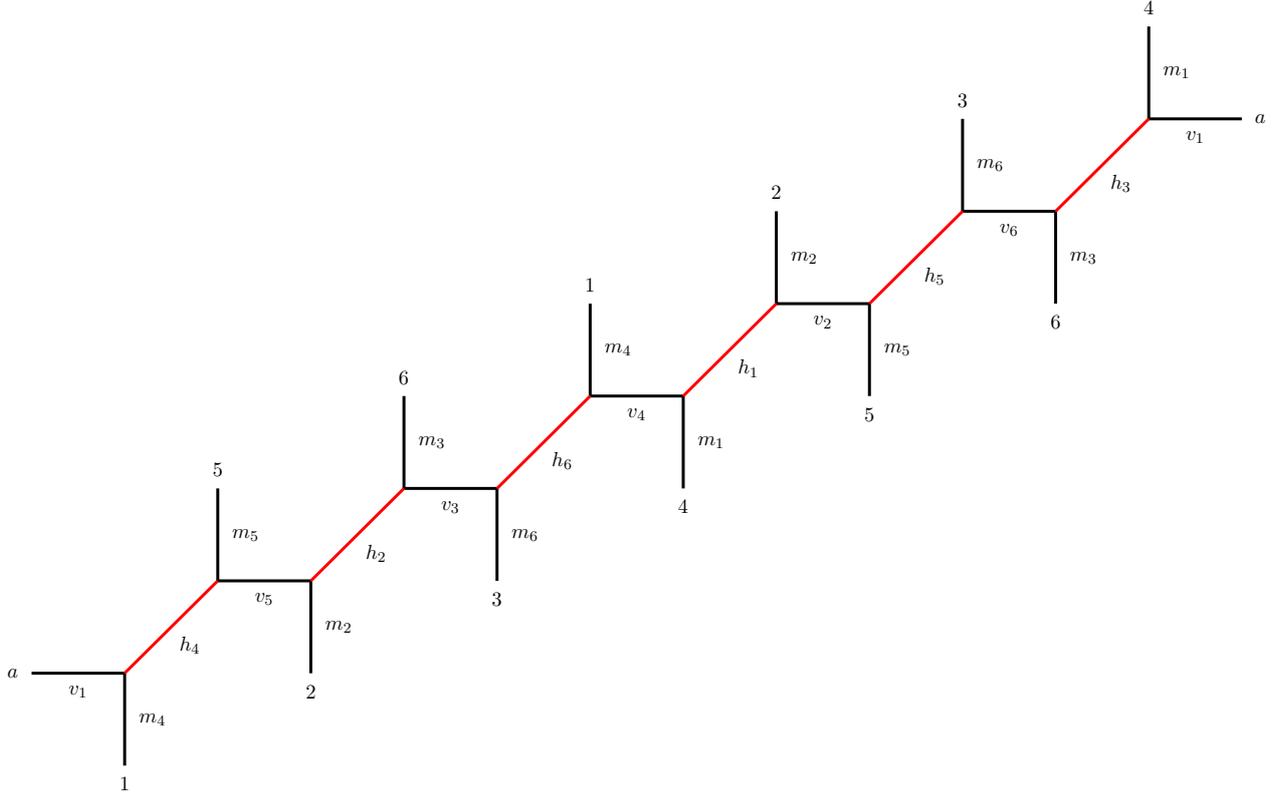
\begin{figure}[htb]
\begin{center}
\scalebox{0.7}{\parbox{24cm}{\begin{tikzpicture}[scale = 1.75]
\draw[ultra thick] (0,0) -- (1,0);
\draw[ultra thick] (2,1)--(3,1);
\draw[ultra thick] (4,2)--(5,2);
\draw[ultra thick] (6,3)--(7,3);
\draw[ultra thick] (8,4)--(9,4);
\draw[ultra thick] (10,5)--(11,5);
\draw[ultra thick] (12,6)--(13,6);
\draw[ultra thick] (1,0) -- (1,-1);
\draw[ultra thick] (2,1) -- (2,2);
\draw[ultra thick] (3,0) -- (3,1);
\draw[ultra thick] (4,2) -- (4,3);
\draw[ultra thick] (5,2) -- (5,1);
\draw[ultra thick] (6,4) -- (6,3);
\draw[ultra thick] (7,3) -- (7,2);
\draw[ultra thick] (8,4) -- (8,5);
\draw[ultra thick] (9,4) -- (9,3);
\draw[ultra thick] (10,5) -- (10,6);
\draw[ultra thick] (11,5) -- (11,4);
\draw[ultra thick] (12,6) -- (12,7);
\draw[ultra thick,red] (1,0) -- (2,1);
\draw[ultra thick,red] (3,1) -- (4,2);
\draw[ultra thick,red] (5,2) -- (6,3);
\draw[ultra thick,red] (7,3) -- (8,4);
\draw[ultra thick,red] (9,4) -- (10,5);
\draw[ultra thick,red] (11,5) -- (12,6);
\node at (0.5,-0.2) {{\small $v_1$}};
\node at (2.5,0.8) {{\small $v_5$}};
\node at (4.5,1.8) {{\small $v_3$}};
\node at (6.5,2.8) {{\small $v_4$}};
\node at (8.5,3.8) {{\small $v_2$}};
\node at (10.5,4.8) {{\small $v_6$}};
\node at (12.5,5.8) {{\small $v_1$}};
\node at (1.7,0.3) {{\small $h_4$}};
\node at (3.7,1.3) {{\small $h_2$}};
\node at (5.7,2.3) {{\small $h_6$}};
\node at (7.7,3.3) {{\small $h_1$}};
\node at (9.7,4.3) {{\small $h_5$}};
\node at (11.7,5.3) {{\small $h_3$}};
\node at (1.3,-0.5) {{\small $m_4$}};
\node at (3.3,0.5) {{\small $m_2$}};
\node at (5.3,1.5) {{\small $m_6$}};
\node at (7.3,2.5) {{\small $m_1$}};
\node at (9.3,3.5) {{\small $m_5$}};
\node at (11.3,4.5) {{\small $m_3$}};
\node at (2.3,1.5) {{\small $m_5$}};
\node at (4.3,2.5) {{\small $m_3$}};
\node at (6.3,3.5) {{\small $m_4$}};
\node at (8.3,4.5) {{\small $m_2$}};
\node at (10.3,5.5) {{\small $m_6$}};
\node at (12.3,6.5) {{\small $m_1$}};
\node at (1,-1.2) {{\small \bf $1$}};
\node at (3,-0.2) {{\small \bf $2$}};
\node at (5,0.8) {{\small \bf $3$}};
\node at (7,1.8) {{\small \bf $4$}};
\node at (9,2.8) {{\small \bf $5$}};
\node at (11,3.8) {{\small \bf $6$}};
\node at (2,2.2) {{\small \bf $5$}};
\node at (4,3.2) {{\small \bf $6$}};
\node at (6,4.2) {{\small \bf $1$}};
\node at (8,5.2) {{\small \bf $2$}};
\node at (10,6.2) {{\small \bf $3$}};
\node at (12,7.2) {{\small \bf $4$}};
\node at (-0.2,0) {{\small \bf $a$}};
\node at (13.2,6) {{\small \bf $a$}};
\end{tikzpicture}}
}
\end{center}
\caption{\sl Diagram after the $SL(2,\mathbb{Z})$ transformation in equation~(\ref{SLTrafo1}). The red lines undergo a flop transition in the following step.}
\label{Fig:SL21}
\end{figure}

To proceed, we cut the $(3,2)$ diagram along the dashed lines shown in figure~\ref{Fig:Cutting23}
and glue them back together in order to obtain figure~\ref{Fig:Gluing23}.
Cutting again along the dashed line and glueing the lines labelled by $a$ together, we arrive at the web drawn in Figure~\ref{Fig:Diag61}.
\FloatBarrier
\noindent
Next, we perform an $SL(2,\mathbb{Z})$ transformation. The latter does not change the length of a given line, however, it changes their orientation. Specifically, the transformation we consider rotates the lines as follows:
\begin{align}
&(1,0)\longrightarrow (1,1)\,,&&(0,1)\longrightarrow (-1,0)\,,&&(1,1)\longrightarrow (0,1)\,,\label{SLTrafo1}
\end{align}
\emph{i.e.} it rotates a horizontal line into a diagonal one, a diagonal into a vertical and a vertical into a horizontal. The resulting diagram is shown in figure~\ref{Fig:SL21}.
In order to be consistent, however, we have to impose two types of consistency conditions
\begin{itemize}
\item The vertical length of the line connecting the nodes 1 with 1 must be equal to the length of the line connecting the nodes 2 with 2 as well as 3 with 3 \emph{etc.}:
\begin{align}
&m_4+h_4+h_2+h_6=m_2+h_2+h_6+h_1=m_6+h_6+h_1+h_5\nonumber\\
&\hspace{1cm}=m_1+h_1+h_5+h_3=m_5+h_5+h_3+h_4=m_3+h_3+h_4+h_2\,.\label{Cond1Diag}
\end{align}
\item The horizontal distance between the nodes 1 and 2 at the bottom of the diagram has to be identical to the horizontal distance between the nodes 1 and 2 at the top of the diagram. The same constraint has to hold for the nodes 2 and 3, 3 and 4, \emph{etc.}
\begin{align}
&h_4+v_5=v_4+h_1\,,&&h_2+v_3=v_2+h_5\,,\nonumber\\
&h_6+v_4=v_6+h_3\,,&&h_1+v_2=v_1+h_4\,,\nonumber\\
&h_5+v_6=v_5+h_2\,,&&h_3+v_1=v_3+h_6\,.\label{Cond2Diag}
\end{align}
\end{itemize}
The consistency conditions Eq.(\ref{Cond1Diag}) and Eq.(\ref{Cond2Diag}) are fully equivalent to the conditions Eq.(\ref{InitC1}) through  Eq.(\ref{InitC6}), which guarantee that the original diagram in Figure~\ref{Fig:Cutting23} is consistent.

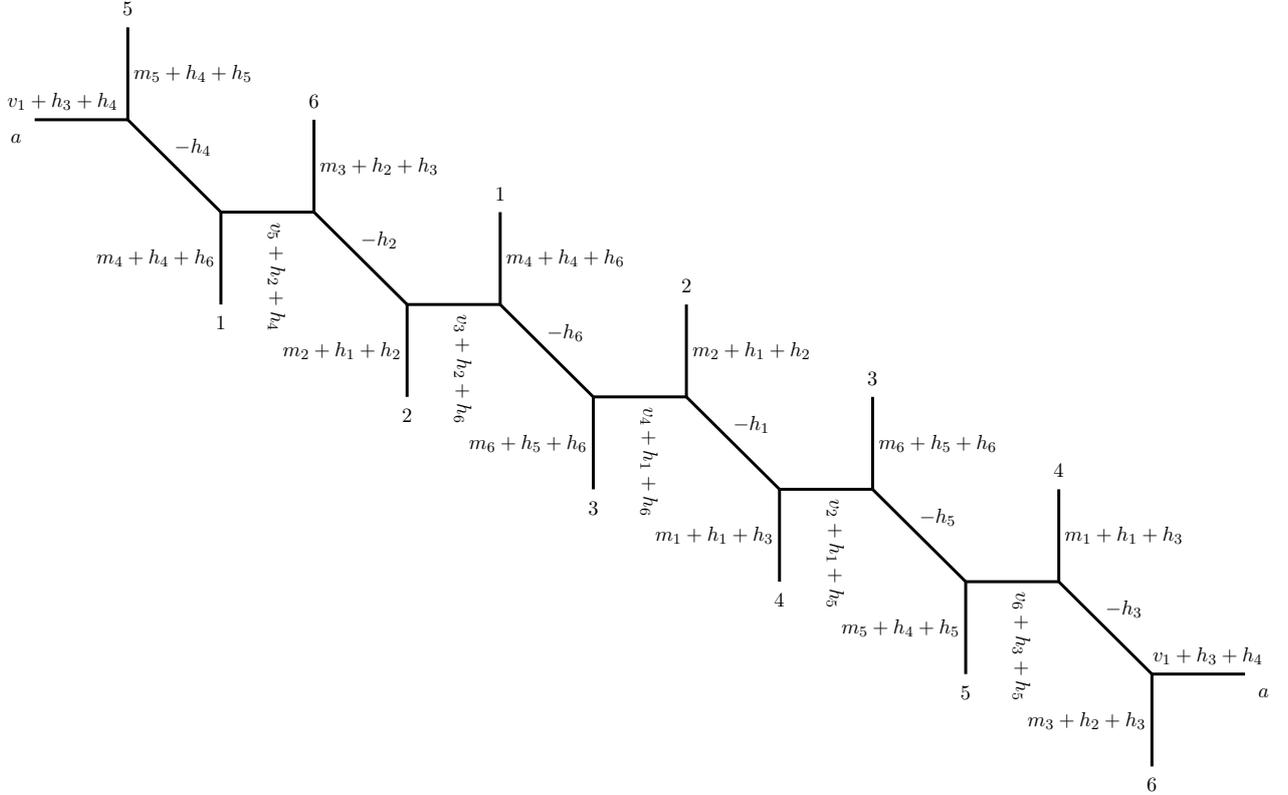
\begin{figure}[htbp]
\begin{center}
\scalebox{0.7}{\parbox{25cm}{\begin{tikzpicture}[scale = 1.75]
\draw[ultra thick] (0,0) -- (1,-1);
\draw[ultra thick] (2,-1) -- (3,-2);
\draw[ultra thick] (4,-2) -- (5,-3);
\draw[ultra thick] (6,-3) -- (7,-4);
\draw[ultra thick] (8,-4) -- (9,-5);
\draw[ultra thick] (10,-5) -- (11,-6);
\draw[ultra thick] (0,0) -- (0,1);
\draw[ultra thick] (2,-1) -- (2,0);
\draw[ultra thick] (4,-2) -- (4,-1);
\draw[ultra thick] (6,-3) -- (6,-2);
\draw[ultra thick] (8,-4) -- (8,-3);
\draw[ultra thick] (10,-5) -- (10,-4);
\draw[ultra thick] (1,-1) -- (1,-2);
\draw[ultra thick] (3,-2) -- (3,-3);
\draw[ultra thick] (5,-3) -- (5,-4);
\draw[ultra thick] (7,-4) -- (7,-5);
\draw[ultra thick] (9,-5) -- (9,-6);
\draw[ultra thick] (11,-6) -- (11,-7);
\draw[ultra thick] (-1,0) -- (0,0);
\draw[ultra thick] (1,-1) -- (2,-1);
\draw[ultra thick] (3,-2) -- (4,-2);
\draw[ultra thick] (5,-3) -- (6,-3);
\draw[ultra thick] (7,-4) -- (8,-4);
\draw[ultra thick] (9,-5) -- (10,-5);
\draw[ultra thick] (11,-6) -- (12,-6);
\node at (-1.2,-0.2) {{\small \bf $a$}};
\node at (12.2,-6.2) {{\small \bf $a$}};
\node at (1,-2.2) {{\small \bf $1$}};
\node at (3,-3.2) {{\small \bf $2$}};
\node at (5,-4.2) {{\small \bf $3$}};
\node at (7,-5.2) {{\small \bf $4$}};
\node at (9,-6.2) {{\small \bf $5$}};
\node at (11,-7.2) {{\small \bf $6$}};
\node at (0,1.2) {{\small \bf $5$}};
\node at (2,0.2) {{\small \bf $6$}};
\node at (4,-0.8) {{\small \bf $1$}};
\node at (6,-1.8) {{\small \bf $2$}};
\node at (8,-2.8) {{\small \bf $3$}};
\node at (10,-3.8) {{\small \bf $4$}};
\node at (0.7,-0.3) {{\small $-h_4$}};
\node at (2.7,-1.3) {{\small $-h_2$}};
\node at (4.7,-2.3) {{\small $-h_6$}};
\node at (6.7,-3.3) {{\small $-h_1$}};
\node at (8.7,-4.3) {{\small $-h_5$}};
\node at (10.7,-5.3) {{\small $-h_3$}};
\node at (0.3,-1.5) {{\small $m_4+h_4+h_6$}};
\node at (2.3,-2.5) {{\small $m_2+h_1+h_2$}};
\node at (4.3,-3.5) {{\small $m_6+h_5+h_6$}};
\node at (6.3,-4.5) {{\small $m_1+h_1+h_3$}};
\node at (8.3,-5.5) {{\small $m_5+h_4+h_5$}};
\node at (10.3,-6.5) {{\small $m_3+h_2+h_3$}};
\node at (0.7,0.5) {{\small $m_5+h_4+h_5$}};
\node at (2.7,-0.5) {{\small $m_3+h_2+h_3$}};
\node at (4.7,-1.5) {{\small $m_4+h_4+h_6$}};
\node at (6.7,-2.5) {{\small $m_2+h_1+h_2$}};
\node at (8.7,-3.5) {{\small $m_6+h_5+h_6$}};
\node at (10.7,-4.5) {{\small $m_1+h_1+h_3$}};
\node at (-0.7,0.2) {{\small $v_1+h_3+h_4$}};
\node[rotate=270] at (1.6,-1.7) {{\small $v_5+h_2+h_4$}};
\node[rotate=270] at (3.6,-2.7) {{\small $v_3+h_2+h_6$}};
\node[rotate=270] at (5.6,-3.7) {{\small $v_4+h_1+h_6$}};
\node[rotate=270] at (7.6,-4.7) {{\small $v_2+h_1+h_5$}};
\node[rotate=270] at (9.6,-5.7) {{\small $v_6+h_3+h_5$}};
\node at (11.6,-5.8) {{\small $v_1+h_3+h_4$}};
\end{tikzpicture}}
}
\end{center}
\caption{\sl The final $(p,q)$-web diagram after six flop transformations.}
\label{Fig:Flop1}
\end{figure}

The diagram in figure~\ref{Fig:SL21} looks already very similar to the $(N,M)=(6,1)$ configuration, however, the identification of the vertical lines at the bottom of the diagram with the lines at the top of the diagram is different. Indeed, the latter are 'shifted' with respect to the former. However, through a series of further transformations, we can bring the diagram into the form of the $(6,1)$ web, with the correct identification of the vertical lines. To this end, as a next step we perform a flop transition on all the lines drawn in red in figure~\ref{Fig:SL21}. Using the general transformations as explained in (\ref{ExampleFlop}) we can draw the new diagram in Figure~\ref{Fig:Flop1}.
\FloatBarrier

\noindent
After undergoing the flop transformations, we can again check the consistency conditions for the diagram in Figure~\ref{Fig:Flop1}. As before, we have to impose two different types of conditions
 \begin{itemize}
\item The vertical length of the line connecting the nodes 1 with 1 must be equal to the length of the line connecting the nodes 2 with 2 as well as 3 with 3 \emph{etc.}:
\begin{align}
&m_4+h_4+h_2+h_6=m_2+h_2+h_6+h_1=m_6+h_6+h_1+h_5\nonumber\\
&\hspace{1cm}=m_1+h_1+h_5+h_3=m_5+h_5+h_3+h_4=m_3+h_3+h_4+h_2\,.\label{Cond1Flop}
\end{align}
These conditions are identical to equation Eq.(\ref{Cond1Diag}). 
\item The horizontal distance between the nodes 1 and 2 at the bottom of the diagram has to be identical to the horizontal distance between the nodes 1 and 2 at the top of the diagram. The same constraint has to hold for the nodes 2 and 3, 3 and 4, etc
\begin{align}
&h_4+v_5=v_4+h_1\,,&&h_2+v_3=v_2+h_5\,,\nonumber\\
&h_6+v_4=v_6+h_3\,,&&h_1+v_2=v_1+h_4\,,\nonumber\\
&h_5+v_6=v_5+h_2\,,&&h_3+v_1=v_3+h_6\,.\label{Cond2Flop}
\end{align}
These conditions are identical to equation (\ref{Cond2Diag}).
\end{itemize}
Thus, the consistency conditions Eq.(\ref{Cond1Flop}) and Eq.(\ref{Cond2Flop}) are still fully equivalent to the conditions Eq.(\ref{InitC1}) through (\ref{InitC6}). This is to be expected, since the flop transformation does not change the overall length in the vertical or horizontal direction, as can be seen from Eq.(\ref{ExampleFlop}).

As a next step, we perform another $SL(2,\mathbb{Z})$ transformation to change the orientation of all lines in Figure~\ref{Fig:Flop1}. Indeed, we transform
\begin{align}
&(1,0)\longrightarrow (1,-1)\,,&&(0,1)\longrightarrow (0,1)\,,&&(1,1)\longrightarrow (1,0)\,,\label{SLTrafo2}
\end{align}
which changes horizontal lines into diagonal ones and vice versa, while leaving the vertical lines invariant. The resulting diagram is shown in Figure~\ref{Fig:SL22}.
\begin{figure}[hptb]
\begin{center}
\scalebox{0.7}{\parbox{25cm}{\begin{tikzpicture}[scale = 1.75]
\draw[ultra thick] (4,2) -- (4,3);
\draw[ultra thick] (5,2) -- (5,1);
\draw[ultra thick] (6,3) -- (6,4);
\draw[ultra thick] (7,3) -- (7,2);
\draw[ultra thick] (8,4) -- (8,5);
\draw[ultra thick] (9,4) -- (9,3);
\draw[ultra thick] (10,5) -- (10,6);
\draw[ultra thick] (11,5) -- (11,4);
\draw[ultra thick] (12,6) -- (12,7);
\draw[ultra thick] (13,6) -- (13,5);
\draw[ultra thick] (14,7) -- (14,8);
\draw[ultra thick] (15,7) -- (15,6);
\draw[ultra thick] (4,2) -- (5,2);
\draw[ultra thick] (6,3) -- (7,3);
\draw[ultra thick] (8,4) -- (9,4);
\draw[ultra thick] (10,5) -- (11,5);
\draw[ultra thick] (12,6) -- (13,6);
\draw[ultra thick] (14,7) -- (15,7);
\draw[ultra thick] (3,1) -- (4,2);
\draw[ultra thick,red] (5,2) -- (6,3);
\draw[ultra thick,red] (7,3) -- (8,4);
\draw[ultra thick,red] (9,4) -- (10,5);
\draw[ultra thick,red] (11,5) -- (12,6);
\draw[ultra thick,red] (13,6) -- (14,7);
\draw[ultra thick] (15,7) -- (16,8);
\node[rotate=45] at (3.3,1.6) {{\small $v_1+h_3+h_4$}};
\node[rotate=45] at (5.3,2.6) {{\small $v_5+h_2+h_4$}};
\node[rotate=45] at (7.3,3.6) {{\small $v_3+h_2+h_6$}};
\node[rotate=45] at (9.3,4.6) {{\small $v_4+h_1+h_6$}};
\node[rotate=45] at (11.3,5.6) {{\small $v_2+h_1+h_5$}};
\node[rotate=45] at (13.3,6.6) {{\small $v_6+h_3+h_5$}};
\node[rotate=45] at (15.3,7.6) {{\small $v_1+h_3+h_4$}};
\node at (4.5,1.8) {{\small $-h_4$}};
\node at (6.5,2.8) {{\small $-h_2$}};
\node at (8.5,3.8) {{\small $-h_6$}};
\node at (10.5,4.8) {{\small $-h_1$}};
\node at (12.5,5.8) {{\small $-h_5$}};
\node at (14.5,6.8) {{\small $-h_3$}};
\node at (5.7,1.4) {{\small $m_4+h_4+h_6$}};
\node at (7.7,2.4) {{\small $m_2+h_1+h_2$}};
\node at (9.7,3.4) {{\small $m_6+h_5+h_6$}};
\node at (11.7,4.5) {{\small $m_1+h_1+h_3$}};
\node at (13.7,5.5) {{\small $m_5+h_4+h_5$}};
\node at (15.7,6.5) {{\small $m_3+h_2+h_3$}};
\node at (3.3,2.6) {{\small $m_5+h_4+h_5$}};
\node at (5.3,3.6) {{\small $m_3+h_2+h_3$}};
\node at (7.3,4.6) {{\small $m_4+h_4+h_6$}};
\node at (9.3,5.6) {{\small $m_2+h_1+h_2$}};
\node at (11.3,6.6) {{\small $m_6+h_5+h_6$}};
\node at (13.3,7.6) {{\small $m_1+h_1+h_3$}};
\node at (2.8,0.8) {{\small \bf $a$}};
\node at (16.2,8.2) {{\small \bf $a$}};
\node at (5,0.8) {{\small \bf $1$}};
\node at (7,1.8) {{\small \bf $2$}};
\node at (9,2.8) {{\small \bf $3$}};
\node at (11,3.8) {{\small \bf $4$}};
\node at (13,4.8) {{\small \bf $5$}};
\node at (15,5.8) {{\small \bf $6$}};
\node at (4,3.2) {{\small \bf $5$}};
\node at (6,4.2) {{\small \bf $6$}};
\node at (8,5.2) {{\small \bf $1$}};
\node at (10,6.2) {{\small \bf $2$}};
\node at (12,7.2) {{\small \bf $3$}};
\node at (14,8.2) {{\small \bf $4$}};
\end{tikzpicture}}}
\end{center}
\caption{\sl Diagram after another $SL(2,\mathbb{Z})$ transformation. The red lines are to undergo a flop transition in the next step.}
\label{Fig:SL22}
\end{figure}
Since the $SL(2,\mathbb{Z})$ transformation does not modify the length of the individual lines (but only changes their orientation), the consistency conditions are also not altered, but remain Eq.(\ref{Cond1Flop}) and Eq.(\ref{Cond2Flop}).

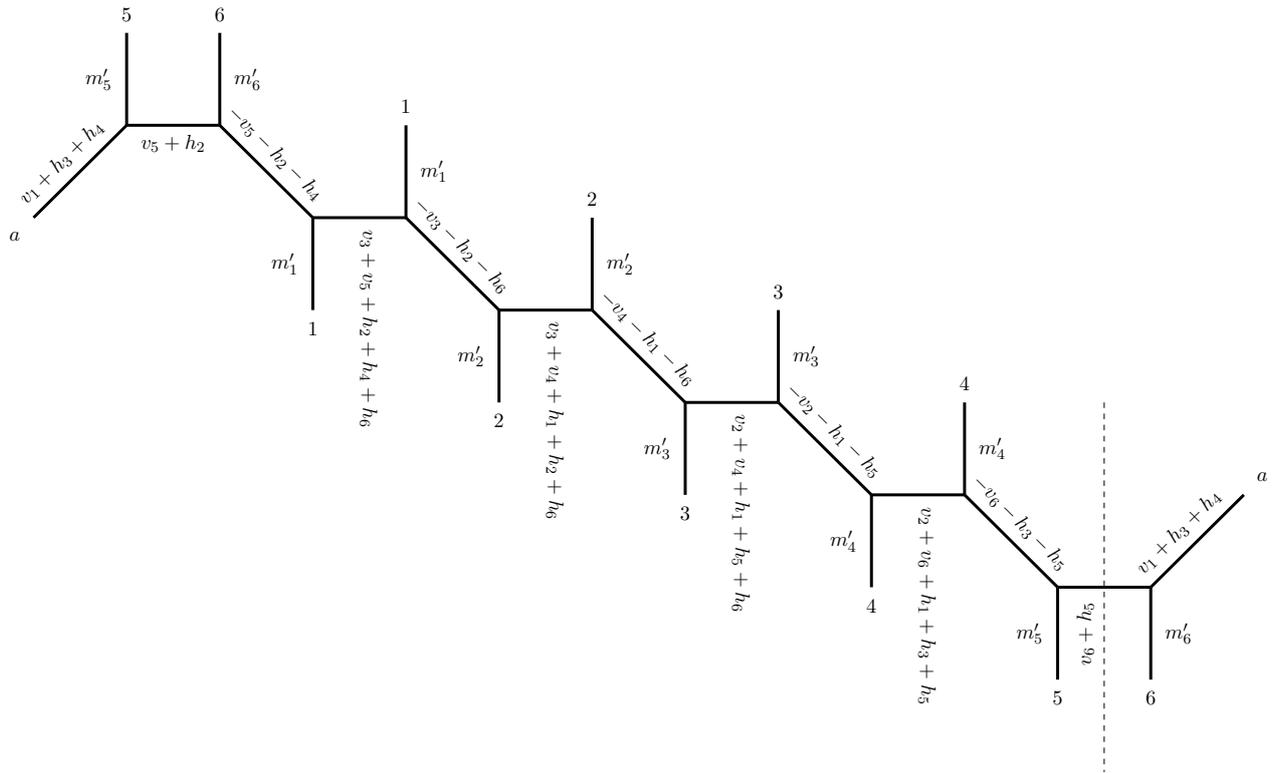
\begin{figure}[hptb]
\begin{center}
\scalebox{0.7}{\parbox{25cm}{\begin{tikzpicture}[scale = 1.75]
\draw[ultra thick] (-2,-1) -- (-1,0);
\draw[ultra thick] (0,0) -- (1,-1);
\draw[ultra thick] (2,-1) -- (3,-2);
\draw[ultra thick] (4,-2) -- (5,-3);
\draw[ultra thick] (6,-3) -- (7,-4);
\draw[ultra thick] (8,-4) -- (9,-5);
\draw[ultra thick] (10,-5) -- (11,-4);
\draw[ultra thick] (-1,0) -- (-1,1);
\draw[ultra thick] (0,0) -- (0,1);
\draw[ultra thick] (2,-1) -- (2,0);
\draw[ultra thick] (4,-2) -- (4,-1);
\draw[ultra thick] (6,-3) -- (6,-2);
\draw[ultra thick] (8,-4) -- (8,-3);
\draw[ultra thick] (1,-1) -- (1,-2);
\draw[ultra thick] (3,-2) -- (3,-3);
\draw[ultra thick] (5,-3) -- (5,-4);
\draw[ultra thick] (7,-4) -- (7,-5);
\draw[ultra thick] (9,-5) -- (9,-6);
\draw[ultra thick] (10,-5) -- (10,-6);
\draw[ultra thick] (-1,0) -- (0,0);
\draw[ultra thick] (1,-1) -- (2,-1);
\draw[ultra thick] (3,-2) -- (4,-2);
\draw[ultra thick] (5,-3) -- (6,-3);
\draw[ultra thick] (7,-4) -- (8,-4);
\draw[ultra thick] (9,-5) -- (10,-5);
\node at (-2.2,-1.2) {{\small \bf $a$}};
\node at (11.2,-3.8) {{\small \bf $a$}};
\node at (1,-2.2) {{\small \bf $1$}};
\node at (3,-3.2) {{\small \bf $2$}};
\node at (5,-4.2) {{\small \bf $3$}};
\node at (7,-5.2) {{\small \bf $4$}};
\node at (9,-6.2) {{\small \bf $5$}};
\node at (10,-6.2) {{\small \bf $6$}};
\node at (-1,1.2) {{\small \bf $5$}};
\node at (0,1.2) {{\small \bf $6$}};
\node at (2,0.2) {{\small \bf $1$}};
\node at (4,-0.8) {{\small \bf $2$}};
\node at (6,-1.8) {{\small \bf $3$}};
\node at (8,-2.8) {{\small \bf $4$}};
\node[rotate=45] at (-1.7,-0.4) {{\small $v_1+h_3+h_4$}};
\node[rotate=315] at (0.6,-0.3) {{\small $-v_5-h_2-h_4$}};
\node[rotate=315] at (2.6,-1.3) {{\small $-v_3-h_2-h_6$}};
\node[rotate=315] at (4.6,-2.3) {{\small $-v_4-h_1-h_6$}};
\node[rotate=315] at (6.6,-3.3) {{\small $-v_2-h_1-h_5$}};
\node[rotate=315] at (8.6,-4.3) {{\small $-v_6-h_3-h_5$}};
\node[rotate=45] at (10.3,-4.4) {{\small $v_1+h_3+h_4$}};
\node at (-0.5,-0.2) {{\small $v_5+h_2$}};
\node[rotate=90] at (9.3,-5.5) {{\small $v_6+h_5$}};
\node[rotate=270] at (1.6,-2.2) {{\small $v_3+v_5+h_2+h_4+h_6$}};
\node[rotate=270] at (3.6,-3.2) {{\small $v_3+v_4+h_1+h_2+h_6$}};
\node[rotate=270] at (5.6,-4.2) {{\small $v_2+v_4+h_1+h_5+h_6$}};
\node[rotate=270] at (7.6,-5.2) {{\small $v_2+v_6+h_1+h_3+h_5$}};
\node at (-1.3,0.5) {{\small $m'_5$}};
\node at (0.3,0.5) {{\small $m'_6$}};
\node at (2.3,-0.5) {{\small $m'_1$}};
\node at (4.3,-1.5) {{\small $m'_2$}};
\node at (6.3,-2.5) {{\small $m'_3$}};
\node at (8.3,-3.5) {{\small $m'_4$}};
\node at (0.7,-1.5) {{\small $m'_1$}};
\node at (2.7,-2.5) {{\small $m'_2$}};
\node at (4.7,-3.5) {{\small $m'_3$}};
\node at (6.7,-4.5) {{\small $m'_4$}};
\node at (8.7,-5.5) {{\small $m'_5$}};
\node at (10.3,-5.5) {{\small $m'_6$}};
\draw[dashed] (9.5,-3) -- (9.5,-7);
\end{tikzpicture}}
}
\end{center}
\caption{\sl Diagram after five further flop transformations. The quantities $m'_{1,2,3,4,5,6}$ are given in Eq.(\ref{TwistLength}). The diagram will be cut along the dashed line in the following step.}
\label{Fig:Flop2}
\end{figure}

As a next step we perform flop transitions on all the lines marked red in figure~\ref{Fig:SL22}. Using the map of parameters in Eq.(\ref{ExampleFlop}), the new diagram takes the form of Figure~\ref{Fig:Flop2},
\FloatBarrier

\noindent
where the vertical lengths $m'_{1,2,3,4,5,6}$ are given by
{\allowdisplaybreaks\begin{align}
&m'_1=m_4+v_3+v_5+2(h_2+h_4+h_6)\,,\nonumber\\
&m_2'=m_2+v_3+v_4+2(h_1+h_2+h_6)\,,\nonumber\\
&m'_3=m_6+v_2+v_4+2(h_1+h_5+h_6)\,,\nonumber\\
&m'_4=m_1+v_2+v_6+2(h_1+h_3+h_5)\,,\nonumber\\
&m'_5=m_5+v_6+h_3+h_4+2h_5\,,\nonumber\\
&m'_6=m_3+v_5+2h_2+h_3+h_4\,.\label{TwistLength}
\end{align}}

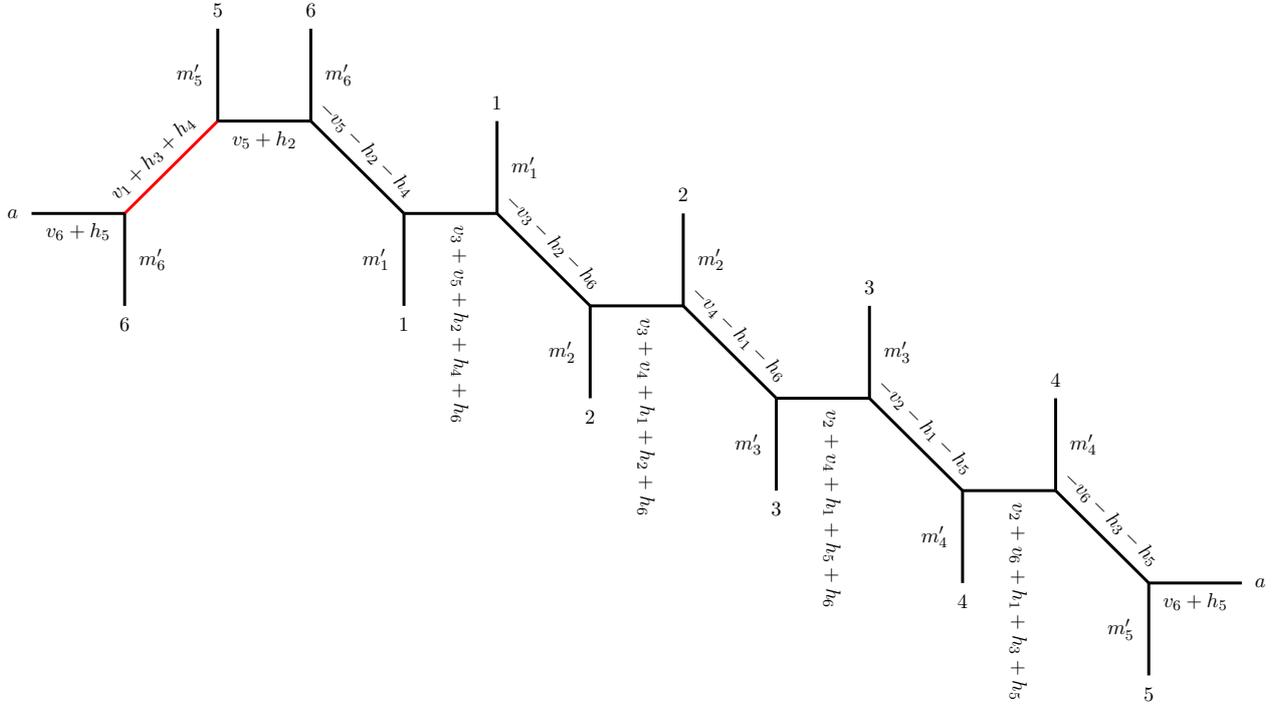
\begin{figure}[htb]
\begin{center}
\scalebox{0.7}{\parbox{25cm}{\begin{tikzpicture}[scale = 1.75]
\draw[ultra thick,red] (-2,-1) -- (-1,0);
\draw[ultra thick] (0,0) -- (1,-1);
\draw[ultra thick] (2,-1) -- (3,-2);
\draw[ultra thick] (4,-2) -- (5,-3);
\draw[ultra thick] (6,-3) -- (7,-4);
\draw[ultra thick] (8,-4) -- (9,-5);
\draw[ultra thick] (-1,0) -- (-1,1);
\draw[ultra thick] (0,0) -- (0,1);
\draw[ultra thick] (2,-1) -- (2,0);
\draw[ultra thick] (4,-2) -- (4,-1);
\draw[ultra thick] (6,-3) -- (6,-2);
\draw[ultra thick] (8,-4) -- (8,-3);
\draw[ultra thick] (-2,-2) -- (-2,-1);
\draw[ultra thick] (1,-1) -- (1,-2);
\draw[ultra thick] (3,-2) -- (3,-3);
\draw[ultra thick] (5,-3) -- (5,-4);
\draw[ultra thick] (7,-4) -- (7,-5);
\draw[ultra thick] (9,-5) -- (9,-6);
\draw[ultra thick] (-3,-1) -- (-2,-1);
\draw[ultra thick] (-1,0) -- (0,0);
\draw[ultra thick] (1,-1) -- (2,-1);
\draw[ultra thick] (3,-2) -- (4,-2);
\draw[ultra thick] (5,-3) -- (6,-3);
\draw[ultra thick] (7,-4) -- (8,-4);
\draw[ultra thick] (9,-5) -- (10,-5);
\node at (-3.2,-1) {{\small \bf $a$}};
\node at (10.2,-5) {{\small \bf $a$}};
\node at (-2,-2.2) {{\small \bf $6$}};
\node at (1,-2.2) {{\small \bf $1$}};
\node at (3,-3.2) {{\small \bf $2$}};
\node at (5,-4.2) {{\small \bf $3$}};
\node at (7,-5.2) {{\small \bf $4$}};
\node at (9,-6.2) {{\small \bf $5$}};
\node at (-1,1.2) {{\small \bf $5$}};
\node at (0,1.2) {{\small \bf $6$}};
\node at (2,0.2) {{\small \bf $1$}};
\node at (4,-0.8) {{\small \bf $2$}};
\node at (6,-1.8) {{\small \bf $3$}};
\node at (8,-2.8) {{\small \bf $4$}};
\node[rotate=45] at (-1.7,-0.4) {{\small $v_1+h_3+h_4$}};
\node[rotate=315] at (0.6,-0.3) {{\small $-v_5-h_2-h_4$}};
\node[rotate=315] at (2.6,-1.3) {{\small $-v_3-h_2-h_6$}};
\node[rotate=315] at (4.6,-2.3) {{\small $-v_4-h_1-h_6$}};
\node[rotate=315] at (6.6,-3.3) {{\small $-v_2-h_1-h_5$}};
\node[rotate=315] at (8.6,-4.3) {{\small $-v_6-h_3-h_5$}};
\node at (-0.5,-0.2) {{\small $v_5+h_2$}};
\node at (9.5,-5.2) {{\small $v_6+h_5$}};
\node at (-2.5,-1.2) {{\small $v_6+h_5$}};
\node[rotate=270] at (1.6,-2.2) {{\small $v_3+v_5+h_2+h_4+h_6$}};
\node[rotate=270] at (3.6,-3.2) {{\small $v_3+v_4+h_1+h_2+h_6$}};
\node[rotate=270] at (5.6,-4.2) {{\small $v_2+v_4+h_1+h_5+h_6$}};
\node[rotate=270] at (7.6,-5.2) {{\small $v_2+v_6+h_1+h_3+h_5$}};
\node at (-1.3,0.5) {{\small $m'_5$}};
\node at (0.3,0.5) {{\small $m'_6$}};
\node at (2.3,-0.5) {{\small $m'_1$}};
\node at (4.3,-1.5) {{\small $m'_2$}};
\node at (6.3,-2.5) {{\small $m'_3$}};
\node at (8.3,-3.5) {{\small $m'_4$}};
\node at (-1.7,-1.5) {{\small $m'_6$}};
\node at (0.7,-1.5) {{\small $m'_1$}};
\node at (2.7,-2.5) {{\small $m'_2$}};
\node at (4.7,-3.5) {{\small $m'_3$}};
\node at (6.7,-4.5) {{\small $m'_4$}};
\node at (8.7,-5.5) {{\small $m'_5$}};
\end{tikzpicture}}
}
\end{center}
\caption{\sl Re-gluing the diagram in Figure~\ref{Fig:Flop2}. The red line will undergo a flop transition in the next step.}
\label{Fig:Cut2}
\vskip0.2cm
\end{figure}

\noindent
As a next step we cut the diagram in Figure~\ref{Fig:Flop2} along the dashed line and glue it again along the diagonal line marked $a$. The result is shown in Figure~\ref{Fig:Cut2},
for which we perform a flop transition for the line shown in red to arrive at Figure~\ref{Fig:Flop3}.
\begin{figure}[htb]
\begin{center}
\scalebox{0.7}{\parbox{25cm}{\begin{tikzpicture}[scale = 1.75]
\draw[ultra thick] (-2,1) -- (-1,0);
\draw[ultra thick] (0,0) -- (1,-1);
\draw[ultra thick] (2,-1) -- (3,-2);
\draw[ultra thick] (4,-2) -- (5,-3);
\draw[ultra thick] (6,-3) -- (7,-4);
\draw[ultra thick] (8,-4) -- (9,-5);
\draw[ultra thick] (0,0) -- (0,1);
\draw[ultra thick] (2,-1) -- (2,0);
\draw[ultra thick] (4,-2) -- (4,-1);
\draw[ultra thick] (6,-3) -- (6,-2);
\draw[ultra thick] (8,-4) -- (8,-3);
\draw[ultra thick] (-1,0) -- (-1,-1);
\draw[ultra thick] (-2,2) -- (-2,1);
\draw[ultra thick] (1,-1) -- (1,-2);
\draw[ultra thick] (3,-2) -- (3,-3);
\draw[ultra thick] (5,-3) -- (5,-4);
\draw[ultra thick] (7,-4) -- (7,-5);
\draw[ultra thick] (9,-5) -- (9,-6);
\draw[ultra thick] (-3,1) -- (-2,1);
\draw[ultra thick] (-1,0) -- (0,0);
\draw[ultra thick] (1,-1) -- (2,-1);
\draw[ultra thick] (3,-2) -- (4,-2);
\draw[ultra thick] (5,-3) -- (6,-3);
\draw[ultra thick] (7,-4) -- (8,-4);
\draw[ultra thick] (9,-5) -- (10,-5);
\node at (-3.2,1) {{\small \bf $a$}};
\node at (10.2,-5) {{\small \bf $a$}};
\node at (-1,-1.2) {{\small \bf $6$}};
\node at (1,-2.2) {{\small \bf $1$}};
\node at (3,-3.2) {{\small \bf $2$}};
\node at (5,-4.2) {{\small \bf $3$}};
\node at (7,-5.2) {{\small \bf $4$}};
\node at (9,-6.2) {{\small \bf $5$}};
\node at (-2,2.2) {{\small \bf $5$}};
\node at (0,1.2) {{\small \bf $6$}};
\node at (2,0.2) {{\small \bf $1$}};
\node at (4,-0.8) {{\small \bf $2$}};
\node at (6,-1.8) {{\small \bf $3$}};
\node at (8,-2.8) {{\small \bf $4$}};
\node[rotate=315] at (-1.4,0.7) {{\small $-v_1-h_3-h_4$}};
\node[rotate=315] at (0.6,-0.3) {{\small $-v_5-h_2-h_4$}};
\node[rotate=315] at (2.6,-1.3) {{\small $-v_3-h_2-h_6$}};
\node[rotate=315] at (4.6,-2.3) {{\small $-v_4-h_1-h_6$}};
\node[rotate=315] at (6.6,-3.3) {{\small $-v_2-h_1-h_5$}};
\node[rotate=315] at (8.6,-4.3) {{\small $-v_6-h_3-h_5$}};
\node[rotate=270] at (-2.4,-0.2) {{\small $v_1+v_6+h_3+h_4+h_5$}};
\node[rotate=270] at (-0.4,-1.2) {{\small $v_1+v_5+h_2+h_3+h_4$}};
\node[rotate=270] at (1.6,-2.2) {{\small $v_3+v_5+h_2+h_4+h_6$}};
\node[rotate=270] at (3.6,-3.2) {{\small $v_3+v_4+h_1+h_2+h_6$}};
\node[rotate=270] at (5.6,-4.2) {{\small $v_2+v_4+h_1+h_5+h_6$}};
\node[rotate=270] at (7.6,-5.2) {{\small $v_2+v_6+h_1+h_3+h_5$}};
\node[rotate=270] at (9.6,-6.2) {{\small $v_1+v_6+h_3+h_4+h_5$}};
\node at (-1.7,1.5) {{\small $m''_5$}};
\node at (0.3,0.5) {{\small $m''_6$}};
\node at (2.3,-0.5) {{\small $m'_1$}};
\node at (4.3,-1.5) {{\small $m'_2$}};
\node at (6.3,-2.5) {{\small $m'_3$}};
\node at (8.3,-3.5) {{\small $m'_4$}};
\node at (-1.3,-0.5) {{\small $m''_6$}};
\node at (0.7,-1.5) {{\small $m'_1$}};
\node at (2.7,-2.5) {{\small $m'_2$}};
\node at (4.7,-3.5) {{\small $m'_3$}};
\node at (6.7,-4.5) {{\small $m'_4$}};
\node at (8.7,-5.5) {{\small $m''_5$}};
\end{tikzpicture}}
}
\end{center}
\caption{\sl Diagram after a further flop transition. The parameters $m'_{1,2,3,4}$ are as in Eq.(\ref{TwistLength}) while $m''_{5,6}$ are defined in Eq.(\ref{TwistLength2}).}
\label{Fig:Flop3}
\vskip0.2cm
\end{figure}
Here, the lines $m''_{5}$ and $m''_6$ have length
\begin{align}
&m''_5=m'_5+v_1+h_3+h_4=m_5+v_1+v_6+2(h_3+h_4+h_5)\,,\nonumber\\
&m''_6=m'_6+v_1+h_3+h_4=m_3+v_1+v_5+2(h_2+h_3+h_4)\,.\label{TwistLength2}
\end{align}
The consistency conditions for the diagram in Figure~\ref{Fig:Flop3} are the same as Eq.(\ref{Cond1Diag}) and Eq.(\ref{Cond2Diag}).

\begin{figure}[htb]
\begin{center}
\scalebox{0.68}{\parbox{25cm}{\begin{tikzpicture}[scale = 1.75]
\draw[ultra thick] (4,2) -- (4,3);
\draw[ultra thick] (5,2) -- (5,1);
\draw[ultra thick] (6,3) -- (6,4);
\draw[ultra thick] (7,3) -- (7,2);
\draw[ultra thick] (8,4) -- (8,5);
\draw[ultra thick] (9,4) -- (9,3);
\draw[ultra thick] (10,5) -- (10,6);
\draw[ultra thick] (11,5) -- (11,4);
\draw[ultra thick] (12,6) -- (12,7);
\draw[ultra thick] (13,6) -- (13,5);
\draw[ultra thick] (14,7) -- (14,8);
\draw[ultra thick] (15,7) -- (15,6);
\draw[ultra thick] (4,2) -- (5,2);
\draw[ultra thick] (6,3) -- (7,3);
\draw[ultra thick] (8,4) -- (9,4);
\draw[ultra thick] (10,5) -- (11,5);
\draw[ultra thick] (12,6) -- (13,6);
\draw[ultra thick] (14,7) -- (15,7);
\draw[ultra thick] (3,1) -- (4,2);
\draw[ultra thick] (5,2) -- (6,3);
\draw[ultra thick] (7,3) -- (8,4);
\draw[ultra thick] (9,4) -- (10,5);
\draw[ultra thick] (11,5) -- (12,6);
\draw[ultra thick] (13,6) -- (14,7);
\draw[ultra thick] (15,7) -- (16,8);
\node[rotate=45] at (3.1,1.4) {{\small $v_1+v_6+h_3+h_4+h_5$}};
\node[rotate=90] at (5.3,3.6) {{\small $v_1+v_5+h_2+h_3+h_4$}};
\node[rotate=90] at (7.3,4.6) {{\small $v_3+v_5+h_2+h_4+h_6$}};
\node[rotate=90] at (9.3,5.6) {{\small $v_3+v_4+h_1+h_2+h_6$}};
\node[rotate=90] at (11.3,6.6) {{\small $v_2+v_4+h_1+h_5+h_6$}};
\node[rotate=90] at (13.3,7.6) {{\small $v_2+v_6+h_1+h_3+h_5$}};
\node[rotate=45] at (15.7,8) {{\small $v_1+v_6+h_3+h_4+h_5$}};
\node[rotate=270] at (4.3,1.2) {{\small $-v_1-h_3-h_4$}};
\node[rotate=270] at (6.5,2.2) {{\small $-v_5-h_2-h_4$}};
\node[rotate=270] at (8.5,3.2) {{\small $-v_3-h_2-h_6$}};
\node[rotate=270] at (10.5,4.2) {{\small $-v_4-h_1-h_6$}};
\node[rotate=270] at (12.5,5.2) {{\small $-v_2-h_1-h_5$}};
\node[rotate=270] at (14.5,6.2) {{\small $-v_6-h_3-h_5$}};
\node at (5.3,1.4) {{\small $m''_6$}};
\node at (7.3,2.4) {{\small $m'_1$}};
\node at (9.3,3.4) {{\small $m'_2$}};
\node at (11.3,4.5) {{\small $m'_3$}};
\node at (13.3,5.5) {{\small $m'_4$}};
\node at (15.3,6.5) {{\small $m''_5$}};
\node at (4.3,2.6) {{\small $m''_5$}};
\node at (6.3,3.6) {{\small $m''_6$}};
\node at (8.3,4.6) {{\small $m'_1$}};
\node at (10.3,5.6) {{\small $m'_2$}};
\node at (12.3,6.6) {{\small $m'_3$}};
\node at (14.3,7.6) {{\small $m'_4$}};
\node at (2.9,0.8) {{\small \bf $a$}};
\node at (16.2,8) {{\small \bf $a$}};
\node at (5,0.8) {{\small \bf $6$}};
\node at (7,1.8) {{\small \bf $1$}};
\node at (9,2.8) {{\small \bf $2$}};
\node at (11,3.8) {{\small \bf $3$}};
\node at (13,4.8) {{\small \bf $4$}};
\node at (15,5.8) {{\small \bf $5$}};
\node at (4,3.2) {{\small \bf $5$}};
\node at (6,4.2) {{\small \bf $6$}};
\node at (8,5.2) {{\small \bf $1$}};
\node at (10,6.2) {{\small \bf $2$}};
\node at (12,7.2) {{\small \bf $3$}};
\node at (14,8.2) {{\small \bf $4$}};
\draw[dashed] (4.5,4) -- (4.5,0);
\end{tikzpicture}}
}
\end{center}
\caption{\sl Diagram after another $SL(2,\mathbb{Z})$ transformation. In the final step, the diagram will be cut along the dashed line.}
\label{Fig:SL3}
\vskip0.2cm
\end{figure}
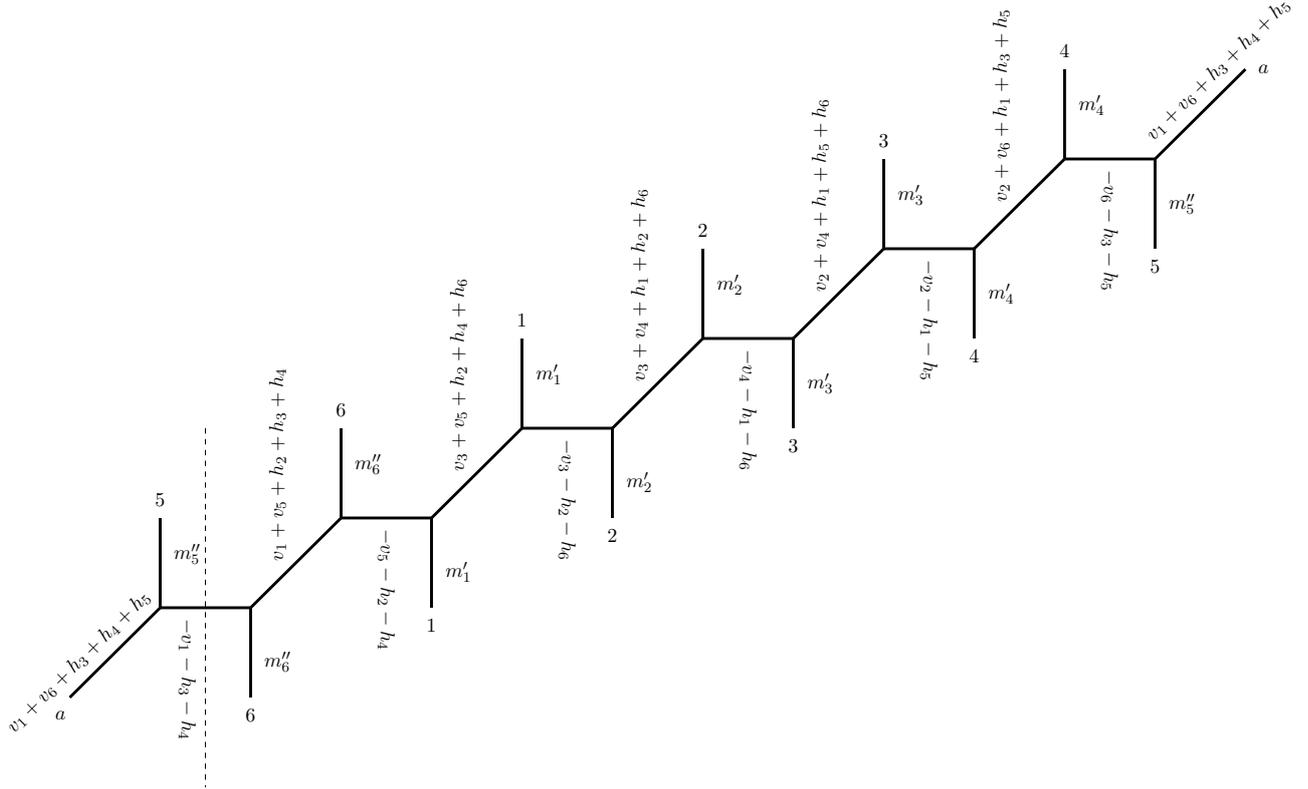

\begin{figure}[htb]
\begin{center}
\scalebox{0.7}{\parbox{25cm}{\begin{tikzpicture}[scale = 1.67]
\draw[ultra thick] (5,2) -- (5,1);
\draw[ultra thick] (6,3) -- (6,4);
\draw[ultra thick] (7,3) -- (7,2);
\draw[ultra thick] (8,4) -- (8,5);
\draw[ultra thick] (9,4) -- (9,3);
\draw[ultra thick] (10,5) -- (10,6);
\draw[ultra thick] (11,5) -- (11,4);
\draw[ultra thick] (12,6) -- (12,7);
\draw[ultra thick] (13,6) -- (13,5);
\draw[ultra thick] (14,7) -- (14,8);
\draw[ultra thick] (15,7) -- (15,6);
\draw[ultra thick] (16,8) -- (16,9);
\draw[ultra thick] (4,2) -- (5,2);
\draw[ultra thick] (6,3) -- (7,3);
\draw[ultra thick] (8,4) -- (9,4);
\draw[ultra thick] (10,5) -- (11,5);
\draw[ultra thick] (12,6) -- (13,6);
\draw[ultra thick] (14,7) -- (15,7);
\draw[ultra thick] (16,8) -- (17,8);
\draw[ultra thick] (5,2) -- (6,3);
\draw[ultra thick] (7,3) -- (8,4);
\draw[ultra thick] (9,4) -- (10,5);
\draw[ultra thick] (11,5) -- (12,6);
\draw[ultra thick] (13,6) -- (14,7);
\draw[ultra thick] (15,7) -- (16,8);
\node[rotate=90] at (5.3,3.6) {{\small $v_1+v_5+h_2+h_3+h_4$}};
\node[rotate=90] at (7.3,4.6) {{\small $v_3+v_5+h_2+h_4+h_6$}};
\node[rotate=90] at (9.3,5.6) {{\small $v_3+v_4+h_1+h_2+h_6$}};
\node[rotate=90] at (11.3,6.6) {{\small $v_2+v_4+h_1+h_5+h_6$}};
\node[rotate=90] at (13.3,7.6) {{\small $v_2+v_6+h_1+h_3+h_5$}};
\node[rotate=90] at (15.3,8.6) {{\small $v_1+v_6+h_3+h_4+h_5$}};
\node at (4.2,2.2) {{\small $-v_1-h_3-h_4$}};
\node[rotate=270] at (6.5,2.2) {{\small $-v_5-h_2-h_4$}};
\node[rotate=270] at (8.5,3.2) {{\small $-v_3-h_2-h_6$}};
\node[rotate=270] at (10.5,4.2) {{\small $-v_4-h_1-h_6$}};
\node[rotate=270] at (12.5,5.2) {{\small $-v_2-h_1-h_5$}};
\node[rotate=270] at (14.5,6.2) {{\small $-v_6-h_3-h_5$}};
\node at (16.8,7.8) {{\small $-v_1-h_3-h_4$}};
\node at (5.3,1.4) {{\small $m''_6$}};
\node at (7.3,2.4) {{\small $m'_1$}};
\node at (9.3,3.4) {{\small $m'_2$}};
\node at (11.3,4.5) {{\small $m'_3$}};
\node at (13.3,5.5) {{\small $m'_4$}};
\node at (15.3,6.5) {{\small $m''_5$}};
\node at (6.3,3.6) {{\small $m''_6$}};
\node at (8.3,4.6) {{\small $m'_1$}};
\node at (10.3,5.6) {{\small $m'_2$}};
\node at (12.3,6.6) {{\small $m'_3$}};
\node at (14.3,7.6) {{\small $m'_4$}};
\node at (16.3,8.6) {{\small $m''_5$}};
\node at (3.9,1.9) {{\small \bf $a$}};
\node at (17.1,8.1) {{\small \bf $a$}};
\node at (5,0.8) {{\small \bf $6$}};
\node at (7,1.8) {{\small \bf $1$}};
\node at (9,2.8) {{\small \bf $2$}};
\node at (11,3.8) {{\small \bf $3$}};
\node at (13,4.8) {{\small \bf $4$}};
\node at (15,5.8) {{\small \bf $5$}};
\node at (6,4.2) {{\small \bf $6$}};
\node at (8,5.2) {{\small \bf $1$}};
\node at (10,6.2) {{\small \bf $2$}};
\node at (12,7.2) {{\small \bf $3$}};
\node at (14,8.2) {{\small \bf $4$}};
\node at (16,9.2) {{\small \bf $5$}};
\end{tikzpicture}}
}
\end{center}
\caption{\sl The final $(p,q)$-web diagram corresponding to $(N,M)=(6,1)$.}
\vskip0.2cm
\label{Fig:Web16}
\end{figure}
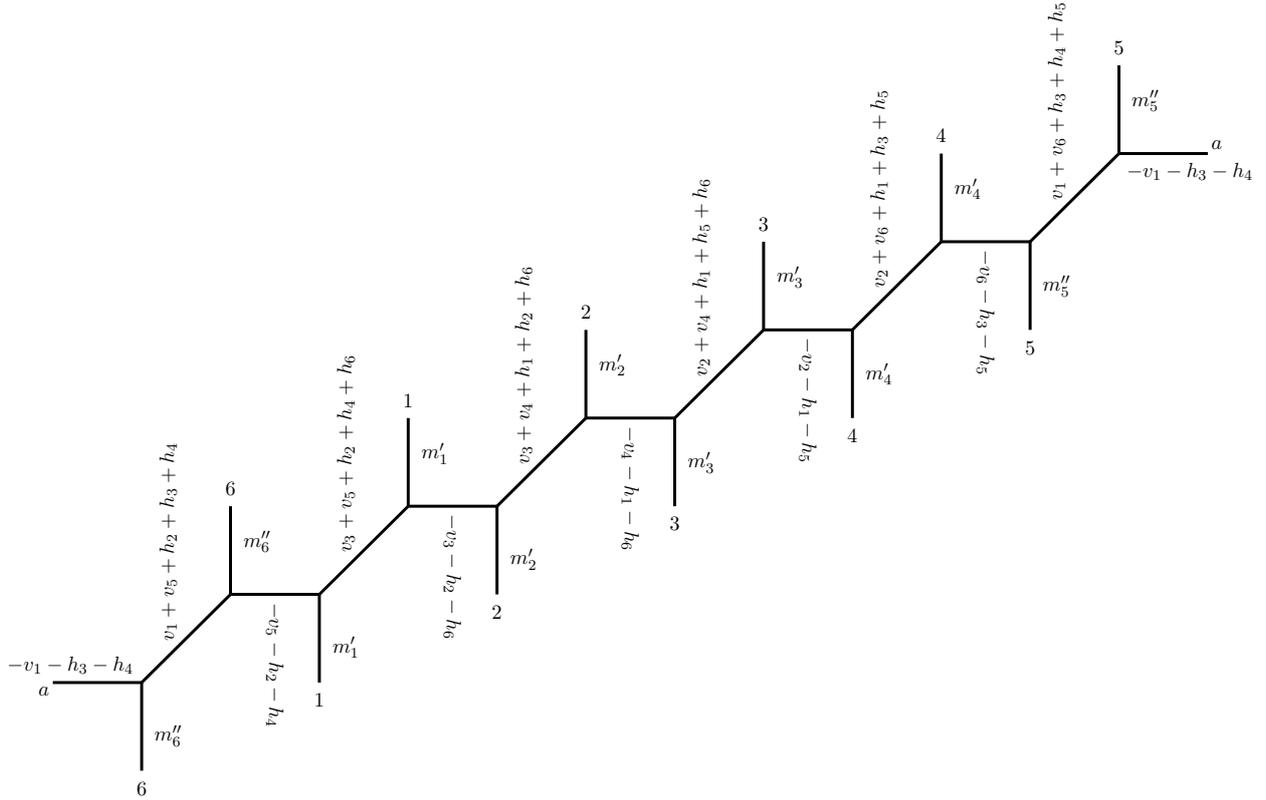

As a next step we perform an $SL(2,\mathbb{Z})$ transformation which acts in the same manner as (\ref{SLTrafo2}), \emph{i.e.}
\begin{align}
&(1,0)\longrightarrow (1,-1)\,,&&(0,1)\longrightarrow (0,1)\,,&&(1,1)\longrightarrow (1,0)\,.\label{SLTrafo3}
\end{align}
Thus, it transforms horizontal lines into diagonal ones and vice versa, while leaving vertical lines invariant. The resulting diagram is shown in Figure~\ref{Fig:SL3}
Finally, in order to obtain the familiar form of the $(N,M)=(6,1)$ web (see \cite{Haghighat:2013gba}), we cut the diagram in Figure~\ref{Fig:SL3} along the dashed line to arrive at Figure~\ref{Fig:Web16}.
Comparing Figure~\ref{Fig:Web16} with Figure~\ref{Fig:SL21}, we see that in the former the vertical lines are directly above each other and not 'rotated' against each other as in the former. Therefore, the diagram in Figure~\ref{Fig:Web16} indeed corresponds to the web $(N,M)=(6,1)$. The consistency conditions are the same as in Eq.(\ref{Cond1Diag}) and Eq.(\ref{Cond2Diag}), which are equivalent to those in Eq.(\ref{InitC1}) through Eq.(\ref{InitC6}) in the original diagram in figure~\ref{Fig:WebToric23}. Thus, by a series of flop transitions, we have been able to relate the Calabi-Yau threefold $X_{3,2}$ to $X_{6,1}$. 

In the F-theory compactification, the Calabi-Yau threefold $X_{3,2}$ gives rise to a little string theory with ${\cal N}=(1,0)$ supersymmetry, while the Calabi-Yau threefold $X_{6,1}$ gives rise to little string theory with $(2,0)$ supersymmetry. Hence, we observe that, in the extended moduli space, theories with different amount of supersymmetries can be mapped to each other via flop transitions. However, as we shall see with the help of further examples below, this phenomenon of supersymmetry enhancement appears only for little string theories labelled by $(N,M)$ with $\text{gcd}(N,M)=1$. 
Much like the number of 5-branes \footnote{It is worth to recall that single NS5-brane on $\mathbb{R}^4/\mathbb{Z}_6$ orbifold singularity is a nontrivial interacting theory, while single NS5-brane on $\mathbb{R}^4$ is a free, decoupled theory \cite{Rey:1991uu, Rey:1989xj}. }, the number of residual supersymmetries can change over the extended moduli space, and this crucially depends on the double elliptic fibration structure of the Calabi-Yau threefold $X_{M,N}$ .

Finally, we remark that a situation exhibiting a similar phenomenon of supersymmetry enhancement was observed in \cite{Chou:1997ba} in a  different context, namely, M-theory compactifications on compact Calabi-Yau threefolds. In this framework, flop transitions of birationally equivalent Calabi-Yau threefolds extend the Coulomb branch of five-dimensional ${\cal N}=1$ theories. It was demonstrated in \cite{Chou:1997ba} through several lower-rank examples that the supercharges are well defined everywhere in this extended moduli space, but that the energy and the charges of BPS configurations are not differentiable everywhere: indeed, the number of residual supersymmetries jumps across the walls of the K\"ahler cones (\emph{i.e.} the locus where the flop transition takes place). 

While the compact Calabi-Yau manifolds studied in \cite{Chou:1997ba} are of a different type than the $X_{N,M}$ discussed in the present work, the underlying moduli spaces share interesting common features: they both have finite volume and are parametrised by real-valued scalar fields, namely the scalar fields of five-dimensional ${\cal N}=1$ vector multiplets in the case of \cite{Chou:1997ba} and the scalar fields of six-dimensional $\mathcal{N}=(1,0)$ tensor multiplets in the present case. While these properties alone do not lead to supersymmetry enhancement across the walls of the extended moduli space (as we shall see explicitly from the example in section~\ref{Sect:46Example}), it raises the question how generic this phenomenon is for general Calabi-Yau compactifications. For the class $X_{N,M}$ of Calabi-Yau manifolds, (based on the discussion of section~\ref{Sect:RelNewtonPoly}) we can identify $\text{gcd}(N,M)=1$ as a sufficient condition for supersymmetry enhancement. It would be interesting to give an exhaustive list of necessary and sufficient conditions for generic classes of Calabi-Yau manifolds and for more general classes of little string theories. We leave this question for further research.

\FloatBarrier
\subsubsection{Simple Check of the Duality}\label{Sect:Sp2Check32}
As a simple check of the proposed relation between the partition functions of the $(N,M)=(3,2)$ and the $(6,1)$ brane webs, we consider the particular case
\begin{align}
&m=m_1=m_2=m_3=m_4=m_5=m_6\,,\nonumber\\
&v=v_1=v_2=v_3=v_4=v_5=v_6\,,\nonumber\\
&h=h_1=h_2=h_3=h_4=h_5=h_6\,.\label{ModuliIdentification}
\end{align}
which is a solution of Eq.(\ref{InitC1}) through Eq.(\ref{InitC6}) (and therefore also of Eq.(\ref{Cond1Diag}) and Eq.(\ref{Cond2Diag})). Moreover, as has been proposed in \cite{Hohenegger:2016eqy} at this particular region in the moduli space, the Nekrasov-Shatashvili limit of the free energies of a generic $(N,M)$ brane can be related to those of the web $(1,1)$ (see Eq.(\ref{RelMNto11})).
Thus, at Eq.(\ref{ModuliIdentification}), the transformation of the parameters obtained by comparing Figure~\ref{Fig:WebToric23} with Figure~\ref{Fig:Web16} must be a symmetry of the partition function of the brane web $(1,1)$. Our strategy of showing such an invariance is to show that this transformation is part of $Sp(2,\mathbb{Z})$ which was proposed in \cite{Dijkgraaf:1996xw,Hohenegger:2013ala} to be a symmetry of the $(M,N)=(1,1)$ case.

To this end, we start with the $(3,2)$ web and introduce
\begin{align}
&\rho=3m+3h\,,&&\text{and}&&\tau=2m+2v\,.
\end{align}
which entails
\begin{align}
&h=\frac{1}{3}(\rho-3m)\,,&&\text{and}&&v=\frac{1}{2}(\tau-2m)\,.
\end{align}
Furthermore, we introduce the $(1,1)$ period matrix
\begin{align}
\Omega=\left(\begin{array}{cc}\tau/2 & m \\ m & \rho/3\end{array}\right)\,.
\end{align}
Notice here the rescaling of the parameters, as is required to make contact with the configuration $(N,M)=(1,1)$. We now want to compare $\Omega$ to a similar period matrix obtained from the  $(6,1)$ web. For the latter, we introduce in a similar fashion
\begin{align}
&\rho'=6v+6h\,,&&\tau'=m+4v+9h\,,&&m'=2v+3h\,.
\end{align}
which in terms of $(\rho,\tau,m)$ reads
\begin{align}
&\rho'=3\tau+2\rho-12m\,,&&\tau'=2\tau+3\rho-12m\,,&&m'=\rho+\tau-5m\,.
\end{align}
We also introduce the $(1,1)$ period matrix
\begin{align}
\Omega'=\left(\begin{array}{cc}\tau' & m' \\ m' & \rho'/6\end{array}\right)=
\left(
\begin{array}{cc}
 2 \tau+3 \rho-12 m  & \tau+\rho -5 m  \\
 \tau+\rho -5 m & \frac{1}{6} (3 \tau+2 \rho-12 m  ) \\
\end{array}
\right)\,.
\end{align}
The two period matrices $\Omega$ and $\Omega'$ are related through an $Sp(2,\mathbb{Z})$ transformation
\begin{align}
\Omega'=(A\Omega+B)\cdot (C\Omega+D)^{-1}\,,
\end{align}
with
\begin{align}
&A=\left(\begin{array}{cc}2 & -3 \\ 1 & -1\end{array}\right)\,,&&B=\left(\begin{array}{cc}0 & 0 \\ 0 & 0\end{array}\right)\,,&&C=\left(\begin{array}{cc}0 & 0 \\ 0 & 0\end{array}\right)\,,&&D=\left(\begin{array}{cc}-1 & -1 \\ 3 & 2\end{array}\right)\,,\label{SpTrafo61}
\end{align}
which satisfy Eq.(\ref{Sp2ZTrafos}) (notice also that $\text{det}\Omega=\frac{\rho\tau}{6}-m^2=\text{det}\Omega'$). This shows that $\mathcal{Z}_{X_{1,1}}$ is indeed invariant under the change of parameters implied by the above chain of dualities.\footnote{Notice that due to $\text{det}(C\Omega+D)=1$ for (\ref{SpTrafo61}), a weight factor in the $Sp(2,\mathbb{Z})$ transformation is irrelevant.}
\subsubsection{Vertex Identification}
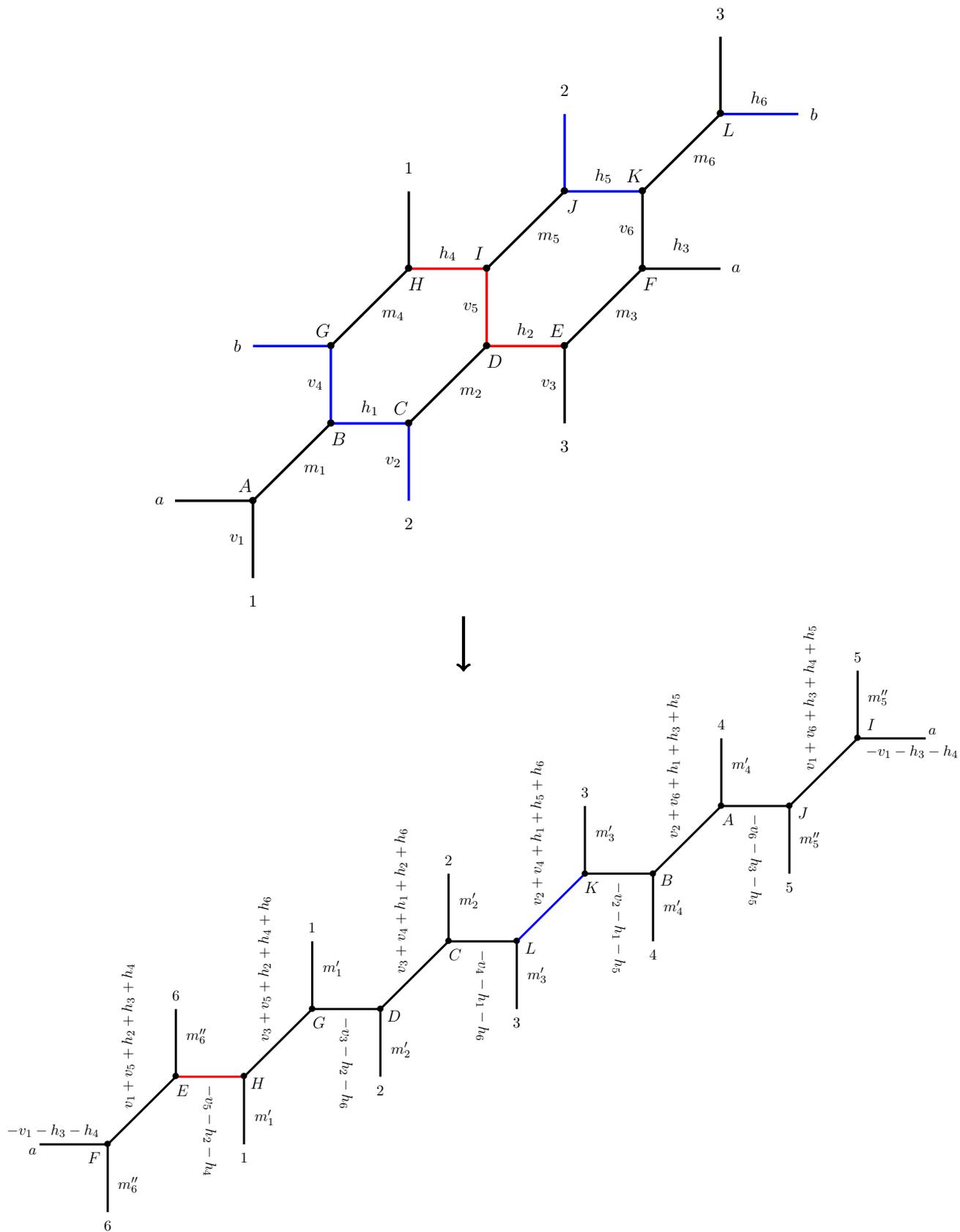
\begin{figure}[htbp]
\begin{center}
\scalebox{0.8}{\parbox{14cm}{\begin{tikzpicture}[scale = 1.75]
\draw[ultra thick] (-6,0) -- (-5,0);
\draw[ultra thick] (-5,-1) -- (-5,0);
\draw[ultra thick] (-5,0) -- (-4,1);
\draw[ultra thick,blue] (-4,1) -- (-3,1);
\draw[ultra thick,blue] (-4,1) -- (-4,2);
\draw[ultra thick,blue] (-3,1) -- (-3,0);
\draw[ultra thick] (-3,1) -- (-2,2);
\draw[ultra thick] (-4,2) -- (-3,3);
\draw[ultra thick,blue] (-5,2) -- (-4,2);
\draw[ultra thick] (-3,3) -- (-3,4);
\draw[ultra thick,red] (-3,3) -- (-2,3);
\draw[ultra thick,red] (-2,2) -- (-2,3);
\draw[ultra thick,red] (-2,2) -- (-1,2);
\draw[ultra thick] (-2,3) -- (-1,4);
\draw[ultra thick] (-1,2) -- (0,3);
\draw[ultra thick] (-1,2) -- (-1,1);

\draw[ultra thick,blue] (-1,4) -- (0,4);
\draw[ultra thick] (0,3) -- (1,3);
\draw[ultra thick] (0,3) -- (0,4);
\draw[ultra thick] (0,4) -- (1,5);
\draw[ultra thick,blue] (1,5) -- (2,5);
\draw[ultra thick] (1,5) -- (1,6);
\draw[ultra thick,blue] (-1,4) -- (-1,5);
\node at (-6.2,0) {{\small \bf $a$}};
\node at (-5.2,2) {{\small \bf $b$}};
\node at (1.2,3) {{\small \bf $a$}};
\node at (2.2,5) {{\small \bf $b$}};
\node at (-5,-1.3) {{\small \bf $1$}};
\node at (-3,-0.3) {{\small \bf $2$}};
\node at (-1,0.7) {{\small \bf $3$}};
\node at (-3,4.3) {{\small \bf $1$}};
\node at (-1,5.3) {{\small \bf $2$}};
\node at (1,6.3) {{\small \bf $3$}};
\node at (-4.2,0.4) {{\small $m_1$}};
\node at (-2.2,1.4) {{\small $m_2$}};
\node at (-0.2,2.4) {{\small $m_3$}};
\node at (-3.2,2.4) {{\small $m_4$}};
\node at (-1.2,3.4) {{\small $m_5$}};
\node at (0.8,4.4) {{\small $m_6$}};
\node at (-3.5,1.2) {{\small $h_1$}};
\node at (-1.5,2.2) {{\small $h_2$}};
\node at (0.5,3.3) {{\small $h_3$}};
\node at (-2.5,3.2) {{\small $h_4$}};
\node at (-0.5,4.2) {{\small $h_5$}};
\node at (1.5,5.2) {{\small $h_6$}};
\node at (-5.2,-0.5) {{\small $v_1$}};
\node at (-3.2,0.5) {{\small $v_2$}};
\node at (-1.2,1.5) {{\small $v_3$}};
\node at (-4.2,1.5) {{\small $v_4$}};
\node at (-2.2,2.5) {{\small $v_5$}};
\node at (-0.2,3.5) {{\small $v_6$}};
\node at (-5,0) {$\bullet$};
\node at (-5.1,0.2) {$A$};
\node at (-4,1) {$\bullet$};
\node at (-3.9,0.8) {$B$};
\node at (-3,1) {$\bullet$};
\node at (-3.1,1.2) {$C$};
\node at (-2,2) {$\bullet$};
\node at (-1.9,1.8) {$D$};
\node at (-1,2) {$\bullet$};
\node at (-1.1,2.2) {$E$};
\node at (0,3) {$\bullet$};
\node at (0.1,2.8) {$F$};
\node at (-4,2) {$\bullet$};
\node at (-4.1,2.2) {$G$};
\node at (-3,3) {$\bullet$};
\node at (-2.9,2.8) {$H$};
\node at (-2,3) {$\bullet$};
\node at (-2.1,3.2) {$I$};
\node at (-1,4) {$\bullet$};
\node at (-0.9,3.8) {$J$};
\node at (0,4) {$\bullet$};
\node at (-0.1,4.2) {$K$};
\node at (1,5) {$\bullet$};
\node at (1.1,4.8) {$L$};
\end{tikzpicture}}
}\\
\begin{tikzpicture}
\draw[ultra thick,->] (0,0) -- (0,-1);
\end{tikzpicture}\\
\vspace{-1cm}
\scalebox{0.7}{\parbox{25cm}{\begin{tikzpicture}[scale = 1.75]
\draw[ultra thick] (5,2) -- (5,1);
\draw[ultra thick] (6,3) -- (6,4);
\draw[ultra thick] (7,3) -- (7,2);
\draw[ultra thick] (8,4) -- (8,5);
\draw[ultra thick] (9,4) -- (9,3);
\draw[ultra thick] (10,5) -- (10,6);
\draw[ultra thick] (11,5) -- (11,4);
\draw[ultra thick] (12,6) -- (12,7);
\draw[ultra thick] (13,6) -- (13,5);
\draw[ultra thick] (14,7) -- (14,8);
\draw[ultra thick] (15,7) -- (15,6);
\draw[ultra thick] (16,8) -- (16,9);
\draw[ultra thick] (4,2) -- (5,2);
\draw[ultra thick,red] (6,3) -- (7,3);
\draw[ultra thick] (8,4) -- (9,4);
\draw[ultra thick] (10,5) -- (11,5);
\draw[ultra thick] (12,6) -- (13,6);
\draw[ultra thick] (14,7) -- (15,7);
\draw[ultra thick] (16,8) -- (17,8);
\draw[ultra thick] (5,2) -- (6,3);
\draw[ultra thick] (7,3) -- (8,4);
\draw[ultra thick] (9,4) -- (10,5);
\draw[ultra thick,blue] (11,5) -- (12,6);
\draw[ultra thick] (13,6) -- (14,7);
\draw[ultra thick] (15,7) -- (16,8);
\node[rotate=90] at (5.3,3.6) {{\small $v_1+v_5+h_2+h_3+h_4$}};
\node[rotate=90] at (7.3,4.6) {{\small $v_3+v_5+h_2+h_4+h_6$}};
\node[rotate=90] at (9.3,5.6) {{\small $v_3+v_4+h_1+h_2+h_6$}};
\node[rotate=90] at (11.3,6.6) {{\small $v_2+v_4+h_1+h_5+h_6$}};
\node[rotate=90] at (13.3,7.6) {{\small $v_2+v_6+h_1+h_3+h_5$}};
\node[rotate=90] at (15.3,8.6) {{\small $v_1+v_6+h_3+h_4+h_5$}};
\node at (4.2,2.2) {{\small $-v_1-h_3-h_4$}};
\node[rotate=270] at (6.5,2.2) {{\small $-v_5-h_2-h_4$}};
\node[rotate=270] at (8.5,3.2) {{\small $-v_3-h_2-h_6$}};
\node[rotate=270] at (10.5,4.2) {{\small $-v_4-h_1-h_6$}};
\node[rotate=270] at (12.5,5.2) {{\small $-v_2-h_1-h_5$}};
\node[rotate=270] at (14.5,6.2) {{\small $-v_6-h_3-h_5$}};
\node at (16.8,7.8) {{\small $-v_1-h_3-h_4$}};
\node at (5.3,1.4) {{\small $m''_6$}};
\node at (7.3,2.4) {{\small $m'_1$}};
\node at (9.3,3.4) {{\small $m'_2$}};
\node at (11.3,4.5) {{\small $m'_3$}};
\node at (13.3,5.5) {{\small $m'_4$}};
\node at (15.3,6.5) {{\small $m''_5$}};
\node at (6.3,3.6) {{\small $m''_6$}};
\node at (8.3,4.6) {{\small $m'_1$}};
\node at (10.3,5.6) {{\small $m'_2$}};
\node at (12.3,6.6) {{\small $m'_3$}};
\node at (14.3,7.6) {{\small $m'_4$}};
\node at (16.3,8.6) {{\small $m''_5$}};
\node at (3.9,1.9) {{\small \bf $a$}};
\node at (17.1,8.1) {{\small \bf $a$}};
\node at (5,0.8) {{\small \bf $6$}};
\node at (7,1.8) {{\small \bf $1$}};
\node at (9,2.8) {{\small \bf $2$}};
\node at (11,3.8) {{\small \bf $3$}};
\node at (13,4.8) {{\small \bf $4$}};
\node at (15,5.8) {{\small \bf $5$}};
\node at (6,4.2) {{\small \bf $6$}};
\node at (8,5.2) {{\small \bf $1$}};
\node at (10,6.2) {{\small \bf $2$}};
\node at (12,7.2) {{\small \bf $3$}};
\node at (14,8.2) {{\small \bf $4$}};
\node at (16,9.2) {{\small \bf $5$}};
\node at (5,2) {$\bullet$};
\node at (4.8,1.8) {$F$};
\node at (6,3) {$\bullet$};
\node at (6.1,2.8) {$E$};
\node at (7,3) {$\bullet$};
\node at (7.2,2.9) {$H$};
\node at (8,4) {$\bullet$};
\node at (8.1,3.8) {$G$};
\node at (9,4) {$\bullet$};
\node at (9.2,3.9) {$D$};
\node at (10,5) {$\bullet$};
\node at (10.1,4.8) {$C$};
\node at (11,5) {$\bullet$};
\node at (11.2,4.9) {$L$};
\node at (12,6) {$\bullet$};
\node at (12.1,5.8) {$K$};
\node at (13,6) {$\bullet$};
\node at (13.2,5.9) {$B$};
\node at (14,7) {$\bullet$};
\node at (14.1,6.8) {$A$};
\node at (15,7) {$\bullet$};
\node at (15.2,6.9) {$J$};
\node at (16,8) {$\bullet$};
\node at (16.2,8.2) {$I$};
\end{tikzpicture}}
}
\end{center}
\caption{\sl Identifying vertices in the $(3,2)$ and $(6,1)$ webs.}
\label{Fig:Compare}
\end{figure}

We compare the diagrams in Figure~\ref{Fig:Cutting23} and Figure~\ref{Fig:Web16}, where we identified vertices that are mapped into each other (see Figure~\ref{Fig:Compare}).
We notice that the length of the horizontal and diagonal lines in the $(6,1)$ diagram can be read off as the distances (using only horizontal and vertical lines) between the same vertices in the $(3,2)$ web. As an example, we have highlighted the distance between the vertices $E$ and $H$ as well as $L$ and $K$ respectively. Notice that this rule does not only apply to neighboring vertices: \emph{e.g.} the (shortest)~\footnote{Counting in the opposite direction in the $(6,1)$ web corresponds to $\rho'+h_4$, where $\rho'$ is the length of the horizontal circle in the $(6,1)$ web.} distance between vertices $H$ and $I$ is (again we only use horizontal and vertical lines)
\begin{align}
h_4=v_5+h_2+h_4-v_1-v_5-h_2-h_3-h_4+v_1+h_3+h_4\,.
\end{align}
In section~\ref{Sect:GenericNM} we discuss how to generalise this observation to a general pattern that allows reconstructing the web $(\tfrac{NM}{k},k)$ starting from a labelling of the vertices of the web $(N,M)$.
\subsection{Relation between $(N,M)=(2,5)$ and $(10,1)$}
The next example is the web $(N,M)=(2,5)$.
\subsubsection{Duality Transformation}
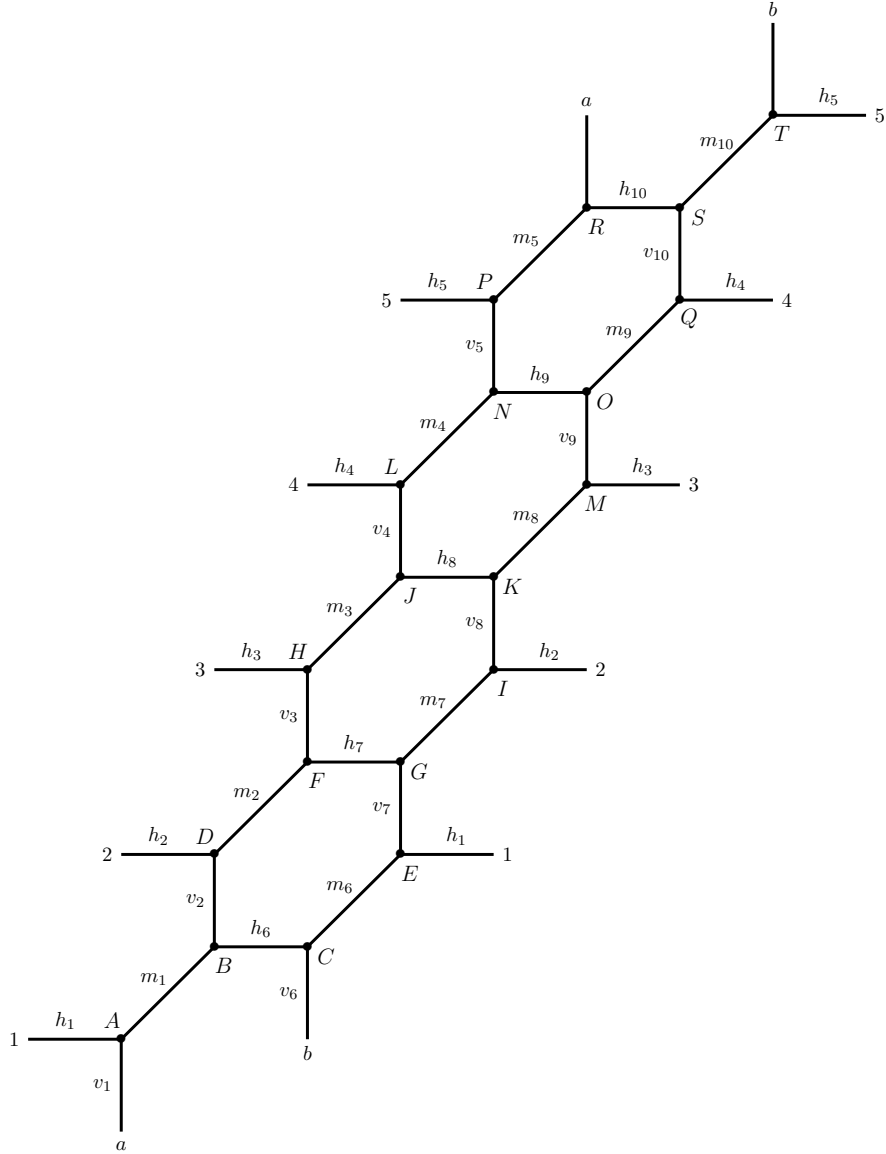
\begin{figure}[htbp]
\begin{center}
\scalebox{0.7}{\parbox{16cm}{\begin{tikzpicture}[scale = 1.75]
\draw[ultra thick] (0,0) -- (1,1);
\draw[ultra thick] (1,2) -- (2,3);
\draw[ultra thick] (2,4) -- (3,5);
\draw[ultra thick] (3,6) -- (4,7);
\draw[ultra thick] (4,8) -- (5,9);
\draw[ultra thick] (2,1) -- (3,2);
\draw[ultra thick] (3,3) -- (4,4);
\draw[ultra thick] (4,5) -- (5,6);
\draw[ultra thick] (5,7) -- (6,8);
\draw[ultra thick] (6,9) -- (7,10);
\draw[ultra thick] (0,-1) -- (0,0);
\draw[ultra thick] (1,1) -- (1,2);
\draw[ultra thick] (2,3) -- (2,4);
\draw[ultra thick] (3,5) -- (3,6);
\draw[ultra thick] (4,7) -- (4,8);
\draw[ultra thick] (5,9) -- (5,10);
\draw[ultra thick] (2,1) -- (2,0);
\draw[ultra thick] (3,2) -- (3,3);
\draw[ultra thick] (4,4) -- (4,5);
\draw[ultra thick] (5,6) -- (5,7);
\draw[ultra thick] (6,8) -- (6,9);
\draw[ultra thick] (7,10) -- (7,11);
\draw[ultra thick] (-1,0) -- (0,0);
\draw[ultra thick] (0,2) -- (1,2);
\draw[ultra thick] (1,4) -- (2,4);
\draw[ultra thick] (2,6) -- (3,6);
\draw[ultra thick] (3,8) -- (4,8);
\draw[ultra thick] (1,1) -- (2,1);
\draw[ultra thick] (2,3) -- (3,3);
\draw[ultra thick] (3,5) -- (4,5);
\draw[ultra thick] (4,7) -- (5,7);
\draw[ultra thick] (5,9) -- (6,9);
\draw[ultra thick] (3,2) -- (4,2);
\draw[ultra thick] (4,4) -- (5,4);
\draw[ultra thick] (5,6) -- (6,6);
\draw[ultra thick] (6,8) -- (7,8);
\draw[ultra thick] (7,10) -- (8,10);
\node at (0,0) {$\bullet$};
\node at (-0.1,0.2) {$A$};
\node at (1,1) {$\bullet$};
\node at (1.1,0.8) {$B$};
\node at (2,1) {$\bullet$};
\node at (2.2,0.9) {$C$};
\node at (1,2) {$\bullet$};
\node at (0.9,2.2) {$D$};
\node at (3,2) {$\bullet$};
\node at (3.1,1.8) {$E$};
\node at (2,3) {$\bullet$};
\node at (2.1,2.8) {$F$};
\node at (3,3) {$\bullet$};
\node at (3.2,2.9) {$G$};
\node at (2,4) {$\bullet$};
\node at (1.9,4.2) {$H$};
\node at (4,4) {$\bullet$};
\node at (4.1,3.8) {$I$};
\node at (3,5) {$\bullet$};
\node at (3.1,4.8) {$J$};
\node at (4,5) {$\bullet$};
\node at (4.2,4.9) {$K$};
\node at (3,6) {$\bullet$};
\node at (2.9,6.2) {$L$};
\node at (5,6) {$\bullet$};
\node at (5.1,5.8) {$M$};
\node at (4,7) {$\bullet$};
\node at (4.1,6.8) {$N$};
\node at (5,7) {$\bullet$};
\node at (5.2,6.9) {$O$};
\node at (4,8) {$\bullet$};
\node at (3.9,8.2) {$P$};
\node at (6,8) {$\bullet$};
\node at (6.1,7.8) {$Q$};
\node at (5,9) {$\bullet$};
\node at (5.1,8.8) {$R$};
\node at (6,9) {$\bullet$};
\node at (6.2,8.9) {$S$};
\node at (7,10) {$\bullet$};
\node at (7.1,9.8) {$T$};
%
\node at (-0.2,-0.5) {{\small \bf $v_1$}};
\node at (0.8,1.5) {{\small \bf $v_2$}};
\node at (1.8,3.5) {{\small \bf $v_3$}};
\node at (2.8,5.5) {{\small \bf $v_4$}};
\node at (3.8,7.5) {{\small \bf $v_5$}};
\node at (1.8,0.5) {{\small \bf $v_6$}};
\node at (2.8,2.5) {{\small \bf $v_7$}};
\node at (3.8,4.5) {{\small \bf $v_8$}};
\node at (4.8,6.5) {{\small \bf $v_9$}};
\node at (5.75,8.5) {{\small \bf $v_{10}$}};
%
\node at (-0.6,0.2) {{\small \bf $h_1$}};
\node at (0.4,2.2) {{\small \bf $h_2$}};
\node at (1.4,4.2) {{\small \bf $h_3$}};
\node at (2.4,6.2) {{\small \bf $h_4$}};
\node at (3.4,8.2) {{\small \bf $h_5$}};
\node at (1.5,1.2) {{\small \bf $h_6$}};
\node at (2.5,3.2) {{\small \bf $h_7$}};
\node at (3.5,5.2) {{\small \bf $h_8$}};
\node at (4.5,7.2) {{\small \bf $h_9$}};
\node at (5.5,9.2) {{\small \bf $h_{10}$}};
\node at (3.6,2.2) {{\small \bf $h_1$}};
\node at (4.6,4.2) {{\small \bf $h_2$}};
\node at (5.6,6.2) {{\small \bf $h_3$}};
\node at (6.6,8.2) {{\small \bf $h_4$}};
\node at (7.6,10.2) {{\small \bf $h_5$}};
%
\node at (0.35,0.65) {{\small \bf $m_1$}};
\node at (1.35,2.65) {{\small \bf $m_2$}};
\node at (2.35,4.65) {{\small \bf $m_3$}};
\node at (3.35,6.65) {{\small \bf $m_4$}};
\node at (4.35,8.65) {{\small \bf $m_5$}};
%
\node at (2.35,1.65) {{\small \bf $m_6$}};
\node at (3.35,3.65) {{\small \bf $m_7$}};
\node at (4.35,5.65) {{\small \bf $m_8$}};
\node at (5.35,7.65) {{\small \bf $m_9$}};
\node at (6.4,9.7) {{\small \bf $m_{10}$}};
\node at (0,-1.15) {{\small \bf $a$}};
\node at (2,-0.15) {{\small \bf $b$}};
\node at (5,10.15) {{\small \bf $a$}};
\node at (7,11.15) {{\small \bf $b$}};
\node at (-1.15,0) {{\small \bf $1$}};
\node at (-0.15,2) {{\small \bf $2$}};
\node at (0.85,4) {{\small \bf $3$}};
\node at (1.85,6) {{\small \bf $4$}};
\node at (2.85,8) {{\small \bf $5$}};
\node at (4.15,2) {{\small \bf $1$}};
\node at (5.15,4) {{\small \bf $2$}};
\node at (6.15,6) {{\small \bf $3$}};
\node at (7.15,8) {{\small \bf $4$}};
\node at (8.15,10) {{\small \bf $5$}};
\end{tikzpicture}}
}
\end{center}
\caption{\sl Parametrisation of the $(N,M)=(2,5)$ web and a labelling of the vertices.}
\label{Fig:Web52Paravertices}
\end{figure}

\begin{figure}[htb]
\begin{center}
\scalebox{0.7}{\parbox{25cm}{\begin{tikzpicture}[scale = 1.1]
\draw[ultra thick] (5,2) -- (5,1);
\draw[ultra thick] (6,3) -- (6,4);
\draw[ultra thick] (7,3) -- (7,2);
\draw[ultra thick] (8,4) -- (8,5);
\draw[ultra thick] (9,4) -- (9,3);
\draw[ultra thick] (10,5) -- (10,6);
\draw[ultra thick] (11,5) -- (11,4);
\draw[ultra thick] (12,6) -- (12,7);
\draw[ultra thick] (13,6) -- (13,5);
\draw[ultra thick] (14,7) -- (14,8);
\draw[ultra thick] (15,7) -- (15,6);
\draw[ultra thick] (16,8) -- (16,9);
\draw[ultra thick] (17,8) -- (17,7);
\draw[ultra thick] (18,9) -- (18,10);
\draw[ultra thick] (19,9) -- (19,8);
\draw[ultra thick] (20,10) -- (20,11);
\draw[ultra thick] (21,10) -- (21,9);
\draw[ultra thick] (22,11) -- (22,12);
\draw[ultra thick] (23,11) -- (23,10);
\draw[ultra thick] (24,12) -- (24,13);
\draw[ultra thick] (4,2) -- (5,2);
\draw[ultra thick] (6,3) -- (7,3);
\draw[ultra thick] (8,4) -- (9,4);
\draw[ultra thick] (10,5) -- (11,5);
\draw[ultra thick] (12,6) -- (13,6);
\draw[ultra thick] (14,7) -- (15,7);
\draw[ultra thick] (16,8) -- (17,8);
\draw[ultra thick] (18,9) -- (19,9);
\draw[ultra thick] (20,10) -- (21,10);
\draw[ultra thick] (22,11) -- (23,11);
\draw[ultra thick] (24,12) -- (25,12);
\draw[ultra thick] (5,2) -- (6,3);
\draw[ultra thick] (7,3) -- (8,4);
\draw[ultra thick] (9,4) -- (10,5);
\draw[ultra thick] (11,5) -- (12,6);
\draw[ultra thick] (13,6) -- (14,7);
\draw[ultra thick] (15,7) -- (16,8);
\draw[ultra thick] (17,8) -- (18,9);
\draw[ultra thick] (19,9) -- (20,10);
\draw[ultra thick] (21,10) -- (22,11);
\draw[ultra thick] (23,11) -- (24,12);
\node[rotate=45] at (5.3,2.8) {{\small $\overline{m}_{10}$}};
\node[rotate=45] at (7.3,3.8) {{\small $\overline{m}_9$}};
\node[rotate=45] at (9.3,4.8) {{\small $\overline{m}_8$}};
\node[rotate=45] at (11.3,5.8) {{\small $\overline{m}_7$}};
\node[rotate=45] at (13.3,6.8) {{\small $\overline{m}_6$}};
\node[rotate=45] at (15.3,7.8) {{\small $\overline{m}_5$}};
\node[rotate=45] at (17.3,8.8) {{\small $\overline{m}_4$}};
\node[rotate=45] at (19.3,9.8) {{\small $\overline{m}_3$}};
\node[rotate=45] at (21.3,10.8) {{\small $\overline{m}_2$}};
\node[rotate=45] at (23.3,11.8) {{\small $\overline{m}_{1}$}};
\node at (4.5,2.3) {{\small $\overline{h}_{10}$}};
\node at (6.5,3.3) {{\small $\overline{h}_9$}};
\node at (8.5,4.3) {{\small $\overline{h}_8$}};
\node at (10.5,5.3) {{\small $\overline{h}_7$}};
\node at (12.5,6.3) {{\small $\overline{h}_6$}};
\node at (14.5,7.3) {{\small $\overline{h}_5$}};
\node at (16.5,8.3) {{\small $\overline{h}_4$}};
\node at (18.5,9.3) {{\small $\overline{h}_3$}};
\node at (20.5,10.3) {{\small $\overline{h}_2$}};
\node at (22.5,11.3) {{\small $\overline{h}_{1}$}};
\node at (24.5,12.3) {{\small $\overline{h}_{10}$}};
\node at (4.7,1.2) {{\small $\overline{v}_{10}$}};
\node at (6.7,2.2) {{\small $\overline{v}_9$}};
\node at (8.7,3.2) {{\small $\overline{v}_8$}};
\node at (10.7,4.2) {{\small $\overline{v}_7$}};
\node at (12.7,5.2) {{\small $\overline{v}_6$}};
\node at (14.7,6.2) {{\small $\overline{v}_5$}};
\node at (16.7,7.2) {{\small $\overline{v}_4$}};
\node at (18.7,8.2) {{\small $\overline{v}_3$}};
\node at (20.7,9.2) {{\small $\overline{v}_2$}};
\node at (22.7,10.2) {{\small $\overline{v}_1$}};
\node at (6.3,3.8) {{\small $\overline{v}_{10}$}};
\node at (8.3,4.8) {{\small $\overline{v}_9$}};
\node at (10.3,5.8) {{\small $\overline{v}_8$}};
\node at (12.3,6.8) {{\small $\overline{v}_7$}};
\node at (14.3,7.8) {{\small $\overline{v}_6$}};
\node at (16.3,8.8) {{\small $\overline{v}_5$}};
\node at (18.3,9.8) {{\small $\overline{v}_4$}};
\node at (20.3,10.8) {{\small $\overline{v}_3$}};
\node at (22.3,11.8) {{\small $\overline{v}_{2}$}};
\node at (24.3,12.8) {{\small $\overline{v}_1$}};
\node at (3.9,1.9) {{\small \bf $a$}};
\node at (25.2,12.1) {{\small \bf $a$}};
\node at (5,0.8) {{\small \bf $10$}};
\node at (7,1.8) {{\small \bf $9$}};
\node at (9,2.8) {{\small \bf $8$}};
\node at (11,3.8) {{\small \bf $7$}};
\node at (13,4.8) {{\small \bf $6$}};
\node at (15,5.8) {{\small \bf $5$}};
\node at (17,6.8) {{\small \bf $4$}};
\node at (19,7.8) {{\small \bf $3$}};
\node at (21,8.8) {{\small \bf $2$}};
\node at (23,9.8) {{\small \bf $1$}};
\node at (6,4.2) {{\small \bf $10$}};
\node at (8,5.2) {{\small \bf $9$}};
\node at (10,6.2) {{\small \bf $8$}};
\node at (12,7.2) {{\small \bf $7$}};
\node at (14,8.2) {{\small \bf $6$}};
\node at (16,9.2) {{\small \bf $5$}};
\node at (18,10.2) {{\small \bf $4$}};
\node at (20,11.2) {{\small \bf $3$}};
\node at (22,12.2) {{\small \bf $2$}};
\node at (24,13.2) {{\small \bf $1$}};
%
\node at (5,2) {$\bullet$};
\node at (5.3,1.9) {$I$};
\node at (6,3) {$\bullet$};
\node at (6.1,2.7) {$G$};
\node at (7,3) {$\bullet$};
\node at (7.3,2.9) {$J$};
\node at (8,4) {$\bullet$};
\node at (8.1,3.7) {$H$};
\node at (9,4) {$\bullet$};
\node at (9.3,3.9) {$Q$};
\node at (10,5) {$\bullet$};
\node at (10.1,4.7) {$O$};
\node at (11,5) {$\bullet$};
\node at (11.3,4.9) {$R$};
\node at (12,6) {$\bullet$};
\node at (12.1,5.7) {$P$};
\node at (13,6) {$\bullet$};
\node at (13.3,5.9) {$E$};
\node at (14,7) {$\bullet$};
\node at (14.1,6.7) {$C$};
\node at (15,7) {$\bullet$};
\node at (15.3,6.9) {$F$};
\node at (16,8) {$\bullet$};
\node at (16.1,7.7) {$D$};
\node at (17,8) {$\bullet$};
\node at (17.3,7.9) {$M$};
\node at (18,9) {$\bullet$};
\node at (18.1,8.7) {$K$};
\node at (19,9) {$\bullet$};
\node at (19.3,8.9) {$N$};
\node at (20,10) {$\bullet$};
\node at (20.1,9.7) {$L$};
\node at (21,10) {$\bullet$};
\node at (21.3,9.9) {$T$};
\node at (22,11) {$\bullet$};
\node at (22.1,10.7) {$S$};
\node at (23,11) {$\bullet$};
\node at (23.3,10.9) {$B$};
\node at (24,12) {$\bullet$};
\node at (24.1,11.7) {$A$};
\end{tikzpicture}}
}
\end{center}
\caption{\sl Parametrisation of the $(N,M)=(10,1)$ web and a labelling of the vertices that is compatible with the one in Figure~\ref{Fig:Web52Paravertices}.}
\label{Fig:Web101Paravertices}
\end{figure}
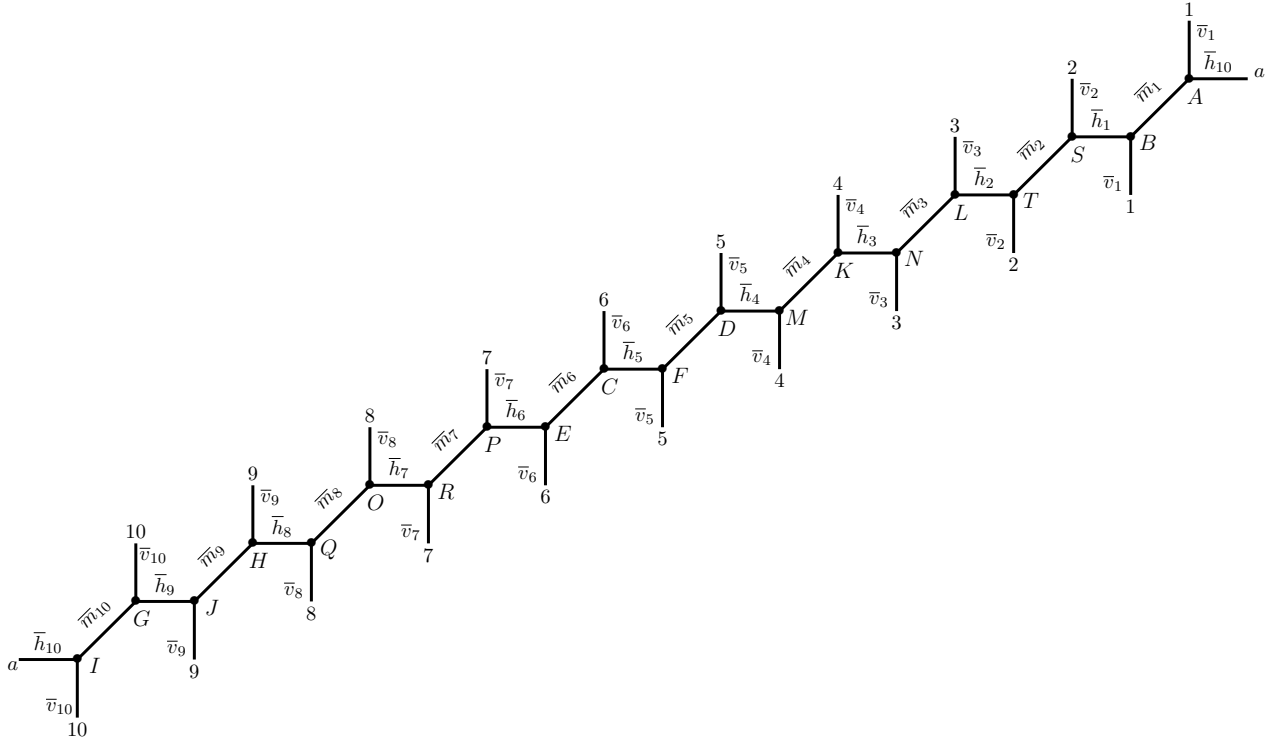

A parametrisation of the $(N,M)=(2,5)$ web is given in Figure~\ref{Fig:Web52Paravertices}. Out of the 30 parameters $(h_i,v_i,m_i)$ with $i=1,\ldots,10$, only 12 are independent. The remaining ones are fixed by consistency relations, as discussed above.
Following a similar sequence of flop-transitions, cutting and re-gluing the diagram, as well as $SL(2,\mathbb{Z})$ transformations, we can dualise the $(N,M)=(2,5)$ to the $(10,1)$ web with the parametrisation given in Figure~\ref{Fig:Web101Paravertices}.
Explicitly, the parameters $\bar{v}_i$ are given in terms of $(h_i,v_i,m_i)$ as
{\allowdisplaybreaks\bea
\overline{v}_{1}&=&m_{1}+4(v_{1}+v_{2}+v_{4}+v_{8}+v_{10})+3(h_{2}+h_{4}+h_{8}+h_{10})\,,\nonumber\\\nonumber
\overline{v}_{2}&=&m_{10}+4(v_{2}+v_{4}+v_{6}+v_{8}+v_{10})+3(h_{2}+h_{4}+h_{6}+h_{8})\,,\\\nonumber
\overline{v}_{3}&=&m_{4}+4(v_{2}+v_{4}+v_{5}+v_{6}+v_{8})+3(h_{2}+h_{5}+h_{6}+h_{8})\,,\\\nonumber
\overline{v}_{4}&=&m_{8}+4(v_{2}+v_{5}+v_{6}+v_{8}+v_{9})+3(h_{2}+h_{5}+h_{6}+h_{9})\,,\\\nonumber
\overline{v}_{5}&=&m_{2}+4(v_{2}+v_{3}+v_{5}+v_{6}+v_{9})+3(h_{3}+h_{5}+h_{6}+h_{9})\,,\\\nonumber
\overline{v}_{6}&=&m_{6}+4(v_{3}+v_{5}+v_{6}+v_{7}+v_{9})+3(h_{3}+h_{5}+h_{7}+h_{9})\,,\\\nonumber
\overline{v}_{7}&=&m_{5}+4(v_{1}+v_{3}+v_{5}+v_{7}+v_{9})+3(h_{1}+h_{3}+h_{7}+h_{9})\,,\\\nonumber
\overline{v}_{8}&=&m_{9}+4(v_{1}+v_{3}+v_{7}+v_{9}+v_{10})+3(h_{1}+h_{3}+h_{7}+h_{10})\,,\\\nonumber
\overline{v}_{9}&=&m_{3}+4(v_{1}+v_{3}+v_{4}+v_{7}+v_{10})+3(h_{1}+h_{4}+h_{7}+h_{10})\,,\\
\overline{v}_{10}&=&m_{7}+4(v_{1}+v_{4}+v_{7}+v_{8}+v_{10})+3(h_{1}+h_{4}+h_{8}+h_{10})\,,
\eea}
while the parameters $\overline{h}$ are
{\allowdisplaybreaks
\begin{align}
&\overline{h}_{1}=-(v_{2}+v_{4}+v_{8}+v_{10}+h_{2}+h_{4}+h_{8})\,,&&
\overline{h}_{2}=-(v_{2}+v_{4}+v_{6}+v_{8}+h_{2}+h_{6}+h_{8})\,,\nonumber\\
&\overline{h}_{3}=-(v_{2}+v_{5}+v_{6}+v_{8}+h_{2}+h_{5}+h_{6})\,,&&
\overline{h}_{4}=-(v_{2}+v_{5}+v_{6}+v_{9}+h_{5}+h_{6}+h_{9})\,,\nonumber\\
&\overline{h}_{5}=-(v_{3}+v_{5}+v_{6}+v_{9}+h_{3}+h_{5}+h_{9})\,,&&
\overline{h}_{6}=-(v_{3}+v_{5}+v_{7}+v_{9}+h_{3}+h_{7}+h_{9})\,,\nonumber\\
&\overline{h}_{7}=-(v_{1}+v_{3}+v_{7}+v_{9}+h_{1}+h_{3}+h_{7})\,,&&
\overline{h}_{8}=-(v_{1}+v_{3}+v_{7}+v_{10}+h_{1}+h_{7}+h_{10})\,,\nonumber\\
&\overline{h}_{9}=-(v_{1}+v_{4}+v_{7}+v_{10}+h_{1}+h_{4}+h_{10})\,,&&
\overline{h}_{10}=-(v_{1}+v_{4}+v_{8}+v_{10}+h_{4}+h_{8}+h_{10})\,,
\end{align}}
and the parameters $\overline{m}$
{\allowdisplaybreaks\bea\nonumber
\overline{m}_{1}&=&(v_{1}+v_{2}+v_{4}+v_{8}+v_{10})+(h_{2}+h_{4}+h_{8}+h_{10})\,,\\\nonumber
\overline{m}_{2}&=&(v_{2}+v_{4}+v_{6}+v_{8}+v_{10})+(h_{2}+h_{4}+h_{6}+h_{8})\,,\\\nonumber
\overline{m}_{3}&=&(v_{2}+v_{4}+v_{5}+v_{6}+v_{8})+(h_{2}+h_{5}+h_{6}+h_{8})\,,\\\nonumber
\overline{m}_{4}&=&(v_{2}+v_{5}+v_{6}+v_{8}+v_{9})+(h_{2}+h_{5}+h_{6}+h_{9})\,,\\\nonumber
\overline{m}_{5}&=&(v_{2}+v_{3}+v_{5}+v_{6}+v_{9})+(h_{3}+h_{5}+h_{6}+h_{9})\,,\\\nonumber
\overline{m}_{6}&=&(v_{3}+v_{5}+v_{6}+v_{7}+v_{9})+(h_{3}+h_{5}+h_{7}+h_{9})\,,\\\nonumber
\overline{m}_{7}&=&(v_{1}+v_{3}+v_{5}+v_{7}+v_{9})+(h_{1}+h_{3}+h_{7}+h_{9})\,,\\\nonumber
\overline{m}_{8}&=&(v_{1}+v_{3}+v_{7}+v_{9}+v_{10})+(h_{1}+h_{3}+h_{7}+h_{10})\,,\\\nonumber
\overline{m}_{9}&=&(v_{1}+v_{3}+v_{4}+v_{7}+v_{10})+(h_{1}+h_{4}+h_{7}+h_{10})\,,\\
\overline{m}_{10}&=&(v_{1}+v_{4}+v_{7}+v_{8}+v_{10})+(h_{1}+h_{4}+h_{8}+h_{10})\,.
\eea}
Using the assignment of vertices in Figure~\ref{Fig:Web52Paravertices} and Figure~\ref{Fig:Web101Paravertices}, the parameters $\overline{h}$ and $\overline{m}$ can be read off from the two vertex diagrams in the same manner as in the case of $(N,M)=(3,2)$.

The fact that the diagram in Figure~\ref{Fig:Web52Paravertices} can be transformed into Figure~\ref{Fig:Web101Paravertices} through a sequence of flops and $SL(2,\mathbb{Z})$ transformations shows that the web diagram $(N,M)=(2,5)$ is indeed dual to the web diagram $(N,M)=(10,1)$.
\subsubsection{Simple Check of the Duality}
As in Section~\ref{Sect:Sp2Check32} for the case $(N,M)=(3,2)$, we can perform a simple check of the duality found above. Indeed, we consider the following choice of the parameters
\begin{align}
m_{1}&=m_{2}=\cdots=m_{10}=m\,,\nonumber\\
h_{1}&=h_{2}=\cdots=h_{10}=h=\frac{\rho}{2}-m\,,\nonumber\\
v_{1}&=v_{2}=\cdots=v_{10}=v=\frac{\tau}{5}-m\,,
\end{align}
which are compatible with all the consistency conditions. Furthermore, we introduce the three parameters $(\tau',\rho',m')$ for the $(10,1)$ web 
\begin{align}
&\tau'=5\tau+8\rho-40m\,,&&\rho'=2\tau+5\rho-20m\,,&&m'=\tau+2\rho-9m\,.
\end{align}
and define the $(1,1)$ period matrices
\begin{align}
&\Omega=\left(\begin{array}{cc}\tau/5 & m \\ m & \rho/2\end{array}\right)\,,&&\text{and}&&\Omega'=\left(\begin{array}{cc} \tau' & m' \\ m' & \rho'/10\end{array}\right)\,,
\end{align}
These satisfy the relation
\begin{align}
\Omega'=(A\Omega+B)\cdot (C\Omega+D)^{-1}\,,
\end{align}
for the matrices
\begin{align}
&A=\left(\begin{array}{cc} 5 & -4 \\ 1 & -1\end{array}\right)\,,&&B=\left(\begin{array}{cc}0 & 0 \\ 0 & 0\end{array}\right)\,,&&C=\left(\begin{array}{cc}0 & 0 \\ 0 & 0\end{array}\right)\,,&&D=\left(\begin{array}{cc}1 & 1 \\ -4 & -5\end{array}\right)\,,
\end{align}
which satisfy as well the relations (\ref{Sp2ZTrafos}). This shows that $(\rho,\tau,m)$ and $(\rho',\tau',m')$ are related through an $Sp(2,\mathbb{Z})$ transformation. Therefore, $\mathcal{Z}_{X_{1,1}}$ is indeed invariant under the change of parameters implied by the above chain of dualities.

\subsection{Relation between $(N,M)=(3,4)$ and $(12,1)$}
The next example is the case $(N,M)=(3,4)$.
\subsubsection{Duality Transformation}
\begin{figure}[htbp]
\begin{center}
\scalebox{0.68}{\parbox{18.8cm}{\begin{tikzpicture}[scale = 1.75]
\draw[ultra thick] (0,0) -- (1,1);
\draw[ultra thick] (1,2) -- (2,3);
\draw[ultra thick] (2,4) -- (3,5);
\draw[ultra thick] (3,6) -- (4,7);
\draw[ultra thick] (2,1) -- (3,2);
\draw[ultra thick] (3,3) -- (4,4);
\draw[ultra thick] (4,5) -- (5,6);
\draw[ultra thick] (5,7) -- (6,8);
\draw[ultra thick] (4,2) -- (5,3);
\draw[ultra thick] (5,4) -- (6,5);
\draw[ultra thick] (6,6) -- (7,7);
\draw[ultra thick] (7,8) -- (8,9);
\draw[ultra thick] (0,-1) -- (0,0);
\draw[ultra thick] (1,1) -- (1,2);
\draw[ultra thick] (2,3) -- (2,4);
\draw[ultra thick] (3,5) -- (3,6);
\draw[ultra thick] (4,7) -- (4,8);
\draw[ultra thick] (2,1) -- (2,0);
\draw[ultra thick] (3,2) -- (3,3);
\draw[ultra thick] (4,4) -- (4,5);
\draw[ultra thick] (5,6) -- (5,7);
\draw[ultra thick] (6,8) -- (6,9);
\draw[ultra thick] (4,2) -- (4,1);
\draw[ultra thick] (5,3) -- (5,4);
\draw[ultra thick] (6,5) -- (6,6);
\draw[ultra thick] (7,7) -- (7,8);
\draw[ultra thick] (8,9) -- (8,10);
\draw[ultra thick] (-1,0) -- (0,0);
\draw[ultra thick] (0,2) -- (1,2);
\draw[ultra thick] (1,4) -- (2,4);
\draw[ultra thick] (2,6) -- (3,6);
\draw[ultra thick] (1,1) -- (2,1);
\draw[ultra thick] (2,3) -- (3,3);
\draw[ultra thick] (3,5) -- (4,5);
\draw[ultra thick] (4,7) -- (5,7);
\draw[ultra thick] (3,2) -- (4,2);
\draw[ultra thick] (4,4) -- (5,4);
\draw[ultra thick] (5,6) -- (6,6);
\draw[ultra thick] (6,8) -- (7,8);
\draw[ultra thick] (5,3) -- (6,3);
\draw[ultra thick] (6,5) -- (7,5);
\draw[ultra thick] (7,7) -- (8,7);
\draw[ultra thick] (8,9) -- (9,9);
\node at (0,0) {$\bullet$};
\node at (-0.1,0.2) {$A_1$};
\node at (1,1) {$\bullet$};
\node at (1.1,0.8) {$A_2$};
\node at (2,1) {$\bullet$};
\node at (2.2,0.9) {$A_3$};
\node at (3,2) {$\bullet$};
\node at (3.1,1.8) {$A_4$};
\node at (4,2) {$\bullet$};
\node at (4.2,1.9) {$A_5$};
\node at (5,3) {$\bullet$};
\node at (5.1,2.8) {$A_6$};
\node at (1,2) {$\bullet$};
\node at (0.9,2.2) {$A_7$};
\node at (2,3) {$\bullet$};
\node at (2.15,2.8) {$A_{8}$};
\node at (3,3) {$\bullet$};
\node at (3.25,2.9) {$A_{9}$};
\node at (4,4) {$\bullet$};
\node at (4.15,3.8) {$A_{10}$};
\node at (5,4) {$\bullet$};
\node at (5.25,3.9) {$A_{11}$};
\node at (6,5) {$\bullet$};
\node at (6.15,4.8) {$A_{12}$};
\node at (2,4) {$\bullet$};
\node at (1.85,4.2) {$A_{13}$};
\node at (3,5) {$\bullet$};
\node at (3.15,4.8) {$A_{14}$};
\node at (4,5) {$\bullet$};
\node at (4.25,4.9) {$A_{15}$};
\node at (5,6) {$\bullet$};
\node at (5.15,5.8) {$A_{26}$};
\node at (6,6) {$\bullet$};
\node at (6.25,5.9) {$A_{17}$};
\node at (7,7) {$\bullet$};
\node at (7.15,6.8) {$A_{18}$};
\node at (3,6) {$\bullet$};
\node at (2.85,6.2) {$A_{19}$};
\node at (4,7) {$\bullet$};
\node at (4.15,6.8) {$A_{20}$};
\node at (5,7) {$\bullet$};
\node at (5.25,6.9) {$A_{21}$};
\node at (6,8) {$\bullet$};
\node at (6.15,7.8) {$A_{22}$};
\node at (7,8) {$\bullet$};
\node at (7.25,7.9) {$A_{23}$};
\node at (8,9) {$\bullet$};
\node at (8.15,8.8) {$A_{24}$};
\node at (-0.2,-0.5) {{\small \bf $v_1$}};
\node at (0.8,1.5) {{\small \bf $v_2$}};
\node at (1.8,3.5) {{\small \bf $v_3$}};
\node at (2.8,5.5) {{\small \bf $v_4$}};
\node at (3.8,7.5) {{\small \bf $v_1$}};
\node at (1.8,0.5) {{\small \bf $v_5$}};
\node at (2.8,2.5) {{\small \bf $v_6$}};
\node at (3.8,4.5) {{\small \bf $v_7$}};
\node at (4.8,6.5) {{\small \bf $v_8$}};
\node at (5.8,8.5) {{\small \bf $v_5$}};
\node at (3.8,1.5) {{\small \bf $v_9$}};
\node at (4.75,3.5) {{\small \bf $v_{10}$}};
\node at (5.75,5.5) {{\small \bf $v_{11}$}};
\node at (6.75,7.5) {{\small \bf $v_{12}$}};
\node at (7.8,9.5) {{\small \bf $v_9$}};
%
\node at (-0.7,0.2) {{\small \bf $h_{1}$}};
\node at (0.3,2.2) {{\small \bf $h_{4}$}};
\node at (1.3,4.2) {{\small \bf $h_{7}$}};
\node at (2.3,6.2) {{\small \bf $h_{10}$}};
\node at (1.5,1.2) {{\small \bf $h_2$}};
\node at (2.5,3.2) {{\small \bf $h_5$}};
\node at (3.5,5.2) {{\small \bf $h_8$}};
\node at (4.5,7.2) {{\small \bf $h_{11}$}};
\node at (3.5,2.2) {{\small \bf $h_3$}};
\node at (4.5,4.2) {{\small \bf $h_6$}};
\node at (5.5,6.2) {{\small \bf $h_9$}};
\node at (6.5,8.2) {{\small \bf $h_{12}$}};
\node at (5.5,3.2) {{\small \bf $h_{1}$}};
\node at (6.5,5.2) {{\small \bf $h_{4}$}};
\node at (7.5,7.2) {{\small \bf $h_{7}$}};
\node at (8.5,9.2) {{\small \bf $h_{10}$}};
%
\node at (0.35,0.65) {{\small \bf $m_1$}};
\node at (1.35,2.65) {{\small \bf $m_2$}};
\node at (2.35,4.65) {{\small \bf $m_3$}};
\node at (3.35,6.65) {{\small \bf $m_4$}};
%
\node at (2.35,1.65) {{\small \bf $m_5$}};
\node at (3.35,3.65) {{\small \bf $m_6$}};
\node at (4.35,5.65) {{\small \bf $m_7$}};
\node at (5.35,7.65) {{\small \bf $m_8$}};
%
\node at (4.35,2.7) {{\small \bf $m_{9}$}};
\node at (5.4,4.7) {{\small \bf $m_{10}$}};
\node at (6.4,6.7) {{\small \bf $m_{11}$}};
\node at (7.4,8.7) {{\small \bf $m_{12}$}};
\node at (0,-1.15) {{\small \bf $a$}};
\node at (2,-0.15) {{\small \bf $b$}};
\node at (4,0.85) {{\small \bf $c$}};
\node at (4,8.15) {{\small \bf $a$}};
\node at (6,9.15) {{\small \bf $b$}};
\node at (8,10.15) {{\small \bf $c$}};
\node at (-1.15,0) {{\small \bf $1$}};
\node at (-0.15,2) {{\small \bf $2$}};
\node at (0.85,4) {{\small \bf $3$}};
\node at (1.85,6) {{\small \bf $4$}};
\node at (6.15,3) {{\small \bf $1$}};
\node at (7.15,5) {{\small \bf $2$}};
\node at (8.15,7) {{\small \bf $3$}};
\node at (9.15,9) {{\small \bf $4$}};
\end{tikzpicture}}
}
\end{center}
\caption{\sl Parametrisation of the $(N,M)=(3,4)$ web along with an assignment of vertices.}
\label{Fig:Web43vertices}
\end{figure}
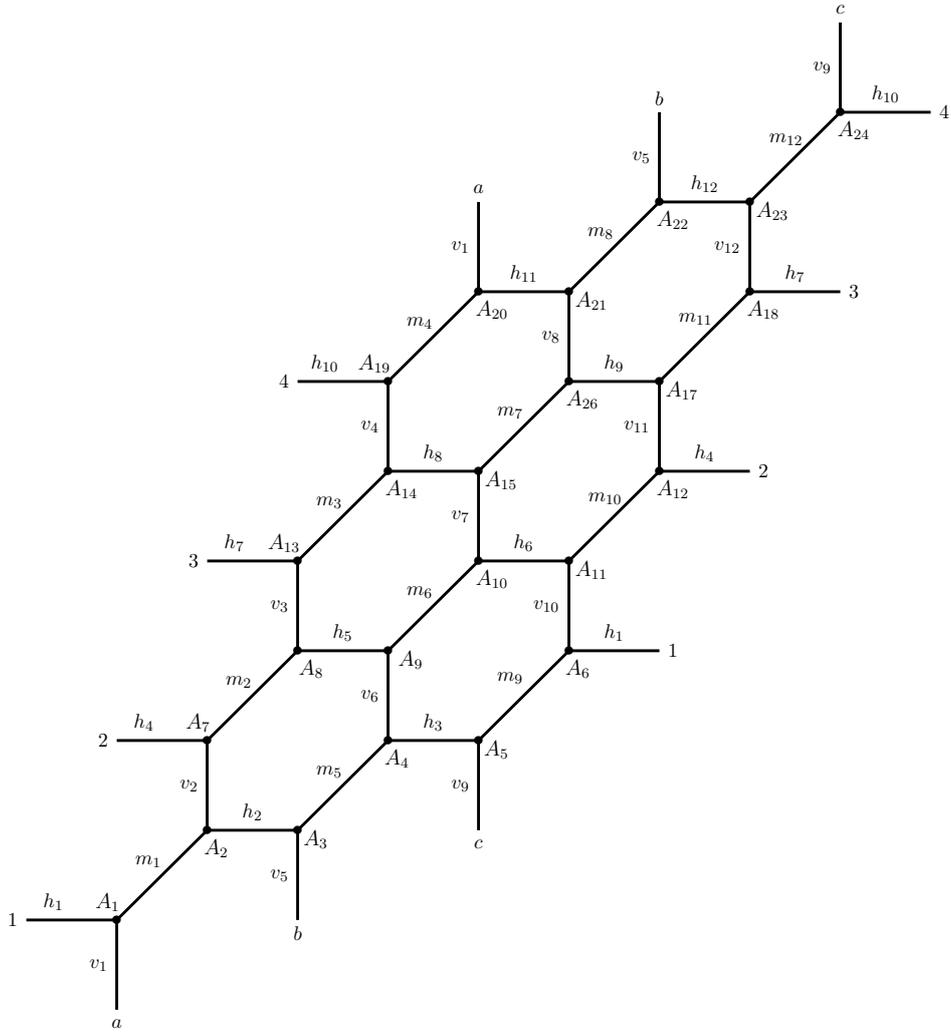

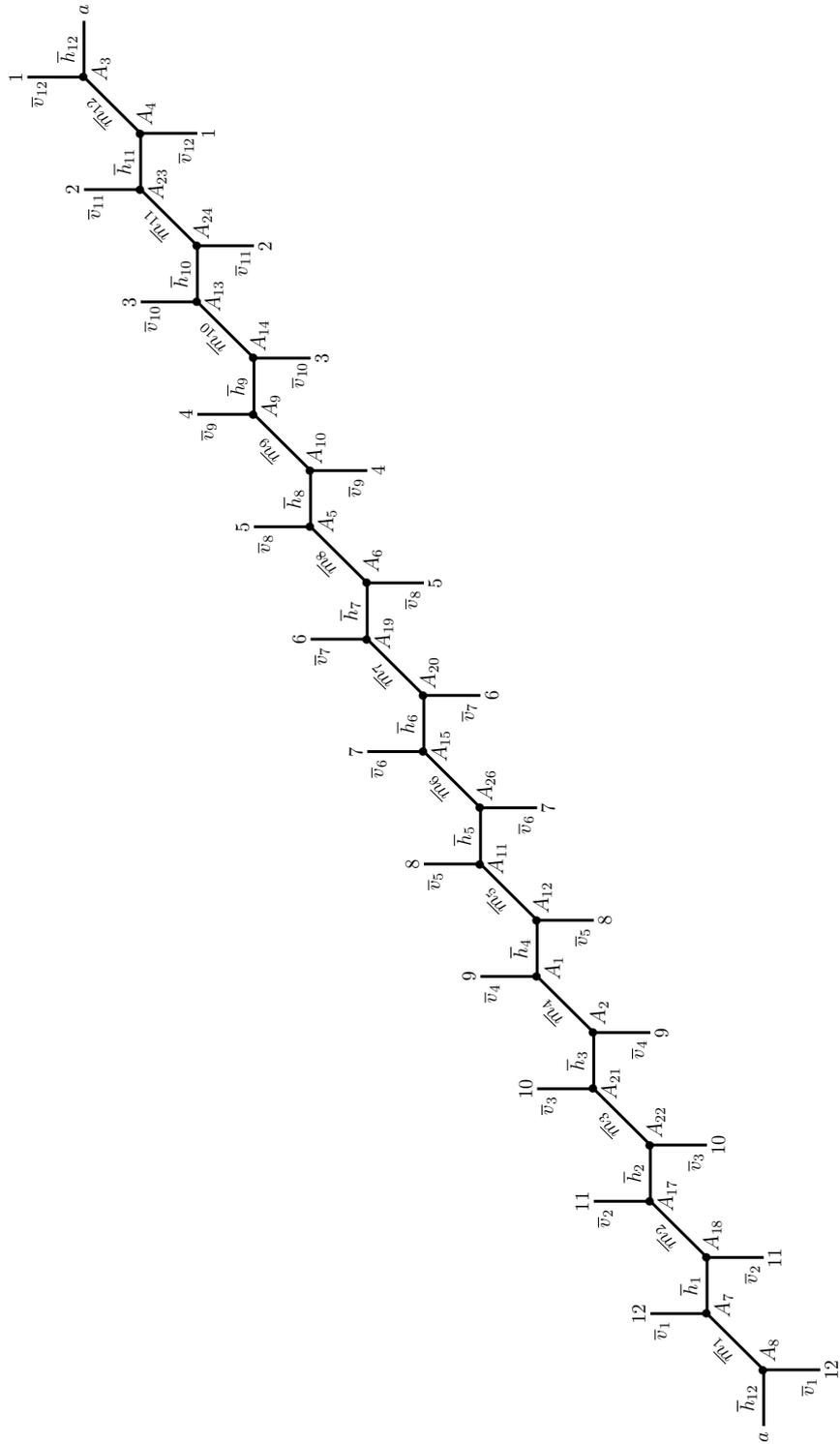
\begin{figure}[htbp]
\begin{center}
\rotatebox{90}{\scalebox{0.7}{\parbox{29.5cm}{\begin{tikzpicture}[scale = 1.1]
\draw[ultra thick] (2,1) -- (2,2);
\draw[ultra thick] (1,0) -- (1,-1);
\draw[ultra thick] (4,2) -- (4,3);
\draw[ultra thick] (3,1) -- (3,0);
\draw[ultra thick] (5,2) -- (5,1);
\draw[ultra thick] (6,3) -- (6,4);
\draw[ultra thick] (7,3) -- (7,2);
\draw[ultra thick] (8,4) -- (8,5);
\draw[ultra thick] (9,4) -- (9,3);
\draw[ultra thick] (10,5) -- (10,6);
\draw[ultra thick] (11,5) -- (11,4);
\draw[ultra thick] (12,6) -- (12,7);
\draw[ultra thick] (13,6) -- (13,5);
\draw[ultra thick] (14,7) -- (14,8);
\draw[ultra thick] (15,7) -- (15,6);
\draw[ultra thick] (16,8) -- (16,9);
\draw[ultra thick] (17,8) -- (17,7);
\draw[ultra thick] (18,9) -- (18,10);
\draw[ultra thick] (19,9) -- (19,8);
\draw[ultra thick] (20,10) -- (20,11);
\draw[ultra thick] (21,10) -- (21,9);
\draw[ultra thick] (22,11) -- (22,12);
\draw[ultra thick] (23,11) -- (23,10);
\draw[ultra thick] (24,12) -- (24,13);
\draw[ultra thick] (0,0) -- (1,0);
\draw[ultra thick] (2,1) -- (3,1);
\draw[ultra thick] (4,2) -- (5,2);
\draw[ultra thick] (6,3) -- (7,3);
\draw[ultra thick] (8,4) -- (9,4);
\draw[ultra thick] (10,5) -- (11,5);
\draw[ultra thick] (12,6) -- (13,6);
\draw[ultra thick] (14,7) -- (15,7);
\draw[ultra thick] (16,8) -- (17,8);
\draw[ultra thick] (18,9) -- (19,9);
\draw[ultra thick] (20,10) -- (21,10);
\draw[ultra thick] (22,11) -- (23,11);
\draw[ultra thick] (24,12) -- (25,12);
\draw[ultra thick] (1,0) -- (2,1);
\draw[ultra thick] (3,1) -- (4,2);
\draw[ultra thick] (5,2) -- (6,3);
\draw[ultra thick] (7,3) -- (8,4);
\draw[ultra thick] (9,4) -- (10,5);
\draw[ultra thick] (11,5) -- (12,6);
\draw[ultra thick] (13,6) -- (14,7);
\draw[ultra thick] (15,7) -- (16,8);
\draw[ultra thick] (17,8) -- (18,9);
\draw[ultra thick] (19,9) -- (20,10);
\draw[ultra thick] (21,10) -- (22,11);
\draw[ultra thick] (23,11) -- (24,12);
\node[rotate=45] at (1.35,0.75) {{\small $\overline{m}_{1}$}};
\node[rotate=45] at (3.35,1.75) {{\small $\overline{m}_{2}$}};
\node[rotate=45] at (5.35,2.75) {{\small $\overline{m}_{3}$}};
\node[rotate=45] at (7.35,3.75) {{\small $\overline{m}_4$}};
\node[rotate=45] at (9.35,4.75) {{\small $\overline{m}_5$}};
\node[rotate=45] at (11.35,5.75) {{\small $\overline{m}_6$}};
\node[rotate=45] at (13.35,6.75) {{\small $\overline{m}_7$}};
\node[rotate=45] at (15.35,7.75) {{\small $\overline{m}_8$}};
\node[rotate=45] at (17.35,8.75) {{\small $\overline{m}_{9}$}};
\node[rotate=45] at (19.35,9.75) {{\small $\overline{m}_{10}$}};
\node[rotate=45] at (21.35,10.75) {{\small $\overline{m}_{11}$}};
\node[rotate=45] at (23.35,11.75) {{\small $\overline{m}_{12}$}};
\node at (0.5,0.3) {{\small $\overline{h}_{12}$}};
\node at (2.5,1.3) {{\small $\overline{h}_{1}$}};
\node at (4.5,2.3) {{\small $\overline{h}_{2}$}};
\node at (6.5,3.3) {{\small $\overline{h}_3$}};
\node at (8.5,4.3) {{\small $\overline{h}_4$}};
\node at (10.5,5.3) {{\small $\overline{h}_5$}};
\node at (12.5,6.3) {{\small $\overline{h}_6$}};
\node at (14.5,7.3) {{\small $\overline{h}_7$}};
\node at (16.5,8.3) {{\small $\overline{h}_8$}};
\node at (18.5,9.3) {{\small $\overline{h}_9$}};
\node at (20.5,10.3) {{\small $\overline{h}_{10}$}};
\node at (22.5,11.3) {{\small $\overline{h}_{11}$}};
\node at (24.5,12.3) {{\small $\overline{h}_{12}$}};
\node at (0.7,-0.8) {{\small $\overline{v}_{1}$}};
\node at (2.7,0.2) {{\small $\overline{v}_{2}$}};
\node at (4.7,1.2) {{\small $\overline{v}_{3}$}};
\node at (6.7,2.2) {{\small $\overline{v}_4$}};
\node at (8.7,3.2) {{\small $\overline{v}_5$}};
\node at (10.7,4.2) {{\small $\overline{v}_6$}};
\node at (12.7,5.2) {{\small $\overline{v}_7$}};
\node at (14.7,6.2) {{\small $\overline{v}_8$}};
\node at (16.7,7.2) {{\small $\overline{v}_9$}};
\node at (18.7,8.2) {{\small $\overline{v}_{10}$}};
\node at (20.7,9.2) {{\small $\overline{v}_{11}$}};
\node at (22.7,10.2) {{\small $\overline{v}_{12}$}};
\node at (1.7,1.8) {{\small $\overline{v}_{1}$}};
\node at (3.7,2.8) {{\small $\overline{v}_{2}$}};
\node at (5.7,3.8) {{\small $\overline{v}_{3}$}};
\node at (7.7,4.8) {{\small $\overline{v}_{4}$}};
\node at (9.7,5.8) {{\small $\overline{v}_{5}$}};
\node at (11.7,6.8) {{\small $\overline{v}_{6}$}};
\node at (13.7,7.8) {{\small $\overline{v}_{7}$}};
\node at (15.7,8.8) {{\small $\overline{v}_{8}$}};
\node at (17.7,9.8) {{\small $\overline{v}_{9}$}};
\node at (19.7,10.8) {{\small $\overline{v}_{10}$}};
\node at (21.7,11.8) {{\small $\overline{v}_{11}$}};
\node at (23.7,12.8) {{\small $\overline{v}_{12}$}};
\node at (-0.2,0) {{\small \bf $a$}};
\node at (25.2,12) {{\small \bf $a$}};
\node at (1,-1.2) {{\small \bf $12$}};
\node at (3,-0.2) {{\small \bf $11$}};
\node at (5,0.8) {{\small \bf $10$}};
\node at (7,1.8) {{\small \bf $9$}};
\node at (9,2.8) {{\small \bf $8$}};
\node at (11,3.8) {{\small \bf $7$}};
\node at (13,4.8) {{\small \bf $6$}};
\node at (15,5.8) {{\small \bf $5$}};
\node at (17,6.8) {{\small \bf $4$}};
\node at (19,7.8) {{\small \bf $3$}};
\node at (21,8.8) {{\small \bf $2$}};
\node at (23,9.8) {{\small \bf $1$}};
\node at (2,2.2) {{\small \bf $12$}};
\node at (4,3.2) {{\small \bf $11$}};
\node at (6,4.2) {{\small \bf $10$}};
\node at (8,5.2) {{\small \bf $9$}};
\node at (10,6.2) {{\small \bf $8$}};
\node at (12,7.2) {{\small \bf $7$}};
\node at (14,8.2) {{\small \bf $6$}};
\node at (16,9.2) {{\small \bf $5$}};
\node at (18,10.2) {{\small \bf $4$}};
\node at (20,11.2) {{\small \bf $3$}};
\node at (22,12.2) {{\small \bf $2$}};
\node at (24,13.2) {{\small \bf $1$}};
%
\node at (1,0) {$\bullet$};
\node at (1.35,-0.1) {$A_{8}$};
\node at (2,1) {$\bullet$};
\node at (2.15,0.7) {$A_{7}$};
\node at (3,1) {$\bullet$};
\node at (3.4,0.9) {$A_{18}$};
\node at (4,2) {$\bullet$};
\node at (4.1,1.7) {$A_{17}$};
\node at (5,2) {$\bullet$};
\node at (5.4,1.9) {$A_{22}$};
\node at (6,3) {$\bullet$};
\node at (6.1,2.7) {$A_{21}$};
\node at (7,3) {$\bullet$};
\node at (7.35,2.9) {$A_{2}$};
\node at (8,4) {$\bullet$};
\node at (8.15,3.7) {$A_{1}$};
\node at (9,4) {$\bullet$};
\node at (9.4,3.9) {$A_{12}$};
\node at (10,5) {$\bullet$};
\node at (10.1,4.7) {$A_{11}$};
\node at (11,5) {$\bullet$};
\node at (11.4,4.9) {$A_{26}$};
\node at (12,6) {$\bullet$};
\node at (12.1,5.7) {$A_{15}$};
\node at (13,6) {$\bullet$};
\node at (13.4,5.9) {$A_{20}$};
\node at (14,7) {$\bullet$};
\node at (14.1,6.7) {$A_{19}$};
\node at (15,7) {$\bullet$};
\node at (15.4,6.9) {$A_{6}$};
\node at (16,8) {$\bullet$};
\node at (16.1,7.7) {$A_{5}$};
\node at (17,8) {$\bullet$};
\node at (17.4,7.9) {$A_{10}$};
\node at (18,9) {$\bullet$};
\node at (18.15,8.7) {$A_{9}$};
\node at (19,9) {$\bullet$};
\node at (19.4,8.9) {$A_{14}$};
\node at (20,10) {$\bullet$};
\node at (20.1,9.7) {$A_{13}$};
\node at (21,10) {$\bullet$};
\node at (21.4,9.9) {$A_{24}$};
\node at (22,11) {$\bullet$};
\node at (22.1,10.7) {$A_{23}$};
\node at (23,11) {$\bullet$};
\node at (23.35,10.9) {$A_{4}$};
\node at (24,12) {$\bullet$};
\node at (24.15,11.7) {$A_{3}$};
\end{tikzpicture}}}
}
\end{center}
\caption{\sl Parametrisation of the web $(N,M)=(12,1)$web with an assignment of vertices compatible with Figure~\ref{Fig:Web43vertices}.}
\label{Fig:Web112vertices}
\end{figure}

A parametrisation of the web $(N,M)=(3,4)$ is given in Figure~\ref{Fig:Web43vertices}, which also shows an assignment of vertices. Of the 36 parameters in the web in Figure~\ref{Fig:Web43vertices}, only 14 are independent on account of the consistency conditions.  
After going through a sequence of flop-transitions, cutting and re-gluing as well as $SL(2,\mathbb{Z})$ transformations, we can bring this diagram to the form of the $(N,M)=(12,1)$ web with the parametrisation given in Figure~\ref{Fig:Web112vertices}.
Specifically, the parameters $\overline{m}$ are given by
{\allowdisplaybreaks
\begin{align}
&\overline{m}_{1}=h_{2}+h_{7}+h_{12}+v_{2}+v_{3}+v_{5}+v_{12}\,,&&
\overline{m}_{2}=h_{2}+h_{4}+h_{12}+v_{2}+v_{5}+v_{11}+v_{12}\nonumber\\
&\overline{m}_{3}=h_{2}+h_{4}+h_{9}+v_{2}+v_{5}+v_{8}+v_{11}\,,&&
\overline{m}_{4}=h_{4}+h_{9}+h_{11}+v_{1}+v_{2}+v_{8}+v_{11}\nonumber\\
&\overline{m}_{5}=h_{1}+h_{9}+h_{11}+v_{1}+v_{8}+v_{10}+v_{11}\,,&&
\overline{m}_{6}=h_{1}+h_{6}+h_{11}+v_{1}+v_{7}+v_{8}+v_{10}\,,\nonumber\\
&\overline{m}_{7}=h_{1}+h_{6}+h_{8}+v_{1}+v_{4}+v_{7}+v_{10}\,,&&
\overline{m}_{8}=h_{6}+h_{8}+h_{10}+v_{4}+v_{7}+v_{9}+v_{10}\,,\nonumber\\
&\overline{m}_{9}=h_{3}+h_{8}+h_{10}+v_{4}+v_{6}+v_{7}+v_{9}\,,&&
\overline{m}_{10}=h_{3}+h_{5}+h_{10}+v_{3}+v_{4}+v_{6}+v_{9}\,,\nonumber\\
&\overline{m}_{11}=h_{3}+h_{5}+h_{7}+v_{3}+v_{6}+v_{9}+v_{12}\,,&&
\overline{m}_{12}=h_{5}+h_{7}+h_{12}+v_{3}+v_{5}+v_{6}+v_{12}\,,
\end{align}}
while the parameters $\bar{h}$ are
{\allowdisplaybreaks
\begin{align}
&\overline{h}_{1}=-(h_{2}+h_{12}+v_{2}+v_{5}+v_{12})\,,&&
\overline{h}_{2}=-(h_{2}+h_{4}+v_{2}+v_{5}+v_{11})\,,\nonumber\\
&\overline{h}_{3}=-(h_{4}+h_{9}+v_{2}+v_{8}+v_{11})\,,&&
\overline{h}_{4}=-(h_{9}+h_{11}+v_{1}+v_{8}+v_{11})\,,\nonumber\\
&\overline{h}_{5}=-(h_{1}+h_{11}+v_{1}+v_{8}+v_{10})\,,&&
\overline{h}_{6}=-(h_{1}+h_{6}+v_{1}+v_{7}+v_{10})\,,\nonumber\\
&\overline{h}_{7}=-(h_{6}+h_{8}+v_{4}+v_{7}+v_{10})\,,&&
\overline{h}_{8}=-(h_{8}+h_{10}+v_{4}+v_{7}+v_{9})\,,\nonumber\\
&\overline{h}_{9}=-(h_{3}+h_{10}+v_{4}+v_{6}+v_{9})\,,&&
\overline{h}_{10}=-(h_{3}+h_{5}+v_{3}+v_{6}+v_{9})\,,\nonumber\\
&\overline{h}_{11}=-(h_{5}+h_{7}+v_{3}+v_{6}+v_{12})\,,&&
\overline{h}_{12}=-(h_{7}+h_{12}+v_{3}+v_{5}+v_{12})\,.
\end{align}}
and the parameters $\bar{v}$ are
{\allowdisplaybreaks
\bea\nonumber
\overline{v}_{1}&=&m_{2}+2(h_{2}+h_{7}+h_{12})+3(v_{2}+v_{3}+v_{5}+v_{12})\,,\\\nonumber
\overline{v}_{2}&=&m_{11}+2(h_{2}+h_{4}+h_{12})+3(v_{2}+v_{5}+v_{11}+v_{12})\,,\\\nonumber
\overline{v}_{3}&=&m_{8}+2(h_{2}+h_{4}+h_{9})+3(v_{2}+v_{5}+v_{8}+v_{11})\,,\\\nonumber
\overline{v}_{4}&=&m_{1}+2(h_{4}+h_{9}+h_{11})+3(v_{1}+v_{2}+v_{8}+v_{11})\,,\\\nonumber
\overline{v}_{5}&=&m_{10}+2(h_{1}+h_{9}+h_{11})+3(v_{1}+v_{8}+v_{10}+v_{11})\,,\\\nonumber
\overline{v}_{6}&=&m_{7}+2(h_{1}+h_{6}+h_{11})+3(v_{1}+v_{7}+v_{8}+v_{10})\,,\\\nonumber
\overline{v}_{7}&=&m_{4}+2(h_{1}+h_{6}+h_{8})+3(v_{1}+v_{4}+v_{7}+v_{10})\,,\\\nonumber
\overline{v}_{8}&=&m_{9}+2(h_{6}+h_{8}+h_{10})+3(v_{4}+v_{7}+v_{9}+v_{10})\,,\\\nonumber
\overline{v}_{9}&=&m_{6}+2(h_{3}+h_{8}+h_{10})+3(v_{4}+v_{6}+v_{7}+v_{9})\,,\\\nonumber
\overline{v}_{10}&=&m_{3}+2(h_{3}+h_{5}+h_{10})+3(v_{3}+v_{4}+v_{6}+v_{9})\,,\\\nonumber
\overline{v}_{11}&=&m_{12}+2(h_{3}+h_{5}+h_{7})+3(v_{3}+v_{6}+v_{9}+v_{12})\,,\\\nonumber
\overline{v}_{12}&=&m_{5}+2(h_{5}+h_{7}+h_{12})+3(v_{3}+v_{5}+v_{6}+v_{12})\,.
\eea
}
As in the cases before, using the assignment of vertices in Figure~\ref{Fig:Web43vertices} and Figure~\ref{Fig:Web112vertices} the parameters $\overline{h}$ and $\overline{m}$ can be directly read off. Furthermore, the transformation above shows that the web diagrams $(N,M)=(3,4)$ and $(12,1)$ are indeed dual to each other.
\subsubsection{Simple Check of Duality}
In order to perform a check of the duality worked out above, we consider
\begin{align}
m_{1}&=m_{2}=\cdots=m_{12}=m\,,\nonumber\\
h_{1}&=h_{2}=\cdots=h_{12}=h=\frac{\rho}{3}-m\,,\nonumber\\
v_{1}&=v_{2}=\cdots=v_{12}=v=\frac{\tau}{4}-m\,.
\end{align}
Furthermore, we introduce the parameters in the $(12,1)$ web
\begin{align}
&\tau'=4\tau+3\rho-24m\,,&&\rho'=3\tau+4\rho-24m\,,&&m'=\tau+\rho-7m\,.\label{RelOmPrime34}
\end{align}
Next we also introduce the period matrices
\begin{align}
&\Omega=\left(\begin{array}{cc}\tau/4 & m \\ m & \rho/3\end{array}\right)\,,&&\text{and} &&\Omega'=\left(\begin{array}{cc} \tau' & m' \\ m' & \rho'/12\end{array}\right)\,,
\end{align}
which are related to each other by
\begin{align}
\Omega'=(A\Omega+B)\cdot (C\Omega+D)^{-1}\,,
\end{align}
with the matrices
\begin{align}
&A=\left(\begin{array}{cc} 4 & -3 \\ 1 & -1\end{array}\right)\,,&&B=\left(\begin{array}{cc}0 & 0 \\ 0 & 0\end{array}\right)\,,&&C=\left(\begin{array}{cc}0 & 0 \\ 0 & 0\end{array}\right)\,,&&D=\left(\begin{array}{cc}1 & 1 \\ -3 & -4\end{array}\right)\,,
\end{align}
which satisfy indeed the relations (\ref{Sp2ZTrafos}). This shows that $(\rho,\tau,m)$ and $(\rho',\tau',m')$ are related through an $Sp(2,\mathbb{Z})$ transformation. Therefore, $\mathcal{Z}_{X_{1,1}}$ is indeed invariant under the change of parameters implied by the above chain of dualities.
\FloatBarrier

\subsection{Relation between $(N,M)=(4,6)$ and $(12,2)$}\label{Sect:46Example}
Our final example is the relation between the web diagrams $(N,M)=(4,6)$ and $(12,2)$. Notice that in this case the gcd is $\text{gcd}(4,6)=2$ and it is therefore dualised to the web $(12,2)$.
\subsubsection{Duality Transformation}
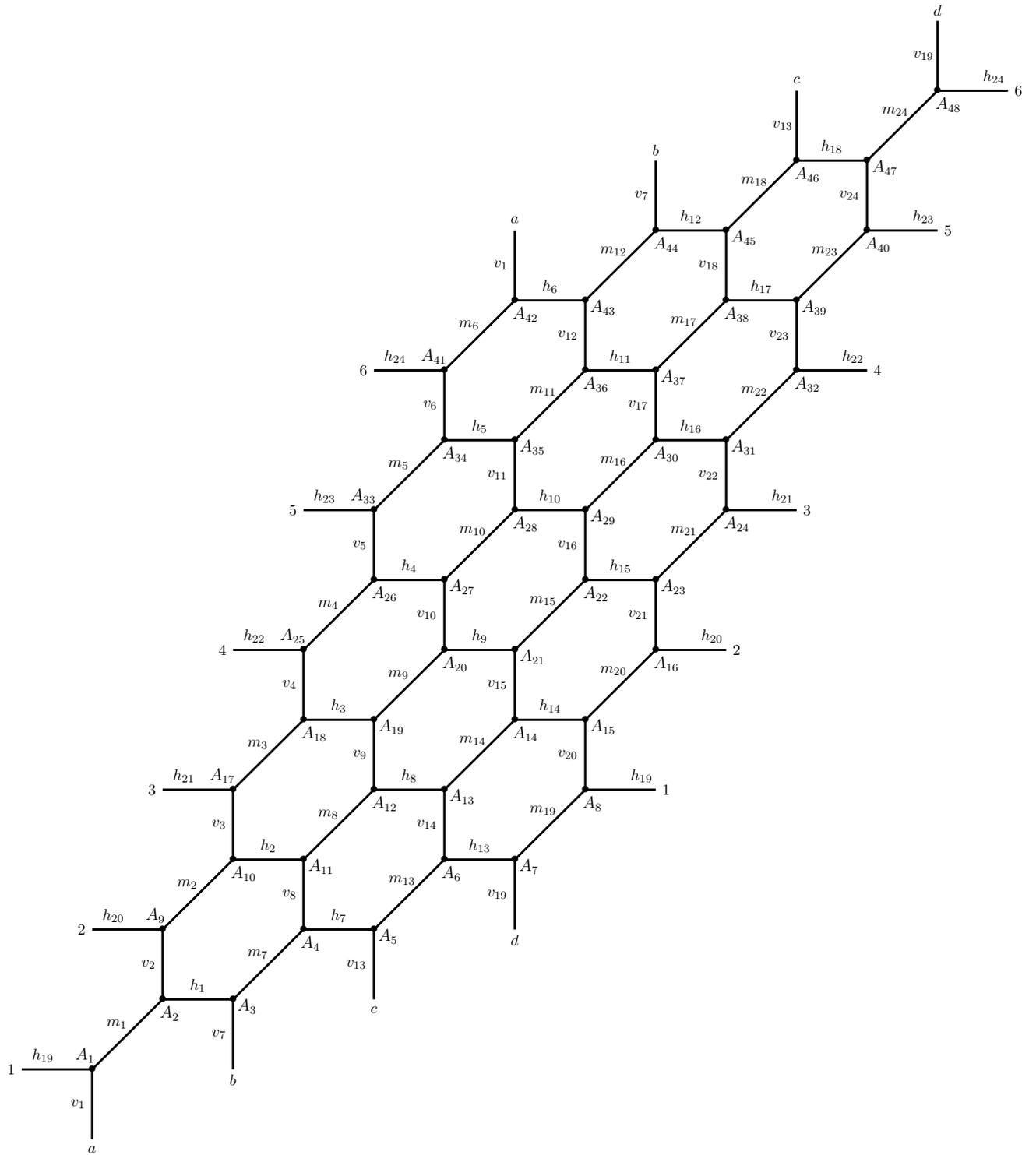
\begin{figure}[htbp]
\begin{center}
\scalebox{0.68}{\parbox{25.8cm}{\begin{tikzpicture}[scale = 1.75]
\draw[ultra thick] (0,0) -- (1,1);
\draw[ultra thick] (1,2) -- (2,3);
\draw[ultra thick] (2,4) -- (3,5);
\draw[ultra thick] (3,6) -- (4,7);
\draw[ultra thick] (4,8) -- (5,9);
\draw[ultra thick] (5,10) -- (6,11);
\draw[ultra thick] (2,1) -- (3,2);
\draw[ultra thick] (3,3) -- (4,4);
\draw[ultra thick] (4,5) -- (5,6);
\draw[ultra thick] (5,7) -- (6,8);
\draw[ultra thick] (6,9) -- (7,10);
\draw[ultra thick] (7,11) -- (8,12);
\draw[ultra thick] (4,2) -- (5,3);
\draw[ultra thick] (5,4) -- (6,5);
\draw[ultra thick] (6,6) -- (7,7);
\draw[ultra thick] (7,8) -- (8,9);
\draw[ultra thick] (8,10) -- (9,11);
\draw[ultra thick] (9,12) -- (10,13);
\draw[ultra thick] (6,3) -- (7,4);
\draw[ultra thick] (7,5) -- (8,6);
\draw[ultra thick] (8,7) -- (9,8);
\draw[ultra thick] (9,9) -- (10,10);
\draw[ultra thick] (10,11) -- (11,12);
\draw[ultra thick] (11,13) -- (12,14);
\draw[ultra thick] (0,-1) -- (0,0);
\draw[ultra thick] (1,1) -- (1,2);
\draw[ultra thick] (2,3) -- (2,4);
\draw[ultra thick] (3,5) -- (3,6);
\draw[ultra thick] (4,7) -- (4,8);
\draw[ultra thick] (5,9) -- (5,10);
\draw[ultra thick] (6,11) -- (6,12);
\draw[ultra thick] (2,1) -- (2,0);
\draw[ultra thick] (3,2) -- (3,3);
\draw[ultra thick] (4,4) -- (4,5);
\draw[ultra thick] (5,6) -- (5,7);
\draw[ultra thick] (6,8) -- (6,9);
\draw[ultra thick] (7,10) -- (7,11);
\draw[ultra thick] (8,12) -- (8,13);
\draw[ultra thick] (4,2) -- (4,1);
\draw[ultra thick] (5,3) -- (5,4);
\draw[ultra thick] (6,5) -- (6,6);
\draw[ultra thick] (7,7) -- (7,8);
\draw[ultra thick] (8,9) -- (8,10);
\draw[ultra thick] (9,11) -- (9,12);
\draw[ultra thick] (10,13) -- (10,14);
\draw[ultra thick] (6,3) -- (6,2);
\draw[ultra thick] (7,4) -- (7,5);
\draw[ultra thick] (8,6) -- (8,7);
\draw[ultra thick] (9,8) -- (9,9);
\draw[ultra thick] (10,10) -- (10,11);
\draw[ultra thick] (11,12) -- (11,13);
\draw[ultra thick] (12,14) -- (12,15);
\draw[ultra thick] (-1,0) -- (0,0);
\draw[ultra thick] (0,2) -- (1,2);
\draw[ultra thick] (1,4) -- (2,4);
\draw[ultra thick] (2,6) -- (3,6);
\draw[ultra thick] (3,8) -- (4,8);
\draw[ultra thick] (4,10) -- (5,10);
\draw[ultra thick] (1,1) -- (2,1);
\draw[ultra thick] (2,3) -- (3,3);
\draw[ultra thick] (3,5) -- (4,5);
\draw[ultra thick] (4,7) -- (5,7);
\draw[ultra thick] (5,9) -- (6,9);
\draw[ultra thick] (6,11) -- (7,11);
\draw[ultra thick] (3,2) -- (4,2);
\draw[ultra thick] (4,4) -- (5,4);
\draw[ultra thick] (5,6) -- (6,6);
\draw[ultra thick] (6,8) -- (7,8);
\draw[ultra thick] (7,10) -- (8,10);
\draw[ultra thick] (8,12) -- (9,12);
\draw[ultra thick] (5,3) -- (6,3);
\draw[ultra thick] (6,5) -- (7,5);
\draw[ultra thick] (7,7) -- (8,7);
\draw[ultra thick] (8,9) -- (9,9);
\draw[ultra thick] (9,11) -- (10,11);
\draw[ultra thick] (10,13) -- (11,13);
\draw[ultra thick] (7,4) -- (8,4);
\draw[ultra thick] (8,6) -- (9,6);
\draw[ultra thick] (9,8) -- (10,8);
\draw[ultra thick] (10,10) -- (11,10);
\draw[ultra thick] (11,12) -- (12,12);
\draw[ultra thick] (12,14) -- (13,14);
\node at (0,0) {$\bullet$};
\node at (-0.1,0.2) {$A_1$};
\node at (1,1) {$\bullet$};
\node at (1.1,0.8) {$A_2$};
\node at (2,1) {$\bullet$};
\node at (2.2,0.9) {$A_3$};
\node at (3,2) {$\bullet$};
\node at (3.1,1.8) {$A_4$};
\node at (4,2) {$\bullet$};
\node at (4.2,1.9) {$A_5$};
\node at (5,3) {$\bullet$};
\node at (5.1,2.8) {$A_6$};
\node at (6,3) {$\bullet$};
\node at (6.2,2.9) {$A_7$};
\node at (7,4) {$\bullet$};
\node at (7.1,3.8) {$A_8$};
\node at (1,2) {$\bullet$};
\node at (0.9,2.2) {$A_9$};
\node at (2,3) {$\bullet$};
\node at (2.15,2.8) {$A_{10}$};
\node at (3,3) {$\bullet$};
\node at (3.25,2.9) {$A_{11}$};
\node at (4,4) {$\bullet$};
\node at (4.15,3.8) {$A_{12}$};
\node at (5,4) {$\bullet$};
\node at (5.25,3.9) {$A_{13}$};
\node at (6,5) {$\bullet$};
\node at (6.15,4.8) {$A_{14}$};
\node at (7,5) {$\bullet$};
\node at (7.25,4.9) {$A_{15}$};
\node at (8,6) {$\bullet$};
\node at (8.15,5.8) {$A_{16}$};
\node at (2,4) {$\bullet$};
\node at (1.85,4.2) {$A_{17}$};
\node at (3,5) {$\bullet$};
\node at (3.15,4.8) {$A_{18}$};
\node at (4,5) {$\bullet$};
\node at (4.25,4.9) {$A_{19}$};
\node at (5,6) {$\bullet$};
\node at (5.15,5.8) {$A_{20}$};
\node at (6,6) {$\bullet$};
\node at (6.25,5.9) {$A_{21}$};
\node at (7,7) {$\bullet$};
\node at (7.15,6.8) {$A_{22}$};
\node at (8,7) {$\bullet$};
\node at (8.25,6.9) {$A_{23}$};
\node at (9,8) {$\bullet$};
\node at (9.15,7.8) {$A_{24}$};
\node at (3,6) {$\bullet$};
\node at (2.85,6.2) {$A_{25}$};
\node at (4,7) {$\bullet$};
\node at (4.15,6.8) {$A_{26}$};
\node at (5,7) {$\bullet$};
\node at (5.25,6.9) {$A_{27}$};
\node at (6,8) {$\bullet$};
\node at (6.15,7.8) {$A_{28}$};
\node at (7,8) {$\bullet$};
\node at (7.25,7.9) {$A_{29}$};
\node at (8,9) {$\bullet$};
\node at (8.15,8.8) {$A_{30}$};
\node at (9,9) {$\bullet$};
\node at (9.25,8.9) {$A_{31}$};
\node at (10,10) {$\bullet$};
\node at (10.15,9.8) {$A_{32}$};
\node at (4,8) {$\bullet$};
\node at (3.85,8.2) {$A_{33}$};
\node at (5,9) {$\bullet$};
\node at (5.15,8.8) {$A_{34}$};
\node at (6,9) {$\bullet$};
\node at (6.25,8.9) {$A_{35}$};
\node at (7,10) {$\bullet$};
\node at (7.15,9.8) {$A_{36}$};
\node at (8,10) {$\bullet$};
\node at (8.25,9.9) {$A_{37}$};
\node at (9,11) {$\bullet$};
\node at (9.15,10.8) {$A_{38}$};
\node at (10,11) {$\bullet$};
\node at (10.25,10.9) {$A_{39}$};
\node at (11,12) {$\bullet$};
\node at (11.15,11.8) {$A_{40}$};
\node at (5,10) {$\bullet$};
\node at (4.85,10.2) {$A_{41}$};
\node at (6,11) {$\bullet$};
\node at (6.15,10.8) {$A_{42}$};
\node at (7,11) {$\bullet$};
\node at (7.25,10.9) {$A_{43}$};
\node at (8,12) {$\bullet$};
\node at (8.15,11.8) {$A_{44}$};
\node at (9,12) {$\bullet$};
\node at (9.25,11.9) {$A_{45}$};
\node at (10,13) {$\bullet$};
\node at (10.15,12.8) {$A_{46}$};
\node at (11,13) {$\bullet$};
\node at (11.25,12.9) {$A_{47}$};
\node at (12,14) {$\bullet$};
\node at (12.15,13.8) {$A_{48}$};
%
\node at (-0.2,-0.5) {{\small \bf $v_1$}};
\node at (0.8,1.5) {{\small \bf $v_2$}};
\node at (1.8,3.5) {{\small \bf $v_3$}};
\node at (2.8,5.5) {{\small \bf $v_4$}};
\node at (3.8,7.5) {{\small \bf $v_5$}};
\node at (4.8,9.5) {{\small \bf $v_6$}};
\node at (5.8,11.5) {{\small \bf $v_1$}};
\node at (1.8,0.5) {{\small \bf $v_7$}};
\node at (2.8,2.5) {{\small \bf $v_8$}};
\node at (3.8,4.5) {{\small \bf $v_9$}};
\node at (4.75,6.5) {{\small \bf $v_{10}$}};
\node at (5.75,8.5) {{\small \bf $v_{11}$}};
\node at (6.75,10.5) {{\small \bf $v_{12}$}};
\node at (7.8,12.5) {{\small \bf $v_{7}$}};
\node at (3.75,1.5) {{\small \bf $v_{13}$}};
\node at (4.75,3.5) {{\small \bf $v_{14}$}};
\node at (5.75,5.5) {{\small \bf $v_{15}$}};
\node at (6.75,7.5) {{\small \bf $v_{16}$}};
\node at (7.75,9.5) {{\small \bf $v_{17}$}};
\node at (8.75,11.5) {{\small \bf $v_{18}$}};
\node at (9.8,13.5) {{\small \bf $v_{13}$}};
\node at (5.75,2.5) {{\small \bf $v_{19}$}};
\node at (6.75,4.5) {{\small \bf $v_{20}$}};
\node at (7.75,6.5) {{\small \bf $v_{21}$}};
\node at (8.75,8.5) {{\small \bf $v_{22}$}};
\node at (9.75,10.5) {{\small \bf $v_{23}$}};
\node at (10.75,12.5) {{\small \bf $v_{24}$}};
\node at (11.8,14.5) {{\small \bf $v_{19}$}};
%
\node at (-0.7,0.2) {{\small \bf $h_{19}$}};
\node at (0.3,2.2) {{\small \bf $h_{20}$}};
\node at (1.3,4.2) {{\small \bf $h_{21}$}};
\node at (2.3,6.2) {{\small \bf $h_{22}$}};
\node at (3.3,8.2) {{\small \bf $h_{23}$}};
\node at (4.3,10.2) {{\small \bf $h_{24}$}};
\node at (1.5,1.2) {{\small \bf $h_1$}};
\node at (2.5,3.2) {{\small \bf $h_2$}};
\node at (3.5,5.2) {{\small \bf $h_3$}};
\node at (4.5,7.2) {{\small \bf $h_4$}};
\node at (5.5,9.2) {{\small \bf $h_5$}};
\node at (6.5,11.2) {{\small \bf $h_6$}};
\node at (3.5,2.2) {{\small \bf $h_7$}};
\node at (4.5,4.2) {{\small \bf $h_8$}};
\node at (5.5,6.2) {{\small \bf $h_9$}};
\node at (6.5,8.2) {{\small \bf $h_{10}$}};
\node at (7.5,10.2) {{\small \bf $h_{11}$}};
\node at (8.5,12.2) {{\small \bf $h_{12}$}};
\node at (5.5,3.2) {{\small \bf $h_{13}$}};
\node at (6.5,5.2) {{\small \bf $h_{14}$}};
\node at (7.5,7.2) {{\small \bf $h_{15}$}};
\node at (8.5,9.2) {{\small \bf $h_{16}$}};
\node at (9.5,11.2) {{\small \bf $h_{17}$}};
\node at (10.5,13.2) {{\small \bf $h_{18}$}};
\node at (7.8,4.2) {{\small \bf $h_{19}$}};
\node at (8.8,6.2) {{\small \bf $h_{20}$}};
\node at (9.8,8.2) {{\small \bf $h_{21}$}};
\node at (10.8,10.2) {{\small \bf $h_{22}$}};
\node at (11.8,12.2) {{\small \bf $h_{23}$}};
\node at (12.8,14.2) {{\small \bf $h_{24}$}};
%
\node at (0.35,0.65) {{\small \bf $m_1$}};
\node at (1.35,2.65) {{\small \bf $m_2$}};
\node at (2.35,4.65) {{\small \bf $m_3$}};
\node at (3.35,6.65) {{\small \bf $m_4$}};
\node at (4.35,8.65) {{\small \bf $m_5$}};
\node at (5.35,10.65) {{\small \bf $m_6$}};
%
\node at (2.35,1.65) {{\small \bf $m_7$}};
\node at (3.35,3.65) {{\small \bf $m_8$}};
\node at (4.35,5.65) {{\small \bf $m_9$}};
\node at (5.4,7.7) {{\small \bf $m_{10}$}};
\node at (6.4,9.7) {{\small \bf $m_{11}$}};
\node at (7.4,11.7) {{\small \bf $m_{12}$}};
%
\node at (4.4,2.7) {{\small \bf $m_{13}$}};
\node at (5.4,4.7) {{\small \bf $m_{14}$}};
\node at (6.4,6.7) {{\small \bf $m_{15}$}};
\node at (7.4,8.7) {{\small \bf $m_{16}$}};
\node at (8.4,10.7) {{\small \bf $m_{17}$}};
\node at (9.4,12.7) {{\small \bf $m_{18}$}};
%
\node at (6.4,3.7) {{\small \bf $m_{19}$}};
\node at (7.4,5.7) {{\small \bf $m_{20}$}};
\node at (8.4,7.7) {{\small \bf $m_{21}$}};
\node at (9.4,9.7) {{\small \bf $m_{22}$}};
\node at (10.4,11.7) {{\small \bf $m_{23}$}};
\node at (11.4,13.7) {{\small \bf $m_{24}$}};
\node at (0,-1.15) {{\small \bf $a$}};
\node at (2,-0.15) {{\small \bf $b$}};
\node at (4,0.85) {{\small \bf $c$}};
\node at (6,1.85) {{\small \bf $d$}};
\node at (6,12.15) {{\small \bf $a$}};
\node at (8,13.15) {{\small \bf $b$}};
\node at (10,14.15) {{\small \bf $c$}};
\node at (12,15.15) {{\small \bf $d$}};
\node at (-1.15,0) {{\small \bf $1$}};
\node at (-0.15,2) {{\small \bf $2$}};
\node at (0.85,4) {{\small \bf $3$}};
\node at (1.85,6) {{\small \bf $4$}};
\node at (2.85,8) {{\small \bf $5$}};
\node at (3.85,10) {{\small \bf $6$}};
\node at (8.15,4) {{\small \bf $1$}};
\node at (9.15,6) {{\small \bf $2$}};
\node at (10.15,8) {{\small \bf $3$}};
\node at (11.15,10) {{\small \bf $4$}};
\node at (12.15,12) {{\small \bf $5$}};
\node at (13.15,14) {{\small \bf $6$}};
\end{tikzpicture}}
}
\end{center}
\caption{\sl Parametrisation of the web $(N,M)=(4,6)$ along with an assignment of vertices.}
\label{Fig:Web46vertices}
\end{figure}

\begin{figure}[htbp]
  \centering
  \includegraphics[width=6in]{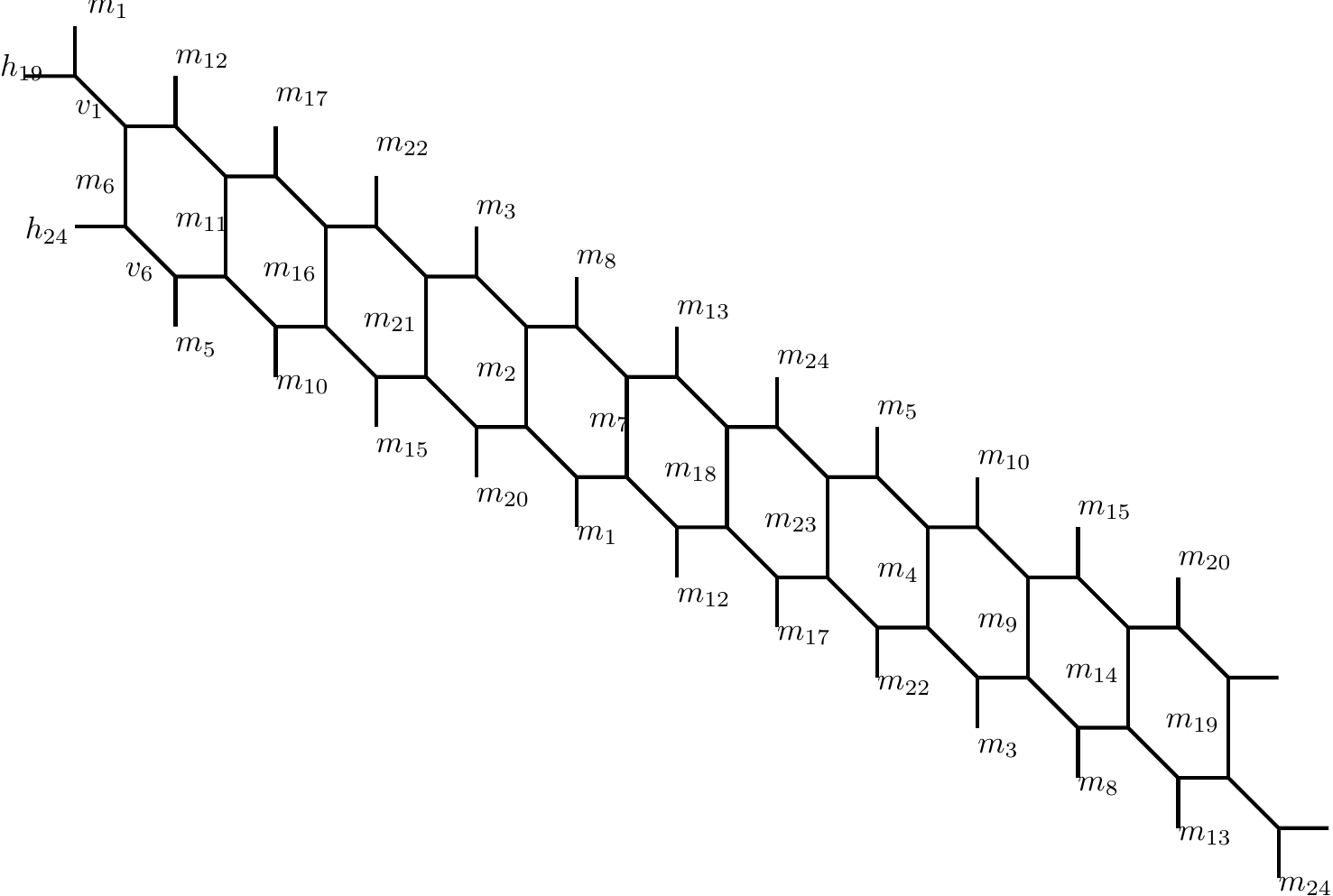}
 \caption{\sl Twisted $(N,M)=(12,2)$ web which is obtained from Figure~\ref{Fig:Web46vertices}  by cutting and re-gluing different lines.}\label{Fig:Web122twisteda}
\end{figure}

\begin{figure}[h]
  \centering
  \includegraphics[width=6in]{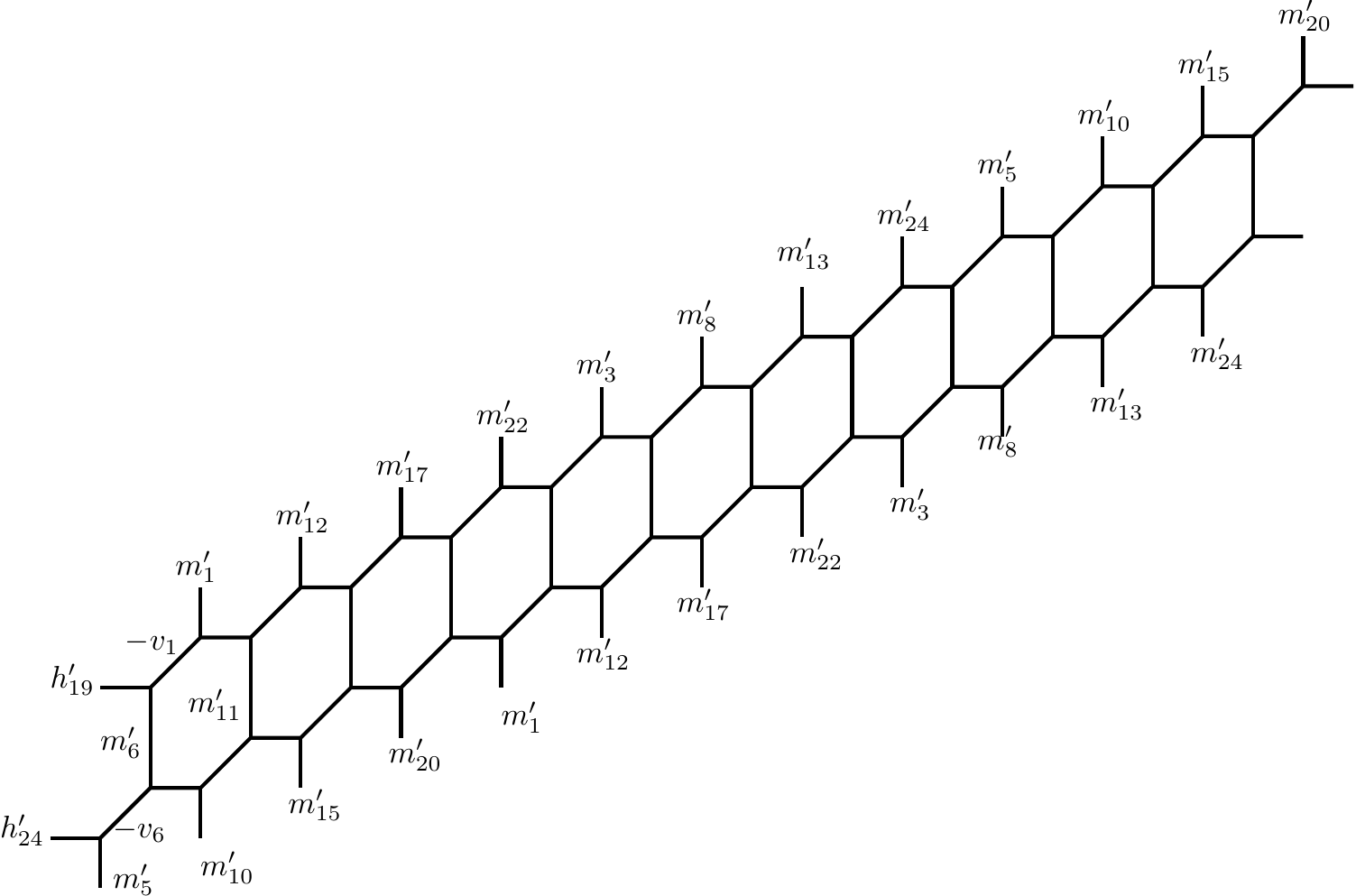}
 \caption{The twisted $(N,M)=(12,2)$ web of Figure~\ref{Fig:Web122twisteda} after flop transitions along the curves $v_i$.}\label{Fig:Web122twistedb}
\end{figure}

\begin{figure}[htb]
\begin{center}
\rotatebox{90}{\scalebox{0.7}{\parbox{29.5cm}{\begin{tikzpicture}[scale = 1.1]
\draw[ultra thick] (2,1) -- (2,2);
\draw[ultra thick] (1,0) -- (1,-1);
\draw[ultra thick] (4,2) -- (4,3);
\draw[ultra thick] (3,1) -- (3,0);
\draw[ultra thick] (5,2) -- (5,1);
\draw[ultra thick] (6,3) -- (6,4);
\draw[ultra thick] (7,3) -- (7,2);
\draw[ultra thick] (8,4) -- (8,5);
\draw[ultra thick] (9,4) -- (9,3);
\draw[ultra thick] (10,5) -- (10,6);
\draw[ultra thick] (11,5) -- (11,4);
\draw[ultra thick] (12,6) -- (12,7);
\draw[ultra thick] (13,6) -- (13,5);
\draw[ultra thick] (14,7) -- (14,8);
\draw[ultra thick] (15,7) -- (15,6);
\draw[ultra thick] (16,8) -- (16,9);
\draw[ultra thick] (17,8) -- (17,7);
\draw[ultra thick] (18,9) -- (18,10);
\draw[ultra thick] (19,9) -- (19,8);
\draw[ultra thick] (20,10) -- (20,11);
\draw[ultra thick] (21,10) -- (21,9);
\draw[ultra thick] (22,11) -- (22,12);
\draw[ultra thick] (23,11) -- (23,10);
\draw[ultra thick] (24,12) -- (24,13);
\draw[ultra thick] (0,0) -- (1,0);
\draw[ultra thick] (2,1) -- (3,1);
\draw[ultra thick] (4,2) -- (5,2);
\draw[ultra thick] (6,3) -- (7,3);
\draw[ultra thick] (8,4) -- (9,4);
\draw[ultra thick] (10,5) -- (11,5);
\draw[ultra thick] (12,6) -- (13,6);
\draw[ultra thick] (14,7) -- (15,7);
\draw[ultra thick] (16,8) -- (17,8);
\draw[ultra thick] (18,9) -- (19,9);
\draw[ultra thick] (20,10) -- (21,10);
\draw[ultra thick] (22,11) -- (23,11);
\draw[ultra thick] (24,12) -- (25,12);
\draw[ultra thick] (1,0) -- (2,1);
\draw[ultra thick] (3,1) -- (4,2);
\draw[ultra thick] (5,2) -- (6,3);
\draw[ultra thick] (7,3) -- (8,4);
\draw[ultra thick] (9,4) -- (10,5);
\draw[ultra thick] (11,5) -- (12,6);
\draw[ultra thick] (13,6) -- (14,7);
\draw[ultra thick] (15,7) -- (16,8);
\draw[ultra thick] (17,8) -- (18,9);
\draw[ultra thick] (19,9) -- (20,10);
\draw[ultra thick] (21,10) -- (22,11);
\draw[ultra thick] (23,11) -- (24,12);
\draw[ultra thick] (3,3) -- (3,4);
\draw[ultra thick] (5,4) -- (5,5);
\draw[ultra thick] (7,5) -- (7,6);
\draw[ultra thick] (9,6) -- (9,7);
\draw[ultra thick] (11,7) -- (11,8);
\draw[ultra thick] (13,8) -- (13,9);
\draw[ultra thick] (15,9) -- (15,10);
\draw[ultra thick] (17,10) -- (17,11);
\draw[ultra thick] (19,11) -- (19,12);
\draw[ultra thick] (21,12) -- (21,13);
\draw[ultra thick] (23,13) -- (23,14);
\draw[ultra thick] (25,14) -- (25,15);
\draw[ultra thick] (1,2) -- (2,2);
\draw[ultra thick] (3,3) -- (4,3);
\draw[ultra thick] (5,4) -- (6,4);
\draw[ultra thick] (7,5) -- (8,5);
\draw[ultra thick] (9,6) -- (10,6);
\draw[ultra thick] (11,7) -- (12,7);
\draw[ultra thick] (13,8) -- (14,8);
\draw[ultra thick] (15,9) -- (16,9);
\draw[ultra thick] (17,10) -- (18,10);
\draw[ultra thick] (19,11) -- (20,11);
\draw[ultra thick] (21,12) -- (22,12);
\draw[ultra thick] (23,13) -- (24,13);
\draw[ultra thick] (25,14) -- (26,14);
\draw[ultra thick] (2,2) -- (3,3);
\draw[ultra thick] (4,3) -- (5,4);
\draw[ultra thick] (6,4) -- (7,5);
\draw[ultra thick] (8,5) -- (9,6);
\draw[ultra thick] (10,6) -- (11,7);
\draw[ultra thick] (12,7) -- (13,8);
\draw[ultra thick] (14,8) -- (15,9);
\draw[ultra thick] (16,9) -- (17,10);
\draw[ultra thick] (18,10) -- (19,11);
\draw[ultra thick] (20,11) -- (21,12);
\draw[ultra thick] (22,12) -- (23,13);
\draw[ultra thick] (24,13) -- (25,14);
\node[rotate=45] at (1.35,0.75) {{\small $\bar{m}_{16}$}};
\node[rotate=45] at (3.35,1.75) {{\small $\bar{m}_{11}$}};
\node[rotate=45] at (5.35,2.75) {{\small $\bar{m}_{6}$}};
\node[rotate=45] at (7.35,3.75) {{\small $\bar{m}_{19}$}};
\node[rotate=45] at (9.35,4.75) {{\small $\bar{m}_{24}$}};
\node[rotate=45] at (11.35,5.75) {{\small $\bar{m}_9$}};
\node[rotate=45] at (13.35,6.75) {{\small $\bar{m}_4$}};
\node[rotate=45] at (15.35,7.75) {{\small $\bar{m}_{23}$}};
\node[rotate=45] at (17.35,8.75) {{\small $\bar{m}_{18}$}};
\node[rotate=45] at (19.35,9.75) {{\small $\bar{m}_7$}};
\node[rotate=45] at (21.35,10.75) {{\small $\bar{m}_2$}};
\node[rotate=45] at (23.35,11.75) {{\small $\bar{m}_{21}$}};
\node[rotate=45] at (2.35,2.75) {{\small $\bar{m}_{1}$}};
\node[rotate=45] at (4.35,3.75) {{\small $\bar{m}_{20}$}};
\node[rotate=45] at (6.35,4.75) {{\small $\bar{m}_{15}$}};
\node[rotate=45] at (8.35,5.75) {{\small $\bar{m}_{10}$}};
\node[rotate=45] at (10.35,6.75) {{\small $\bar{m}_{5}$}};
\node[rotate=45] at (12.35,7.75) {{\small $\bar{m}_{24}$}};
\node[rotate=45] at (14.35,8.75) {{\small $\bar{m}_{13}$}};
\node[rotate=45] at (16.35,9.75) {{\small $\bar{m}_{8}$}};
\node[rotate=45] at (18.35,10.75) {{\small $\bar{m}_{3}$}};
\node[rotate=45] at (20.35,11.75) {{\small $\bar{m}_{22}$}};
\node[rotate=45] at (22.35,12.75) {{\small $\bar{m}_{17}$}};
\node[rotate=45] at (24.35,13.75) {{\small $\bar{m}_{12}$}};
\node at (0.5,0.3) {{\small $\bar{h}_{15}$}};
\node at (2.5,1.3) {{\small $\bar{h}_{10}$}};
\node at (4.5,2.3) {{\small $\bar{h}_{5}$}};
\node at (6.5,3.3) {{\small $\bar{h}_{24}$}};
\node at (8.5,4.3) {{\small $\bar{h}_{13}$}};
\node at (10.5,5.3) {{\small $\bar{h}_8$}};
\node at (12.5,6.3) {{\small $\bar{h}_3$}};
\node at (14.5,7.3) {{\small $\bar{h}_{22}$}};
\node at (16.5,8.3) {{\small $\bar{h}_{17}$}};
\node at (18.5,9.3) {{\small $\bar{h}_{12}$}};
\node at (20.5,10.3) {{\small $\bar{h}_1$}};
\node at (22.5,11.3) {{\small $\bar{h}_{20}$}};
\node at (24.5,12.3) {{\small $\bar{h}_{15}$}};
\node at (1.5,2.3) {{\small $\bar{h}_{6}$}};
\node at (3.5,3.3) {{\small $\bar{h}_{19}$}};
\node at (5.5,4.3) {{\small $\bar{h}_{14}$}};
\node at (7.5,5.3) {{\small $\bar{h}_{9}$}};
\node at (9.5,6.3) {{\small $\bar{h}_{4}$}};
\node at (11.5,7.3) {{\small $\bar{h}_{23}$}};
\node at (13.5,8.3) {{\small $\bar{h}_{18}$}};
\node at (15.5,9.3) {{\small $\bar{h}_{7}$}};
\node at (17.5,10.3) {{\small $\bar{h}_{2}$}};
\node at (19.5,11.3) {{\small $\bar{h}_{21}$}};
\node at (21.5,12.3) {{\small $\bar{h}_{16}$}};
\node at (23.5,13.3) {{\small $\bar{h}_{11}$}};
\node at (25.5,14.3) {{\small $\bar{h}_{6}$}};
\node at (3.3,3.8) {{\small $\bar{v}_{20}$}};
\node at (5.3,4.8) {{\small $\bar{v}_{15}$}};
\node at (7.3,5.8) {{\small $\bar{v}_{10}$}};
\node at (9.3,6.8) {{\small $\bar{v}_5$}};
\node at (11.3,7.8) {{\small $\bar{v}_{24}$}};
\node at (13.3,8.8) {{\small $\bar{v}_{13}$}};
\node at (15.3,9.8) {{\small $\bar{v}_8$}};
\node at (17.3,10.8) {{\small $\bar{v}_3$}};
\node at (19.3,11.8) {{\small $\bar{v}_{22}$}};
\node at (21.3,12.8) {{\small $\bar{v}_{17}$}};
\node at (23.3,13.8) {{\small $\bar{v}_{12}$}};
\node at (25.3,14.8) {{\small $\bar{v}_1$}};
\node at (0.7,-0.8) {{\small $\bar{v}_{20}$}};
\node at (2.7,0.2) {{\small $\bar{v}_{16}$}};
\node at (4.7,1.2) {{\small $\bar{v}_{10}$}};
\node at (6.7,2.2) {{\small $\bar{v}_5$}};
\node at (8.7,3.2) {{\small $\bar{v}_{24}$}};
\node at (10.7,4.2) {{\small $\bar{v}_{13}$}};
\node at (12.7,5.2) {{\small $\bar{v}_8$}};
\node at (14.7,6.2) {{\small $\bar{v}_3$}};
\node at (16.7,7.2) {{\small $\bar{v}_{22}$}};
\node at (18.7,8.2) {{\small $\bar{v}_{17}$}};
\node at (20.7,9.2) {{\small $\bar{v}_{12}$}};
\node at (22.7,10.2) {{\small $\bar{v}_1$}};
\node at (1.7,1.5) {{\small $\bar{v}_{11}$}};
\node at (3.7,2.5) {{\small $\bar{v}_{6}$}};
\node at (5.7,3.5) {{\small $\bar{v}_{19}$}};
\node at (7.7,4.5) {{\small $\bar{v}_{14}$}};
\node at (9.7,5.5) {{\small $\bar{v}_{9}$}};
\node at (11.7,6.5) {{\small $\bar{v}_{4}$}};
\node at (13.7,7.5) {{\small $\bar{v}_{23}$}};
\node at (15.7,8.5) {{\small $\bar{v}_{18}$}};
\node at (17.7,9.5) {{\small $\bar{v}_{7}$}};
\node at (19.7,10.5) {{\small $\bar{v}_{2}$}};
\node at (21.7,11.5) {{\small $\bar{v}_{21}$}};
\node at (23.7,12.5) {{\small $\bar{v}_{16}$}};
\node at (-0.2,0) {{\small \bf $a$}};
\node at (0.8,2) {{\small \bf $b$}};
\node at (25.2,12) {{\small \bf $a$}};
\node at (26.2,14) {{\small \bf $b$}};
\node at (1,-1.2) {{\small \bf $12$}};
\node at (3,-0.2) {{\small \bf $11$}};
\node at (5,0.8) {{\small \bf $10$}};
\node at (7,1.8) {{\small \bf $9$}};
\node at (9,2.8) {{\small \bf $8$}};
\node at (11,3.8) {{\small \bf $7$}};
\node at (13,4.8) {{\small \bf $6$}};
\node at (15,5.8) {{\small \bf $5$}};
\node at (17,6.8) {{\small \bf $4$}};
\node at (19,7.8) {{\small \bf $3$}};
\node at (21,8.8) {{\small \bf $2$}};
\node at (23,9.8) {{\small \bf $1$}};
\node at (3,4.2) {{\small \bf $12$}};
\node at (5,5.2) {{\small \bf $11$}};
\node at (7,6.2) {{\small \bf $10$}};
\node at (9,7.2) {{\small \bf $9$}};
\node at (11,8.2) {{\small \bf $8$}};
\node at (13,9.2) {{\small \bf $7$}};
\node at (15,10.2) {{\small \bf $6$}};
\node at (17,11.2) {{\small \bf $5$}};
\node at (19,12.2) {{\small \bf $4$}};
\node at (21,13.2) {{\small \bf $3$}};
\node at (23,14.2) {{\small \bf $2$}};
\node at (25,15.2) {{\small \bf $1$}};
%
\node at (1,0) {$\bullet$};
\node at (1.4,-0.1) {$A_{16}$};
\node at (2,1) {$\bullet$};
\node at (2.1,0.7) {$A_{35}$};
\node at (3,1) {$\bullet$};
\node at (3.4,0.9) {$A_{22}$};
\node at (4,2) {$\bullet$};
\node at (4.1,1.7) {$A_{41}$};
\node at (5,2) {$\bullet$};
\node at (5.4,1.9) {$A_{28}$};
\node at (6,3) {$\bullet$};
\node at (6.15,2.7) {$A_{7}$};
\node at (7,3) {$\bullet$};
\node at (7.4,2.9) {$A_{34}$};
\node at (8,4) {$\bullet$};
\node at (8.1,3.7) {$A_{13}$};
\node at (9,4) {$\bullet$};
\node at (9.4,3.9) {$A_{48}$};
\node at (10,5) {$\bullet$};
\node at (10.1,4.7) {$A_{19}$};
\node at (11,5) {$\bullet$};
\node at (11.35,4.9) {$A_{6}$};
\node at (12,6) {$\bullet$};
\node at (12.1,5.7) {$A_{25}$};
\node at (13,6) {$\bullet$};
\node at (13.4,5.9) {$A_{12}$};
\node at (14,7) {$\bullet$};
\node at (14.1,6.7) {$A_{39}$};
\node at (15,7) {$\bullet$};
\node at (15.4,6.9) {$A_{18}$};
\node at (16,8) {$\bullet$};
\node at (16.1,7.7) {$A_{45}$};
\node at (17,8) {$\bullet$};
\node at (17.4,7.9) {$A_{32}$};
\node at (18,9) {$\bullet$};
\node at (18.15,8.7) {$A_{3}$};
\node at (19,9) {$\bullet$};
\node at (19.4,8.9) {$A_{38}$};
\node at (20,10) {$\bullet$};
\node at (20.15,9.7) {$A_{9}$};
\node at (21,10) {$\bullet$};
\node at (21.4,9.9) {$A_{44}$};
\node at (22,11) {$\bullet$};
\node at (22.1,10.7) {$A_{23}$};
\node at (23,11) {$\bullet$};
\node at (23.35,10.9) {$A_{2}$};
\node at (24,12) {$\bullet$};
\node at (24.1,11.7) {$A_{29}$};
%
\node at (2,2) {$\bullet$};
\node at (2.4,1.9) {$A_{36}$};
\node at (3,3) {$\bullet$};
\node at (3.1,2.7) {$A_{15}$};
\node at (4,3) {$\bullet$};
\node at (4.4,2.9) {$A_{42}$};
\node at (5,4) {$\bullet$};
\node at (5.1,3.7) {$A_{21}$};
\node at (6,4) {$\bullet$};
\node at (6.35,3.9) {$A_{8}$};
\node at (7,5) {$\bullet$};
\node at (7.15,4.7) {$A_{27}$};
\node at (8,5) {$\bullet$};
\node at (8.4,4.9) {$A_{14}$};
\node at (9,6) {$\bullet$};
\node at (9.1,5.7) {$A_{33}$};
\node at (10,6) {$\bullet$};
\node at (10.4,5.9) {$A_{20}$};
\node at (11,7) {$\bullet$};
\node at (11.1,6.7) {$A_{47}$};
\node at (12,7) {$\bullet$};
\node at (12.4,6.9) {$A_{26}$};
\node at (13,8) {$\bullet$};
\node at (13.15,7.7) {$A_{5}$};
\node at (14,8) {$\bullet$};
\node at (14.4,7.9) {$A_{40}$};
\node at (15,9) {$\bullet$};
\node at (15.1,8.7) {$A_{11}$};
\node at (16,9) {$\bullet$};
\node at (16.4,8.9) {$A_{46}$};
\node at (17,10) {$\bullet$};
\node at (17.1,9.7) {$A_{17}$};
\node at (18,10) {$\bullet$};
\node at (18.35,9.9) {$A_{4}$};
\node at (19,11) {$\bullet$};
\node at (19.1,10.7) {$A_{31}$};
\node at (20,11) {$\bullet$};
\node at (20.4,10.9) {$A_{10}$};
\node at (21,12) {$\bullet$};
\node at (21.1,11.7) {$A_{37}$};
\node at (22,12) {$\bullet$};
\node at (22.4,11.9) {$A_{24}$};
\node at (23,13) {$\bullet$};
\node at (23.1,12.7) {$A_{43}$};
\node at (24,13) {$\bullet$};
\node at (24.4,12.9) {$A_{30}$};
\node at (25,14) {$\bullet$};
\node at (25.15,13.7) {$A_{1}$};
\end{tikzpicture}}}
}
\end{center}
\caption{\sl Parametrisation of the web $(N,M)=(12,2)$ web along with an assignment of vertices.}
\label{Fig:Web122vertices}
\end{figure}
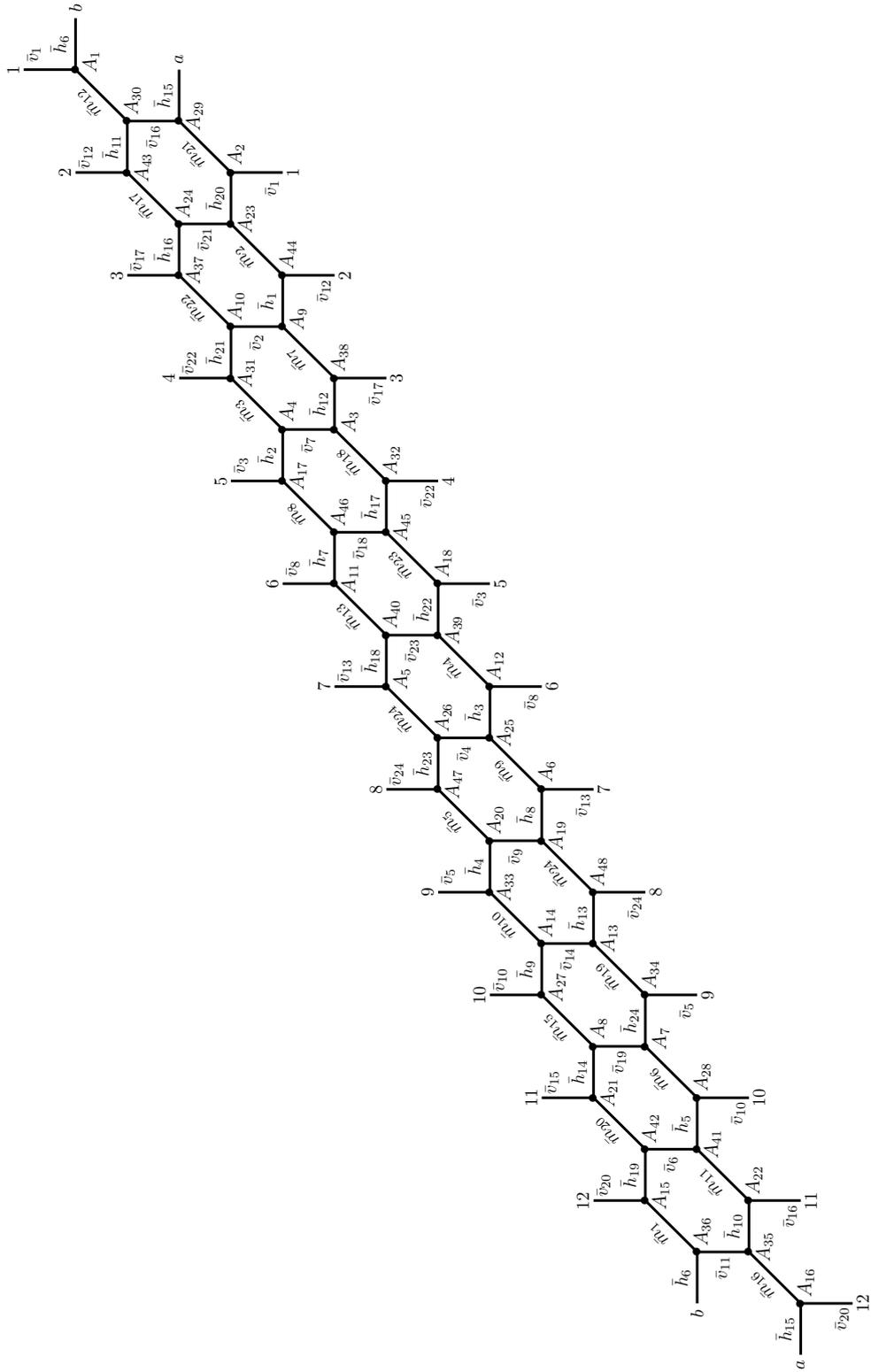

A parametrisation of the web $(N,M)=(4,6)$ is given in Figure~\ref{Fig:Web46vertices}. Out of the 72 parameters in the diagram, only 26 are independent on account of the consistency conditions.
By cutting and re-gluing the diagram in Figure~\ref{Fig:Web46vertices}, we can bring it into a form which is very similar to the web $(N,M)=(12,2)$, except that the lines at the top of the diagram are glued to the lines at the bottom in a twisted fashion (see Figure~\ref{Fig:Web122twisteda}).
Next we perform a flop transition on the curves labelled by parameters $v_{i}$, which is carried out in both 'layers', as shown in the Figure~\ref{Fig:Web122twistedb}.
The new parameters $h'_{i}$ can be determined using \figref{Fig:flop} and are given by
{\allowdisplaybreaks
\begin{align}
&h'_{1}=h_{1}+v_{2}+v_{7}\,,&&h'_{9}=h_{9}+v_{10}+v_{15}\,,&&h'_{17}=h_{17}+v_{18}+v_{23}\,,\nonumber\\
&h'_{2}=h_{2}+v_{3}+v_{8}\,,&&h'_{10}=h_{10}+v_{11}+v_{16}\,,&&h'_{18}=h_{18}+v_{13}+v_{24}\,,\nonumber\\
&h'_{3}=h_{3}+v_{4}+v_{9}\,,&&h'_{11}=h_{11}+v_{12}+v_{17}\,,&&h'_{19}=h_{19}+v_{1}+v_{20}\,,\nonumber\\
&h'_{4}=h_{4}+v_{5}+v_{10}\,,&&h'_{12}=h_{12}+v_{7}+v_{18}\,,&&h'_{20}=h_{20}+v_{21}+v_{2}\,,\nonumber\\
&h'_{5}=h_{5}+v_{6}+v_{11}\,,&&h'_{13}=h_{13}+v_{14}+v_{19}\,,&&h'_{21}=h_{21}+v_{22}+v_{3}\,,\nonumber\\
&h'_{6}=h_{6}+v_{1}+v_{12}\,,&&h'_{14}=h_{14}+v_{15}+v_{20}\,,&&h'_{22}=h_{22}+v_{23}+v_{4}\,,\nonumber\\
&h'_{7}=h_{7}+v_{8}+v_{13}\,,&&h'_{15}=h_{15}+v_{16}+v_{21}\,,&&h'_{23}=h_{23}+v_{24}+v_{5}\,,\nonumber\\
&h'_{8}=h_{8}+v_{9}+v_{14}\,,&&h'_{16}=h_{16}+v_{17}+v_{22}\,,&&h'_{24}=h_{24}+v_{6}+v_{19}\,.
\end{align}}
along with the parameters $m'_{i}$ 
{\allowdisplaybreaks
\begin{align}
&m'_{1}=m_{1}+v_{1}+v_{2}\,,&&m'_{2}=m_{2}+v_{3}+v_{2}\,,&&m'_{3}=m_{3}+v_{3}+v_{4}\,,\nonumber\\
&m'_{4}=m_{4}+v_{5}+v_{4}\,,&&m'_{5}=m_{5}+v_{5}+v_{4}\,,&&m'_{6}=m_{6}+v_{1}+v_{6}\,,\nonumber\\
&m'_{7}=m_{7}+v_{8}+v_{7}\,,&&m'_{8}=m_{8}+v_{8}+v_{9}\,,&&m'_{9}=m_{9}+v_{10}+v_{9}\,,\nonumber\\
&m'_{10}=m_{10}+v_{10}+v_{11}\,,&&m'_{11}=m_{11}+v_{12}+v_{11}\,,&&m'_{12}=m_{12}+v_{12}+v_{7}\,,\nonumber\\
&m'_{13}=m_{13}+v_{13}+v_{14}\,,&&m'_{14}=m_{14}+v_{15}+v_{14}\,,&&m'_{15}=m_{15}+v_{15}+v_{16}\,,\nonumber\\
&m'_{16}=m_{16}+v_{17}+v_{16}\,,&&m'_{17}=m_{17}+v_{17}+v_{18}\,,&&m'_{18}=m_{18}+v_{13}+v_{18}\,,\nonumber\\
&m'_{19}=m_{19}+v_{20}+v_{19}\,,&&m'_{20}=m_{20}+v_{20}+v_{21}\,,&&m'_{21}=m_{21}+v_{22}+v_{21}\,,\nonumber\\
&m'_{22}=m_{22}+v_{22}+v_{23}\,,&&m'_{23}=m_{23}+v_{24}+v_{23}\,,&&m'_{24}=m_{24}+v_{24}+v_{19}\,.
\end{align}}
Now carrying out the flop transition on the curves labelled by $h'_{i}$ in the web shown in \figref{Fig:Web122twistedb} gives the web shown in \figref{Fig:Web122vertices} which is the standard untwisted $(12,2)$ web.
The parameters $\overline{v}_{i}$ in terms of the parameters of the original $(4,6)$ web are given by
{\allowdisplaybreaks
\begin{align}
&\overline{v}_{1}=m'_{1}+h'_{6}+h'_{20}\,,&&\overline{v}_{2}=m'_{2}+h'_{1}+h'_{21}\,,&&\overline{v}_{3}=m'_{3}+h'_{2}+h'_{22}\,,\nonumber\\
&\overline{v}_{4}=m'_{4}+h'_{3}+h'_{23}\,,&&\overline{v}_{5}=m'_{5}+h'_{4}+h'_{24}\,,&&\overline{v}_{6}=m'_{6}+h'_{5}+h'_{19}\,,\nonumber\\
&\overline{v}_{7}=m'_{7}+h'_{12}+h'_{2}\,,&&\overline{v}_{8}=m'_{8}+h'_{7}+h'_{3}\,,&&\overline{v}_{9}=m'_{9}+h'_{8}+h'_{4}\,,\nonumber\\
&\overline{v}_{10}=m'_{10}+h'_{9}+h'_{5}\,,&&\overline{v}_{11}=m'_{11}+h'_{10}+h'_{6}\,,&&\overline{v}_{12}=m'_{12}+h'_{11}+h'_{1}\,,\nonumber\\
&\overline{v}_{13}=m'_{13}+h'_{18}+h'_{8}\,,&&\overline{v}_{14}=m'_{14}+h'_{13}+h'_{9}\,,&&\overline{v}_{15}=m'_{15}+h'_{14}+h'_{10}\,,\nonumber\\
&\overline{v}_{16}=m'_{16}+h'_{15}+h'_{11}\,,&&\overline{v}_{17}=m'_{17}+h'_{6}+h'_{20}\,,&&\overline{v}_{18}=m'_{18}+h'_{17}+h'_{7}\,,\nonumber\\
&\overline{v}_{19}=m'_{19}+h'_{14}+h'_{24}\,,&&\overline{v}_{20}=m'_{20}+h'_{19}+h'_{15}\,,&&\overline{v}_{21}=m'_{21}+h'_{16}+h'_{20}\,,\nonumber\\
&\overline{v}_{22}=m'_{22}+h'_{21}+h'_{17}\,,&&\overline{v}'_{23}=m'_{23}+h'_{22}+h'_{18}\,,&&
\overline{v}_{24}=m'_{24}+h'_{23}+h'_{13}\,,
\end{align}}
while the parameters $\overline{m}_{i}$ are 
{\allowdisplaybreaks
\begin{align}
&\overline{m}_{1}=-v_{1}+h'_{19}+h'_{6}\,,&&\overline{m}_{2}=-v_{2}+h'_{20}+h'_{1}\,,&&\overline{m}_{3}=-v_{3}+h'_{21}+h'_{2}\,,\nonumber\\
&\overline{m}_{4}=-v_{4}+h'_{22}+h'_{3}\,,&&
\overline{m}_{5}=-v_{5}+h'_{23}+h'_{4}\,,&&\overline{m}_{6}=-v_{6}+h'_{26}+h'_{5}\,,\nonumber\\
&\overline{m}_{7}=-v_{7}+h'_{1}+h'_{12}\,,&&\overline{m}_{8}=-v_{8}+h'_{2}+h'_{7}\,,&&\overline{m}_{9}=-v_{9}+h'_{3}+h'_{8}\,,\nonumber\\
&\overline{m}_{10}=-v_{10}+h'_{4}+h'_{9}\,,&&\overline{m}_{11}=-v_{11}+h'_{5}+h'_{10}\,,&&\overline{m}_{12}=-v_{12}+h'_{6}+h'_{11}\,,\nonumber\\
&\overline{m}_{13}=-v_{13}+h'_{7}+h'_{18}\,,&&\overline{m}_{14}=-v_{14}+h'_{8}+h'_{13}\,,&&\overline{m}_{15}=-v_{15}+h'_{9}+h'_{14}\,,\nonumber\\
&\overline{m}_{16}=-v_{16}+h'_{10}+h'_{15}\,,&&\overline{m}_{17}=-v_{17}+h'_{11}+h'_{16}\,,&&\overline{m}_{18}=-v_{18}+h'_{12}+h'_{17}\,,\nonumber\\
&\overline{m}_{19}=-v_{19}+h'_{13}+h'_{26}\,,&&\overline{m}_{20}=-v_{20}+h'_{14}+h'_{19}\,,&&\overline{m}_{21}=-v_{21}+h'_{15}+h'_{20}\,,\nonumber\\
&\overline{m}_{22}=-v_{22}+h'_{16}+h'_{21}\,,&&\overline{m}_{23}=-v_{23}+h'_{17}+h'_{22}\,,&&\overline{m}_{24}=-v_{24}+h'_{18}+h'_{23}\,,
\end{align}}
and the parameters $\overline{h}_{i}$ are 
{\allowdisplaybreaks
\begin{align}
&\overline{h}_{1}=-(h_{1}+v_{2}+v_{7})\,,&&\overline{h}_{2}=-(h_{2}+v_{3}+v_{8})\,,&&\overline{h}_{3}=-(h_{3}+v_{4}+v_{9})\,,\nonumber\\
&\overline{h}_{4}=-(h_{4}+v_{5}+v_{10})\,,&&\overline{h}_{5}=-(h_{5}+v_{6}+v_{11})\,,&&\overline{h}_{6}=-(h_{6}+v_{1}+v_{12})\,,\nonumber\\
&\overline{h}_{7}=-(h_{7}+v_{8}+v_{13})\,,&&\overline{h}_{8}=-(h_{8}+v_{9}+v_{14})\,,&&\overline{h}_{9}=-(h_{9}+v_{10}+v_{15})\,,\nonumber\\
&\overline{h}_{10}=-(h_{10}+v_{11}+v_{16})\,,&&\overline{h}_{11}=-(h_{11}+v_{12}+v_{17})\,,&&\overline{h}_{12}=-(h_{12}+v_{7}+v_{18})\,,\nonumber\\
&\overline{h}_{13}=-(h_{13}+v_{14}+v_{19})\,,&&\overline{h}_{14}=-(h_{14}+v_{15}+v_{20})\,,&&\overline{h}_{15}=-(h_{15}+v_{16}+v_{21})\,,\nonumber\\
&\overline{h}_{16}=-(h_{16}+v_{17}+v_{22})\,,&&\overline{h}_{17}=-(h_{17}+v_{18}+v_{23})\,,&&\overline{h}_{18}=-(h_{18}+v_{13}+v_{24})\,,\nonumber\\
&\overline{h}_{19}=-(h_{19}+v_{1}+v_{20})\,,&&\overline{h}_{20}=-(h_{20}+v_{21}+v_{2})\,,&&\overline{h}_{21}=-(h_{21}+v_{22}+v_{3})\,,\nonumber\\
&\overline{h}_{22}=-(h_{22}+v_{23}+v_{4})\,,&&\overline{h}_{23}=-(h_{23}+v_{24}+v_{5})\,,&&\overline{h}_{24}=-(h_{24}+v_{6}+v_{19})\,.
\end{align}}
As in the previous cases, using the assignment of vertices in Figure~\ref{Fig:Web46vertices} and Figure~\ref{Fig:Web122vertices} the parameters $\overline{h}$ and $\overline{m}$ can be directly read off. Furthermore, the transformation above shows that the web diagrams $(N,M)=(4,6)$ and $(12,2)$ are indeed dual to each other.

Finally, we note that in this case (in contrast to the previous examples) the series of transformations discussed above apparently does not lead to an enhancement of supersymmetry at the level of little string theories: indeed, little string theories with $(N,M)=(2,12)$ still preserve 8 supercharges. This is due to the fact that $\text{gcd}(4,6)=2>1$. 
\subsubsection{Simple Check of Duality}
In order to perform a check of the duality worked out above, we consider $(m_{i},h_{i},v_{i})=(m,h,v)$ such that 
\bea
\tau=6(v+m) \implies v=\frac{\tau}{6}-m\\\nonumber
\rho=4(h+m)\implies h=\frac{\rho}{4}-m\\\nonumber
\eea
In the $(12,2)$ web the parameters $(\rho',\tau',m')$ are given by
\begin{align}
&\tau'=2(\overline{v}+\overline{m})\,,&&\text{with}&&\rho'=12(\overline{h}+\overline{m})\,,
\end{align}
with 
\begin{align}
&\overline{v}=m+2h+6v\,,&&\text{and} &&\overline{h}=-(h+2v)\,,&&\text{and} &&\overline{m}=2h+3v\,.
\end{align}
This gives the following map between $(\tau,\rho,m)$ and $(\tau',\rho',m')$:
\begin{align}
&\tau'=3\tau+2\rho - 24 m\,,&&\rho'=3\rho+2\tau-24m\,,&&m'=\frac{\tau}{2}+\frac{\rho}{2}-5m\,.
\end{align}
Next, we also introduce the period matrices
\begin{align}
&\Omega=\left(\begin{array}{cc}\tau/6 & m \\ m & \rho/4\end{array}\right)\,,&&\text{and}&&\Omega'=\left(\begin{array}{cc} \tau'/2 & m' \\ m' & \rho'/12\end{array}\right)\,,
\end{align}
which are related to each other by
\begin{align}
\Omega'=(A\Omega+B)\cdot (C\Omega+D)^{-1}\,,
\end{align}
with the matrices
\begin{align}
&A=\left(\begin{array}{cc} 3 & -2 \\ 1 & -1\end{array}\right)\,,&&B=\left(\begin{array}{cc}0 & 0 \\ 0 & 0\end{array}\right)\,,&&C=\left(\begin{array}{cc}0 & 0 \\ 0 & 0\end{array}\right)\,,&&D=\left(\begin{array}{cc}1 & 1 \\ -2 & -3\end{array}\right)\,,
\end{align}
which satisfy indeed the relations (\ref{Sp2ZTrafos}). This shows that $(\rho,\tau,m)$ and $(\rho',\tau',m')$ are related through an $Sp(2,\mathbb{Z})$ transformation. Therefore, $\mathcal{Z}_{X_{1,1}}$ is indeed invariant under the change of parameters implied by the above chain of dualities.

\FloatBarrier

\section{Map for arbitrary $(N,M)$}\label{Sect:GenericNM}
In the previous section, we have discussed a number of examples of how to transform a web $(N,M)$ into a web $(\tfrac{MN}{k},k)$, where $k=\text{gcd}(M,N)$. While we have argued that such a transformation always exists, this does not fix the parameters of the new toric diagram $(\tfrac{MN}{k},k)$ in terms of the parameters of the original one. However, based on the examples discussed before, we explain in this section that there exists a clear pattern, which we conjecture to hold for generic $(N,M)$. In the following, without loss of generality, we assume that $M>N$.\footnote{In case $N>M$, a similar procedure applies, except that all $v_i$ are replaced by $h_i$ and vice versa.}

\subsection{Parameter Map}
The key in reconstructing the parameters $(\bar{h}_i,\bar{v}_i,\bar{m}_i)$ (for $i=1,\ldots, MN$) of the web $(\tfrac{MN}{k},k)$ in terms of the parameters $(h_i,v_i,m_i)$ of the original web $(N,M)$ is to find a matching between the vertices of the two webs. In this section, starting from a labelling $(A_1,A_2,\ldots,A_{2MN})$ of the vertices of the web $(N,M)$ we construct a labelling of the vertices in the web $(\tfrac{MN}{k},k)$, in other words, we explain how the vertices are mapped into each other. Furthermore, we provide a rule for the area for the curve connecting any two vertices in the web $(\tfrac{MN}{k},k)$ in terms of the parameters of the original web. 
\subsubsection{Notation: Vertices and Paths}
Before we describe the mapping of the parameters between the webs $(N,M)$ and $(\tfrac{MN}{k},k)$, we need to introduce some notation. We start from a labelling of the vertices of the web $(N,M)$. There are two types of the former, depending on the orientation of the external leg along the direction $(1,1)$, \emph{i.e.}
\begin{align}
&\begin{tikzpicture}
\draw[ultra thick] (-1,0) -- (0,0);
\draw[ultra thick] (0,-1) -- (0,0);
\draw[ultra thick] (0,0) -- (0.7,0.7);
\node at (0,0) {$\bullet$};
\node at (0,-2) {vertex of type 1};
\end{tikzpicture}
&&
\begin{tikzpicture}
\draw[ultra thick] (0,0) -- (1,0);
\draw[ultra thick] (0,0) -- (0,1);
\draw[ultra thick] (-0.7,-0.7) -- (0,0);
\node at (0,0) {$\bullet$};
\node at (0,-2) {vertex of type 2};
\end{tikzpicture}\nonumber
\end{align}
Furthermore, for $n\in\mathbb{N}$ we can define for each vertex $A$ the path $\mathcal{P}(A,n)$ that starts at $A$ and follows $n$ vertical and $n-1$ horizontal lines 
\begin{align}
&\parbox{6cm}{\scalebox{0.68}{\parbox{7cm}{\begin{tikzpicture}[scale = 1.75]
\draw[ultra thick] (-1,0) -- (0,0);
\draw[ultra thick, red] (0,0) -- (0,-1);
\node at (-0.2,-0.5) {{\large \bf $v_1$}};
\draw[ultra thick, red] (0,-1) -- (1,-1);
\node at (0.5,-0.8) {{\large \bf $h_1$}};
\draw[ultra thick, red] (1,-1) -- (1,-2);
\node at (0.8,-1.5) {{\large \bf $\vdots$}};
\draw[ultra thick, red] (1,-2) -- (2,-2);
\node at (1.5,-1.8) {{\large \bf $h_{n-1}$}};
\draw[ultra thick, red] (2,-2) -- (2,-3);
\node at (1.8,-2.5) {{\large \bf $v_n$}};
\draw[ultra thick] (2,-3) -- (3,-3);
\draw[ultra thick] (0,0) -- (1,1);
\draw[ultra thick] (1,-1) -- (2,0);
\draw[ultra thick] (2,-2) -- (3,-1);
\draw[ultra thick] (0,-1) -- (-1,-2);
\draw[ultra thick] (1,-2) -- (0,-3);
\draw[ultra thick] (2,-3) -- (1,-4);
\node at (0,0) {$\bullet$};
\node at (-0.1,0.3) {\large $A$};
\node at (2,-3) {$\bullet$};
\node at (1,-5) {\Large path $\mathcal{P}(A,n)$ for a vertex of type 1};
\end{tikzpicture}}}}
&&
&\parbox{6cm}{\scalebox{0.68}{\parbox{7cm}{\begin{tikzpicture}[scale = 1.75]
\draw[ultra thick] (-1,0) -- (0,0);
\draw[ultra thick, red] (0,0) -- (0,-1);
\node at (-0.2,-0.5) {{\large \bf $v_n$}};
\draw[ultra thick, red] (0,-1) -- (1,-1);
\node at (0.5,-0.8) {{\large \bf $h_{n-1}$}};
\draw[ultra thick, red] (1,-1) -- (1,-2);
\node at (0.8,-1.5) {{\large \bf $\vdots$}};
\draw[ultra thick, red] (1,-2) -- (2,-2);
\node at (1.5,-1.8) {{\large \bf $h_{1}$}};
\draw[ultra thick, red] (2,-2) -- (2,-3);
\node at (1.8,-2.5) {{\large \bf $v_1$}};
\draw[ultra thick] (2,-3) -- (3,-3);
\draw[ultra thick] (0,0) -- (1,1);
\draw[ultra thick] (1,-1) -- (2,0);
\draw[ultra thick] (2,-2) -- (3,-1);
\draw[ultra thick] (0,-1) -- (-1,-2);
\draw[ultra thick] (1,-2) -- (0,-3);
\draw[ultra thick] (2,-3) -- (1,-4);
\node at (0,0) {$\bullet$};
\node at (2,-3) {$\bullet$};
\node at (2.1,-3.3) {\large $A$};
\node at (1,-5) {\Large path $\mathcal{P}(A,n)$ for a vertex of type 2};
\end{tikzpicture}}}}\nonumber
\end{align}
Notice that $\mathcal{P}(A,n)$ connects a vertex of type 1 with a vertex of type 2 or vice versa. Finally, we call $\mathfrak{p}(A,n)$ the (ordered) $(2n-1)$-tuple of parameters of the lines that form the path $\mathcal{P}(A,n)$, \emph{i.e.} specifically with regards to the figures above
\begin{align}
\mathfrak{p}(A,n)=(v_1,h_1,v_2,h_2,\ldots,h_{n-1},v_n)\,.
\end{align}
\subsubsection{Vertex Mapping}
As we have mentioned above, from a generic labelling of the $(N,M)$ web, we conjecture that it is possible to directly reconstruct the labelling of the $(\tfrac{MN}{k},k)$ web. The procedure we propose starts from any vertex of type 1 in the $(N,M)$ web which we call $A_1$~\footnote{As a possible source of confusion, we point out that vertices of type 1 in the web $(N,M)$ will be mapped to vertices of type 2 in the web $(\tfrac{MN}{k},k)$ and vice versa. Thus $A_1$ is of type 1 in the $(N,M)$ web, but will be of type 2 in the $(\tfrac{MN}{k},k)$ web.}. To find (one of) the neighbour(s) of this vertex in the $(\tfrac{MN}{k},k)$ web, we follow $\mathcal{P}(A_1,\tfrac{M}{k})$, \emph{i.e.} a path with $\tfrac{M}{k}$ vertical lines and $\tfrac{M}{k}-1$ horizontal lines, that leads to a vertex of type 2 in the $(N,M)$ web:
\begin{align}
&\phantom{x} \hspace{1.55cm}\text{\bf Web }(N,M) && &&\phantom{x} \hspace{0.4cm}\text{\bf Web }(\tfrac{MN}{k},k)\nonumber\\
&\parbox{4.8cm}{\scalebox{0.68}{\parbox{7cm}{\begin{tikzpicture}[scale = 1.75]
\draw[ultra thick] (-1,0) -- (0,0);
\draw[ultra thick, red] (0,0) -- (0,-1);
\node at (-0.2,-0.5) {{\large \bf $v_1$}};
\draw[ultra thick, red] (0,-1) -- (1,-1);
\node at (0.5,-0.8) {{\large \bf $h_1$}};
\draw[ultra thick, red] (1,-1) -- (1,-2);
\node at (0.8,-1.5) {{\large \bf $v_2$}};
\draw[ultra thick, red] (1,-2) -- (2,-2);
\node at (1.5,-1.8) {{\large \bf $h_2$}};
\draw[ultra thick, red] (2,-2) -- (2,-3);
\node at (1.8,-2.5) {{\large \bf $v_3$}};
\draw[ultra thick] (2,-3) -- (3,-3);
\draw[ultra thick] (0,0) -- (1,1);
\draw[ultra thick] (1,-1) -- (2,0);
\draw[ultra thick] (2,-2) -- (3,-1);
\draw[ultra thick] (0,-1) -- (-1,-2);
\draw[ultra thick] (1,-2) -- (0,-3);
\draw[ultra thick] (2,-3) -- (1,-4);
\node at (0,0) {$\bullet$};
\node at (-0.2,0.3) {\large $A_1$};
\node at (2,-3) {$\bullet$};
\node at (2.2,-3.3) {\large $A_2$};
\end{tikzpicture}}}}
&&
\parbox{1.5cm}{\begin{tikzpicture}\draw[ultra thick,->] (0,0) -- (1.5,0);\end{tikzpicture}}
&&\parbox{4.5cm}{\scalebox{0.68}{\parbox{6.6cm}{\begin{tikzpicture}[scale = 1.75]
\draw[ultra thick] (-1,0) -- (0,0);
\draw[ultra thick] (0,0) -- (0,-1);
\draw[ultra thick,red] (0,0) -- (1,1);
\draw[ultra thick] (1,1) -- (2,1);
\draw[ultra thick] (1,1) -- (1,2);
\node at (1,1) {$\bullet$};
\node at (1.2,0.7) {\large $A_1$};
\node at (0,0) {$\bullet$};
\node at (0.3,-0.2) {\large $A_2$};
\node[rotate=315] at (-0.6,1.6) {{\large \bf $v_1+v_2+v_3+h_1+h_2$}};
\end{tikzpicture}}}}
\nonumber
\end{align}
We conjecture that the parameter connecting the two vertices in the $(\tfrac{MN}{k},k)$ web is the sum of all $v_i$'s and $h_i$'s we have followed, 
\begin{align}
\bar{m}_1=\sum_{i=1}^{\tfrac{2M}{k}-1}\left(\mathfrak{p}(A_1,\tfrac{M}{k})\right)_i\,.
\end{align}
From $A_2$, we follow $\mathcal{P}(A_2,\tfrac{M}{k}-1)$, \emph{i.e.} a path with $\tfrac{M}{k}-1$ vertical lines and $\tfrac{M}{k}-2$ horizontal lines, to a vertex of type 1 (which is different from $A_1$)
\begin{align}
&\phantom{x} \hspace{1.55cm}\text{\bf Web }(N,M) && &&\phantom{x} \hspace{0.8cm}\text{\bf Web }(\tfrac{MN}{k},k)\nonumber\\
&\parbox{4.8cm}{\scalebox{0.68}{\parbox{7cm}{\begin{tikzpicture}[scale = 1.75]
\draw[ultra thick] (-1,0) -- (0,0);
\draw[ultra thick] (0,0) -- (0,-1);
\node at (-0.2,-0.5) {{\large \bf $v_1$}};
\draw[ultra thick] (0,-1) -- (1,-1);
\node at (0.5,-0.8) {{\large \bf $h_1$}};
\draw[ultra thick, blue] (1,-1) -- (1,-2);
\node at (0.8,-1.5) {{\large \bf $v_2$}};
\draw[ultra thick, blue] (1,-2) -- (2,-2);
\node at (1.5,-1.8) {{\large \bf $h_2$}};
\draw[ultra thick, blue] (2,-2) -- (2,-3);
\node at (1.8,-2.5) {{\large \bf $v_3$}};
\draw[ultra thick] (2,-3) -- (3,-3);
\draw[ultra thick] (0,0) -- (1,1);
\draw[ultra thick] (1,-1) -- (2,0);
\draw[ultra thick] (2,-2) -- (3,-1);
\draw[ultra thick] (0,-1) -- (-1,-2);
\draw[ultra thick] (1,-2) -- (0,-3);
\draw[ultra thick] (2,-3) -- (1,-4);
\node at (0,0) {$\bullet$};
\node at (-0.2,0.3) {\large $A_1$};
\node at (2,-3) {$\bullet$};
\node at (2.2,-3.3) {\large $A_2$};
\node at (1,-1) {$\bullet$};
\node at (1.2,-1.2) {{\large \bf $A_3$}};
\end{tikzpicture}}}}
&&
\parbox{1.5cm}{\begin{tikzpicture}\draw[ultra thick,->] (0,0) -- (1.5,0);\end{tikzpicture}}
&&\parbox{4.8cm}{\scalebox{0.68}{\parbox{7cm}{\begin{tikzpicture}[scale = 1.75]
\draw[ultra thick] (-2,-1) -- (-1,0);
\draw[ultra thick] (-1,0) -- (-1,1);
\draw[ultra thick,blue] (-1,0) -- (0,0);
\draw[ultra thick] (0,0) -- (0,-1);
\draw[ultra thick] (0,0) -- (1,1);
\draw[ultra thick] (1,1) -- (2,1);
\draw[ultra thick] (1,1) -- (1,2);
\node at (1,1) {$\bullet$};
\node at (1.2,0.7) {\large $A_1$};
\node at (0,0) {$\bullet$};
\node at (0.3,-0.2) {\large $A_2$};
\node at (-1,0) {$\bullet$};
\node at (-1.3,0.2) {\large $A_3$};
\node[rotate=315] at (-0.6,1.6) {{\large \bf $v_1+v_2+v_3+h_1+h_2$}};
\node[rotate=270] at (-0.5,-1) {{\large \bf $-v_2-v_3-h_2$}};
\end{tikzpicture}}}}
\nonumber
\end{align}
The parameter connecting $A_2$ and $A_3$ in the web $(\tfrac{MN}{k},k)$ show up as the negative sum of all $v_i$'s and $h_i$'s that we have followed, \emph{i.e.}
\begin{align}
\bar{h}_1=-\sum_{i=1}^{\tfrac{2M}{k}-3}\left(\mathfrak{p}(A_2,\tfrac{M}{k}-1)\right)_i
\end{align}
Continuing from $A_3$ along $\mathcal{P}(A_3,\tfrac{M}{k})$, \emph{i.e.} a path with $\tfrac{M}{k}$ vertical lines and $\tfrac{M}{k}-1$ horizontal lines, we arrive at a new vertex $A_4$ of type 2. This procedure can be continued for $\tfrac{2MN}{k}$ steps passing through the vertices $A_1,A_2,\ldots,A_{\tfrac{2MN}{k}}$ in the $(N,M)$ web. After this, we return to our original starting point, since at each step of the construction the vertex reached by the path
\begin{align}
\mathcal{P}(A_1,\tfrac{M}{k})\,\cup\,&\mathcal{P}(A_2,\tfrac{M}{k}-1)\,\cup\,\mathcal{P}(A_3,\tfrac{M}{k})\,\cup\,\mathcal{P}(A_4,\tfrac{M}{k}-1)\,\cup\ldots\nonumber\\
&\ldots \cup\,\mathcal{P}(A_{\tfrac{2MN}{k}-1},\tfrac{M}{k})\,\cup\,\mathcal{P}(A_{\tfrac{2MN}{k}},\tfrac{M}{k}-1)\,,
\end{align}
is $A_1$ itself. We therefore have constructed the first 'layer' of the $(\tfrac{MN}{k},k)$ web, which at this point looks like
\begin{align}
\scalebox{0.7}{\parbox{22cm}{\begin{tikzpicture}[scale = 1.15]
\draw[ultra thick] (2,1) -- (2,2);
\draw[ultra thick] (1,0) -- (1,-1);
\draw[ultra thick] (4,2) -- (4,3);
\draw[ultra thick] (3,1) -- (3,0);
\draw[ultra thick] (5,2) -- (5,1);
\draw[ultra thick] (6,3) -- (6,4);
\draw[ultra thick] (7,3) -- (7,2);
\draw[ultra thick] (12,6) -- (12,7);
\draw[ultra thick] (13,6) -- (13,5);
\draw[ultra thick] (14,7) -- (14,8);
\draw[ultra thick] (15,7) -- (15,6);
\draw[ultra thick] (16,8) -- (16,9);
\draw[ultra thick] (17,8) -- (17,7);
\draw[ultra thick] (18,9) -- (18,10);
\draw[ultra thick] (0,0) -- (1,0);
\draw[ultra thick] (2,1) -- (3,1);
\draw[ultra thick] (4,2) -- (5,2);
\draw[ultra thick] (6,3) -- (7,3);
\draw[ultra thick] (12,6) -- (13,6);
\draw[ultra thick] (14,7) -- (15,7);
\draw[ultra thick] (16,8) -- (17,8);
\draw[ultra thick] (18,9) -- (19,9);
\draw[ultra thick] (1,0) -- (2,1);
\draw[ultra thick] (3,1) -- (4,2);
\draw[ultra thick] (5,2) -- (6,3);
\draw[ultra thick] (7,3) -- (8,4);
\draw[ultra thick] (11,5) -- (12,6);
\draw[ultra thick] (13,6) -- (14,7);
\draw[ultra thick] (15,7) -- (16,8);
\draw[ultra thick] (17,8) -- (18,9);
\node[rotate=45] at (1.8,0.45) {{\small $\bar{m}_{\frac{MN}{k}}$}};
\node[rotate=45] at (3.8,1.45) {{\small $\bar{m}_{\frac{MN}{k}-1}$}};
\node[rotate=45] at (5.8,2.45) {{\small $\bar{m}_{\frac{MN}{k}-2}$}};
\node[rotate=45] at (7.8,3.45) {{\small $\bar{m}_{\frac{MN}{k}-3}$}};
\node[rotate=45] at (11.35,5.75) {{\small $\bar{m}_4$}};
\node[rotate=45] at (13.35,6.75) {{\small $\bar{m}_3$}};
\node[rotate=45] at (15.35,7.75) {{\small $\bar{m}_2$}};
\node[rotate=45] at (17.35,8.75) {{\small $\bar{m}_{1}$}};
\node at (0.7,0.4) {{\small $\bar{h}_{\frac{MN}{k}}$}};
\node at (2.7,1.4) {{\small $\bar{h}_{\frac{MN}{k}-1}$}};
\node at (4.7,2.4) {{\small $\bar{h}_{\frac{MN}{k}-2}$}};
\node at (6.7,3.4) {{\small $\bar{h}_{\frac{MN}{k}-3}$}};
\node at (12.5,6.3) {{\small $\bar{h}_3$}};
\node at (14.5,7.3) {{\small $\bar{h}_2$}};
\node at (16.5,8.3) {{\small $\bar{h}_1$}};
\node at (18.5,9.3) {{\small $\bar{h}_k$}};
\node at (-0.2,0) {{\small \bf $a$}};
\node at (19.2,9) {{\small \bf $a$}};
%
\node at (1,0) {$\bullet$};
\node at (1.55,-0.1) {$A_{\frac{2MN}{k}}$};
\node at (2,1) {$\bullet$};
\node at (1.3,1.2) {$A_{\frac{2MN}{k}-1}$};
\node at (3,1) {$\bullet$};
\node at (3.7,0.7) {$A_{\frac{2MN}{k}-2}$};
\node at (4,2) {$\bullet$};
\node at (3.3,2.2) {$A_{\frac{2MN}{k}-3}$};
\node at (5,2) {$\bullet$};
\node at (5.7,1.7) {$A_{\frac{2MN}{k}-4}$};
\node at (6,3) {$\bullet$};
\node at (5.3,3.2) {$A_{\frac{2MN}{k}-5}$};
\node at (7,3) {$\bullet$};
\node at (7.7,2.7) {$A_{\frac{2MN}{k}-6}$};
\node at (12,6) {$\bullet$};
\node at (12.1,5.7) {$A_{7}$};
\node at (13,6) {$\bullet$};
\node at (13.4,5.9) {$A_{6}$};
\node at (14,7) {$\bullet$};
\node at (14.1,6.7) {$A_{5}$};
\node at (15,7) {$\bullet$};
\node at (15.4,6.9) {$A_{4}$};
\node at (16,8) {$\bullet$};
\node at (16.1,7.7) {$A_{3}$};
\node at (17,8) {$\bullet$};
\node at (17.4,7.9) {$A_{2}$};
\node at (18,9) {$\bullet$};
\node at (18.15,8.7) {$A_{1}$};
\node[rotate=30] at (9.5,4.6) 
{{\Huge $\cdots$}};
\end{tikzpicture}}}
\nonumber
\end{align}
This web configuration is the counterpart of the diagonal path in 
Figure  9.  In order to find the remaining vertices, we return to (the type 1 vertex) $A_1$ in the $(N,M)$ web and follow the diagonal line 
\begin{align}
&\phantom{x} \hspace{1.55cm}\text{\bf Web }(N,M) && &&\phantom{x} \hspace{0.8cm}\text{\bf Web }(\tfrac{MN}{k},k)\nonumber\\
&\parbox{4.8cm}{\scalebox{0.68}{\parbox{7cm}{\begin{tikzpicture}[scale = 1.75]
\draw[ultra thick] (-1,0) -- (0,0);
\draw[ultra thick] (0,0) -- (0,-1);
\node at (-0.2,-0.5) {{\large \bf $v_1$}};
\draw[ultra thick] (0,-1) -- (1,-1);
\node at (0.5,-0.8) {{\large \bf $h_1$}};
\draw[ultra thick] (1,-1) -- (1,-2);
\node at (0.8,-1.5) {{\large \bf $v_2$}};
\draw[ultra thick] (1,-2) -- (2,-2);
\node at (1.5,-1.8) {{\large \bf $h_2$}};
\draw[ultra thick] (2,-2) -- (2,-3);
\node at (1.8,-2.5) {{\large \bf $v_3$}};
\draw[ultra thick] (2,-3) -- (3,-3);
\draw[ultra thick,brown] (0,0) -- (1,1);
\draw[ultra thick] (1,-1) -- (2,0);
\draw[ultra thick] (2,-2) -- (3,-1);
\draw[ultra thick] (0,-1) -- (-1,-2);
\draw[ultra thick] (1,-2) -- (0,-3);
\draw[ultra thick] (2,-3) -- (1,-4);
\node at (0,0) {$\bullet$};
\node at (-0.2,0.3) {\large $A_1$};
\node at (2,-3) {$\bullet$};
\node at (2.2,-3.3) {\large $A_2$};
\node at (1,-1) {$\bullet$};
\node at (1.2,-1.2) {{\large \bf $A_3$}};
\node at (1,1) {$\bullet$};
\node at (1.3,0.7) {{\large \bf $A_{\frac{2MN}{k}+1}$}};
\draw[ultra thick] (1,1) -- (2,1);
\draw[ultra thick] (1,1) -- (1,2);
\end{tikzpicture}}}}
&&
\parbox{1.5cm}{\begin{tikzpicture}\draw[ultra thick,->] (0,0) -- (1.5,0);\end{tikzpicture}}
&&\parbox{4.8cm}{\scalebox{0.68}{\parbox{7cm}{\begin{tikzpicture}[scale = 1.75]
\draw[ultra thick] (-2,-1) -- (-1,0);
\draw[ultra thick] (-1,0) -- (-1,1);
\draw[ultra thick] (-1,0) -- (0,0);
\draw[ultra thick] (0,0) -- (0,-1);
\draw[ultra thick] (0,0) -- (1,1);
\draw[ultra thick] (1,1) -- (2,1);
\draw[ultra thick,brown] (1,1) -- (1,2);
\node at (1,1) {$\bullet$};
\node at (1.2,0.7) {\large $A_1$};
\node at (0,0) {$\bullet$};
\node at (0.3,-0.2) {\large $A_2$};
\node at (-1,0) {$\bullet$};
\node at (-1.3,0.2) {\large $A_3$};
\node at (1,2) {$\bullet$};
\node at (1.6,1.9) {\large $A_{\frac{2MN}{k}+1}$};
\draw[ultra thick] (0,2) -- (1,2);
\draw[ultra thick] (1,2) -- (2,3);
\node[rotate=315] at (-0.6,1.6) {{\large \bf $v_1+v_2+v_3+h_1+h_2$}};
\node[rotate=270] at (-0.5,-1) {{\large \bf $-v_2-v_3-h_2$}};
\node at (0.8,1.5) {\large $\bar{v}_1$};
\end{tikzpicture}}}}
\nonumber
\end{align}
to arrive at the next vertex $A_{\tfrac{2MN}{k}+1}$ in the $(N,M)$ web which is of type 2.\footnote{Notice, if $k=1$, this vertex is in fact $A_2$ and the mapping of vertices is already complete. If $k>1$, the vertex $A_{\frac{2MN}{k}+1}$ is different from the ones already encountered, \emph{i.e.} $A_1,\ldots,A_{\frac{2MN}{k}}$.} The parameter $\bar{v}_1$ is somewhat more involved to read off from the original $(N,M)$ web. Indeed, calling the curve connecting $A_1$ and $A_{\frac{MN}{k}+1}$ in the $(N,M)$ web $\mathcal{M}\left(A_1,A_{\tfrac{2MN}{k}+1}\right)$ (with parameter $m$), we consider the path
\begin{align}
\mathcal{P}(A_1,\tfrac{M}{k}-1)\cup \mathcal{M}\left(A_1,A_{\tfrac{2MN}{k}+1}\right)\cup \mathcal{P}\left(A_{\tfrac{2MN}{k}+1},\tfrac{M}{k}-1\right)\,,
\end{align}
\emph{i.e.} from the two vertices $A_1$ and $A_{\frac{2MN}{k}+1}$ we consider a curve that follows $\tfrac{M}{k}-1$ vertical and $\tfrac{M}{k}-2$ horizontal lines upwards and downwards. For illustrative purposes, we consider the simple example $\frac{M}{k}=4$ for which the total path is given in the following picture
\begin{align}
&\parbox{4.8cm}{\scalebox{0.68}{\parbox{7cm}{\begin{tikzpicture}[scale = 1.75]
\draw[ultra thick] (-1,0) -- (0,0);
\draw[ultra thick,red] (0,0) -- (0,-1);
\node at (-0.2,-0.5) {{\large \bf $v_1$}};
\draw[ultra thick,red] (0,-1) -- (1,-1);
\node at (0.5,-0.8) {{\large \bf $h_1$}};
\draw[ultra thick,red] (1,-1) -- (1,-2);
\node at (0.8,-1.5) {{\large \bf $v_2$}};
\draw[ultra thick,red] (1,-2) -- (2,-2);
\node at (1.5,-1.8) {{\large \bf $h_2$}};
\draw[ultra thick,red] (2,-2) -- (2,-3);
\node at (1.8,-2.5) {{\large \bf $v_3$}};
\draw[ultra thick] (2,-3) -- (3,-3);
\draw[ultra thick,red] (0,0) -- (1,1);
\node at (0.4,0.7) {{\large \bf $m$}};
\draw[ultra thick] (1,-1) -- (2,0);
\draw[ultra thick] (2,-2) -- (3,-1);
\draw[ultra thick] (0,-1) -- (-1,-2);
\draw[ultra thick] (1,-2) -- (0,-3);
\draw[ultra thick] (2,-3) -- (1,-4);
\draw[ultra thick] (1,2) -- (2,3);
\node at (0,0) {$\bullet$};
\node at (-0.2,0.3) {\large $A_1$};
\node at (1,1) {$\bullet$};
\node at (1.4,0.7) {{\large \bf $A_{\frac{2MN}{k}+1}$}};
\draw[ultra thick] (1,1) -- (2,1);
\draw[ultra thick,red] (1,1) -- (1,2);
\draw[ultra thick,red] (0,2) -- (1,2);
\draw[ultra thick,red] (0,2) -- (0,3);
\draw[ultra thick] (-1,1) -- (0,2);
\draw[ultra thick] (0,3) -- (1,4);
\draw[ultra thick,red] (0,3) -- (-1,3);
\draw[ultra thick,red] (-1,3) -- (-1,4);
\draw[ultra thick] (-1,3) -- (-2,2);
\node at (1.2,1.5) {{\large \bf $v_4$}};
\node at (0.2,2.5) {{\large \bf $v_5$}};
\node at (0.5,1.8) {{\large \bf $h_3$}};
\node at (-0.8,3.5) {{\large \bf $v_6$}};
\node at (-0.5,2.8) {{\large \bf $h_4$}};
\end{tikzpicture}}}}\label{DiagramPath}
\end{align}
We then conjecture that the parameter $\bar{v}_1$ is given by the sum over all the parameters along this path, however, with factors that decrease the further we get away from the vertices $A_1$ and $A_{\frac{2MN}{k}+1}$:
\begin{align}
\bar{v}_1=m&+\left(\frac{M}{k}-1\right)\left[\left(\mathfrak{p}(A_1,\tfrac{M}{k}-1)\right)_1+\left(\mathfrak{p}(A_{\frac{2MN}{k}+1},\tfrac{M}{k}-1)\right)_1\right]\nonumber\\
&+\sum_{i=1}^{\tfrac{M}{k}-2}\left(\frac{M}{k}-1-i\right)\left[\left(\mathfrak{p}(A_1,\tfrac{M}{k}-1)\right)_{2i}+\left(\mathfrak{p}(A_{\frac{2MN}{k}+1},\tfrac{M}{k}-1)\right)_{2i}\right]\nonumber\\
&+\sum_{i=1}^{\tfrac{M}{k}-2}\left(\frac{M}{k}-1-i\right)\left[\left(\mathfrak{p}(A_1,\tfrac{M}{k}-1)\right)_{2i+1}+\left(\mathfrak{p}(A_{\frac{2MN}{k}+1},\tfrac{M}{k}-1)\right)_{2i+1}\right]\,,
\end{align}
\emph{e.g.}  for the case displayed in the diagram in Eq.(\ref{DiagramPath})
\begin{align}
\bar{v}_1=m+\left(4-1\right)(v_1+v_4)+\left(4-2\right)(h_1+v_2+h_3+v_5)+\left(4-3\right)(h_2+v_3+h_4+v_6)\,.
\end{align}
From the last vertex $A_{\frac{2MN}{k}+1}$, we can repeat all the previous steps to construct further vertices that make up the second 'layer' of the web $(\tfrac{MN}{k},k)$. Indeed, following $\mathcal{P}(A_{\tfrac{2MN}{k}+1},\tfrac{M}{k}-1)$ we arrive at the next vertex $A_{\tfrac{2MN}{k}+2}$, which is\footnote{For $k>1$ the vertex $A_{\tfrac{2MN}{k}+1}$ is different from the vertices $A_1,\ldots,A_{\tfrac{2MN}{k}}$ that we have previously constructed.} 
\begin{align}
&\phantom{x} \hspace{1.55cm}\text{\bf Web }(N,M) && &&\phantom{x} \hspace{0.8cm}\text{\bf Web }(\tfrac{MN}{k},k)\nonumber\\
&\parbox{4.8cm}{\scalebox{0.68}{\parbox{7cm}{\begin{tikzpicture}[scale = 1.75]
\draw[ultra thick] (-1,0) -- (0,0);
\draw[ultra thick] (0,0) -- (0,-1);
\node at (-0.2,-0.5) {{\large \bf $v_1$}};
\draw[ultra thick] (0,-1) -- (1,-1);
\node at (0.5,-0.8) {{\large \bf $h_1$}};
\draw[ultra thick] (1,-1) -- (1,-2);
\node at (0.8,-1.5) {{\large \bf $v_2$}};
\draw[ultra thick] (1,-2) -- (2,-2);
\node at (1.5,-1.8) {{\large \bf $h_2$}};
\draw[ultra thick] (2,-2) -- (2,-3);
\node at (1.8,-2.5) {{\large \bf $v_3$}};
\draw[ultra thick] (2,-3) -- (3,-3);
\draw[ultra thick] (0,0) -- (1,1);
\draw[ultra thick] (1,-1) -- (2,0);
\draw[ultra thick] (2,-2) -- (3,-1);
\draw[ultra thick] (0,-1) -- (-1,-2);
\draw[ultra thick] (1,-2) -- (0,-3);
\draw[ultra thick] (2,-3) -- (1,-4);
\node at (0,0) {$\bullet$};
\node at (-0.2,0.3) {\large $A_1$};
\node at (2,-3) {$\bullet$};
\node at (2.2,-3.3) {\large $A_2$};
\node at (1,-1) {$\bullet$};
\node at (1.2,-1.2) {{\large \bf $A_3$}};
\node at (1,1) {$\bullet$};
\node at (1.35,0.7) {{\large \bf $A_{\frac{2MN}{k}+1}$}};
\draw[ultra thick] (1,1) -- (2,1);
\draw[ultra thick,blue] (1,1) -- (1,2);
\draw[ultra thick] (1,2) -- (2,3);
\draw[ultra thick,blue] (0,2) -- (1,2);
\draw[ultra thick] (-1,1) -- (0,2);
\draw[ultra thick,blue] (0,2) -- (0,3);
\draw[ultra thick] (0,3) -- (1,4);
\draw[ultra thick] (0,3) -- (-1,3);
\node at (0,3) {$\bullet$};
\node at (0.55,2.9) {{\large \bf $A_{\frac{2MN}{k}+2}$}};
\node at (1.2,1.5) {{\large \bf $v_4$}};
\node at (-0.2,2.5) {{\large \bf $v_5$}};
\node at (0.5,1.8) {{\large \bf $h_3$}};
\end{tikzpicture}}}}
&&
\parbox{1.5cm}{\begin{tikzpicture}\draw[ultra thick,->] (0,0) -- (1.5,0);\end{tikzpicture}}
&&\parbox{4.8cm}{\scalebox{0.68}{\parbox{7cm}{\begin{tikzpicture}[scale = 1.75]
\draw[ultra thick] (-2,-1) -- (-1,0);
\draw[ultra thick] (-1,0) -- (-1,1);
\draw[ultra thick] (-1,0) -- (0,0);
\draw[ultra thick] (0,0) -- (0,-1);
\draw[ultra thick] (0,0) -- (1,1);
\draw[ultra thick] (1,1) -- (2,1);
\draw[ultra thick] (1,1) -- (1,2);
\draw[ultra thick] (0,2) -- (0,3);
\node at (1,1) {$\bullet$};
\node at (1.2,0.7) {\large $A_1$};
\node at (0,0) {$\bullet$};
\node at (0.3,-0.2) {\large $A_2$};
\node at (-1,0) {$\bullet$};
\node at (-1.3,0.2) {\large $A_3$};
\node at (1,2) {$\bullet$};
\node at (1.55,1.9) {\large $A_{\frac{2MN}{k}+1}$};
\node at (0,2) {$\bullet$};
\node at (-0.55,2.2) {\large $A_{\frac{2MN}{k}+2}$};
\draw[ultra thick,blue] (0,2) -- (1,2);
\draw[ultra thick] (1,2) -- (2,3);
\node[rotate=315] at (-0.6,1.6) {{\large \bf $v_1+v_2+v_3+h_1+h_2$}};
\node[rotate=270] at (-0.5,-1) {{\large \bf $-v_2-v_3-h_2$}};
\node at (0.8,1.5) {\large $\bar{v}_1$};
\node[rotate=270] at (0.5,3) {{\large \bf $-v_4-v_5-h_3$}};
\end{tikzpicture}}}}
\nonumber
\end{align}
From here on, we can continue to map out the full $(\tfrac{MN}{k},k)$ web: After a total of $2MN$ steps, this procedure is guaranteed to map all vertices of the $(N,M)$ web to the $(\tfrac{MN}{k},k)$ web. Moreover, while we have no general proof, the values for $(\bar{v}_i,\bar{h}_i,\bar{m}_i)$ we obtain in this manner are compatible with all the examples we have discussed so far.
\subsection{$Sp(2,\mathbb{Z})$ Transformation}
Based on the expressions for $(\bar{v}_i,\bar{h}_i,\bar{m}_i)$ obtained in the previous subsection, we can check whether they correspond to an $Sp(2,\mathbb{Z})$ transformation for the partition function $\mathcal{Z}_{X_{1,1}}$, in the same manner as we did for all the examples in section~\ref{Sect:GenericMapMN}. To this end, we set 
\begin{align}
&v_i=v\,,&&h_i=h\,,&&m_i=m\,,&&\forall i=1,\ldots, MN\,,
\end{align}
and hence identify all vertical, horizontal and diagonal parameters in the original $(N,M)$ web. This leads to
\begin{align}
&\bar{v}_i=\bar{v}\,,&&\bar{h}_i=\bar{h}\,,&&\bar{m}_i=\bar{m}\,,&&\forall i=1,\ldots, MN\,,
\end{align}
with
\begin{align}
\bar{v}&=m+2\left[\sum_{n=1}^{\frac{M}{k}-1}\left(\frac{M}{k}-n\right)\right]v+2\left[\sum_{n=1}^{\frac{M}{k}-2}\left(\frac{M}{k}-1-n\right)\right]h\nonumber\\
&=m+2\left[\left(\frac{M}{k}\right)^2-\frac{1}{2}\frac{M}{k}\left(\frac{M}{k}+1\right)\right]v+2\left[\left(\frac{M}{k}-1\right)^2-\frac{1}{2}\frac{M}{k}\left(\frac{M}{k}-1\right)\right]h\nonumber\\
&=m+\frac{M(M-k)}{k^2}\,v+\left(2+\frac{M(M-3k)}{k^2}\right)h\,,
\end{align}
as well as 
\begin{align}
\bar{h}=-\left(\frac{M}{k}-1\right)v-\left(\frac{M}{k}-2\right)h\,,&&\text{and} &&\bar{m}=\frac{M}{k}\,v+\left(\frac{M}{k}-1\right)h\,.
\end{align}
Furthermore, we introduce
\begin{align}
&\rho=N(h+m)\,,&&\text{and} &&\tau=M(v+m)\,,
\end{align}
along with
\begin{align}
&\rho'=\frac{MN}{k}(\bar{h}+\bar{m})=\frac{MN}{k}(h+v)=\frac{N}{k}\,\tau+\frac{M}{k}\,\rho-\frac{2MN}{k}\,m\,,\nonumber\\
&\tau'=k(\bar{v}+\bar{m})=\frac{1}{k}\left[k^2m+(M-k)^2h+M^2v\right]=\frac{M}{k}\,\tau+\frac{(M-k)^2}{Nk}\,\rho-\frac{2M(M-k)}{k}\,m\,,\nonumber\\
&m'=\bar{m}=\frac{\tau}{k}+\frac{1}{N}\left(\frac{M}{k}-1\right)\rho+\left(1-\frac{2M}{k}\right)m\,.\label{Transformationtautaup}
\end{align}
With these parameters, we define the period matrices for the $(1,1)$ web
\begin{align}
&\Omega=\left(\begin{array}{cc}\tau/M & m \\ m & \rho/N\end{array}\right)\,,&&\text{and} &&\Omega'=\left(\begin{array}{cc}\tau'/k & m' \\ m' & \rho'/\left(\frac{MN}{k}\right)\end{array}\right)\,,
\end{align}
which have identical determinants
\begin{align}
\text{det}\,\Omega=\frac{\rho\tau}{MN}-m^2=\text{det}\,\Omega'\,.
\end{align}
In fact, $\Omega$ and $\Omega'$ are related by an $Sp(2,\mathbb{Z})$ transformation in the following manner
\begin{align}
\Omega'=(A\Omega+B)\cdot (C\Omega+D)^{-1}\,,
\end{align}
with
\begin{align}
&A=\left(\begin{array}{cc} \frac{M}{k} &-\frac{M-k}{k} \\ 1 & -1\end{array}\right)\,,&&B=\left(\begin{array}{cc}0 & 0 \\ 0 & 0\end{array}\right)\,,&&C=\left(\begin{array}{cc}0 & 0 \\ 0 & 0\end{array}\right)\,,&&D=\left(\begin{array}{cc}1 & 1 \\ -\frac{M-k}{k} & -\frac{M}{k}\end{array}\right)\,,
\end{align}
which satisfy as well the relations Eq.(\ref{Sp2ZTrafos}). This shows that $(\rho,\tau,m)$ and $(\rho',\tau',m')$ are related through an $Sp(2,\mathbb{Z})$ transformation. Therefore, $\mathcal{Z}_{X_{M,N}}$ at this special locu is indeed invariant under the change of parameters implied by the above chain of dualities.

\subsection{Extended Moduli Space of Calabi-Yau Threefold}
In the previous sections, we have compiled evidence that the web diagrams $(N,M)$ and $(\tfrac{NM}{k},k)$ with $k=\text{gcd}(M,N)$ are dual to each other. This relation, however, also implies dualities between two different webs $(N,M)$ and $(N',M')$ with $MN=M'N'$ and $\text{gcd}(M,N)=k=\text{gcd}(M',N')$. We can give a characterisation of all such webs in the following manner: Let $M,N\in\mathbb{N}$ with the following prime number decomposition
\begin{align}
&M=\left(\prod_{a=1}^rc_a^{h_a}\right)\,\left(\prod_{b=1}^sm_a^{p_a}\right)\,,&&\text{and} &&N=\left(\prod_{a=1}^rc_a^{h_a}\right)\,\left(\prod_{b=1}^tn_a^{q_a}\right)\,,
\end{align}
with $r,s,t\geq 0$ and $c_a$, $m_a$ and $n_a$ prime numbers with multiplicities $h_a$, $p_a$ and $q_a$, respectively. Furthermore, $m_a\neq n_b$ for $a=1,\ldots,s$ and $b=1,\ldots, t$ such that
\begin{align}
k=\text{gcd}(M,N)=\prod_{a=1}^rc_a^{h_a}\,.
\end{align} 
Finally, we define the disjoint sets
\begin{align}
&\mathcal{M}\subset \left\{m_1^{p_1},m_2^{p_2},\ldots,m_s^{p_s},n_1^{q_1},n_2^{q_2},\ldots,n_t^{q_t}\right\}\,,&&\mathcal{N}\subset \left\{m_1^{p_1},m_2^{p_2},\ldots,m_s^{p_s},n_1^{q_1},n_2^{q_2},\ldots,n_t^{q_t}\right\}\,,
\end{align}
such that
\begin{align}
&\mathcal{M}\cap\mathcal{N}=\{\}\,,&&\text{and} &&\mathcal{M}\cup\mathcal{N}=\left\{m_1^{p_1},m_2^{p_2},\ldots,m_s^{p_s},n_1^{q_1},n_2^{q_2},\ldots,n_t^{q_t}\right\}\,,
\end{align}
and we denote the product of all elements of $\mathcal{M}$ by $\mathfrak{m}$, and the product of all elements of $\mathcal{N}$ by $\mathfrak{n}$, respectively (notice that $\text{gcd}(\mathfrak{m},\mathfrak{n})=1$). With this notation in place, we propose that all webs of the form $(k\,\mathfrak{n},k\,\mathfrak{m})$ are dual to each other, \emph{i.e.} for two different $(\mathfrak{n},\mathfrak{m})$ and $(\mathfrak{n}',\mathfrak{m}')$ we propose\footnote{As an example, we propose that the configurations $(30,1)\sim (15,2)\sim (10,3)\sim (6,5)\sim (5,6)\sim (3,10)\sim (2,15)\sim (1,30)$ are dual to each other.}
\begin{align}
X_{k\,\mathfrak{n},k\,\mathfrak{m}}\sim X_{k\,\mathfrak{n}',k\,\mathfrak{m}'}\,.
\end{align}
In particular, as we have proposed above, the partition functions $\mathcal{Z}_{X_{k\,\mathfrak{n},k\,\mathfrak{m}}}$ and $\mathcal{Z}_{X_{k\,\mathfrak{n}',k\,\mathfrak{m}'}}$ are identical after a judicious change of variables.

Notice that the $3\times 3$ matrix of transformation $U$ given by Eq.(\ref{Transformationtautaup}),
\begin{align}
&\left(\begin{array}{c}\rho'\\ \tau' \\ m'\end{array}\right)=U\,\left(\begin{array}{c}\rho\\ \tau \\ m\end{array}\right)\,,&&U=\left(\begin{array}{ccc}\frac{M}{k} & \frac{N}{k} & -\frac{2N\,M}{k}\\ \frac{(M-k)^2}{Nk} & \frac{M}{k}& -\frac{2M(M-k)}{k}\\ \frac{1}{N}(\frac{M}{k}-1)& \frac{1}{k}&1-\frac{2M}{k}\end{array}\right)\,,
\end{align}
has the property $U^2=1$. This property together with the fact that $\mbox{Tr}\,U=1$ implies that the eigenvalues of $U$ are $\{-1,1,1\}$. We can choose a new set of parameters $\alpha_{1},\alpha_{2}$ and $\alpha_3$ (which are a linear combination of $\rho,\tau$ and $m$), such that under the transformation in Eq.(\ref{Transformationtautaup}),
\begin{align}
&\alpha_{1}\mapsto -\alpha_{1}\,,&&\mbox{and}&&\alpha_{2,3}\mapsto \alpha_{2,3}\,.
\end{align}
The plane $\alpha_{1}=0$ is a wall of the $(\rho,\tau,m)$ K\"ahler cone and the flop transitions responsible for Eq.(\ref{Transformationtautaup}) are reflections on this wall. In terms of $(\rho,\tau,m)$ the parameter $\alpha_{1}$ is given by
\bea
\alpha_1=N\Big(\rho+\frac{2M}{M-k}\tau-\frac{m}{M-k}\Big)\,.
\eea
such that the wall $\alpha_1=0$ is given by,
\bea
(M-k)\rho+2M\,\tau-m=0\,.
\eea
The above codimension-one wall in the $(\rho,\tau,m)$ space has a form very similar to that of Humbert surfaces \cite{Humbert}, although the corresponding invariant is negative.
\section{Conclusions}\label{Sect:Conclusions}
In this paper, we have discussed a two-parameter family of little string theories that generically preserve eight supercharges. These theories are labelled by two integers $(N,M)$ and are realised by a particular decoupling limit of M5-branes probing an orbifold geometry.  The M5-branes are spread out on a circle of radius $\rho$, while compactified on a circle of radius $\tau$ with transverse space an A-type orbifold. The proposed duality that we discussed in the paper indicates that, as a consequence of this double compactification, this system of branes is dual to a setup with $\tfrac{NM}{k}$ M5-branes probing $\text{ALE}_{A_{k-1}}$ (for $k=\text{gcd}(N,M)$). Notice, however, that this new set of branes is compactified with new parameters $(\rho',\tau')$, which are related to $(\rho,\tau)$ according to the duality map proposed in Section~\ref{Sect:GenericNM}. We recall that the duality does not hold in the limit $\rho\to \infty$ (\emph{i.e.} the case that the M5-branes are arranged on $\mathbb{R}$ rather than $S^1$).

The M5-brane system was studied using a dual description in terms of toric Calabi-Yau threefolds $X_{N,M}$. These Calabi-Yau threefolds realise little string theories via F-theory compactification. We have indeed shown explicitly for a number of examples that $X_{N,M}$ and $X_{NM/k,k}$ can be related to each other through a series of flop transitions. They are thus just two different points in the extended moduli space and therefore the theories labelled by $(N,M)$ and $(NM/k,k)$, with $k=\text{gcd}(N,M)$, are dual to each other. While at first sight this seems to only give relations between different members of the two-parameter class of $\mathcal{N}=(1,0)$ little string theories studied in \cite{Bhardwaj:2015oru,Hohenegger:2016eqy}. It in fact also indicates that some of these theories preserve more than 8 supercharges: indeed, for the case $k=\text{gcd}(N,M)=1$ the theory $(N,M)$ is dual to $(NM,1)$ which furnishes a description of the decoupling limit of $NM$ NS5-branes in type IIA string theory with a flat transverse space $\mathbb{R}^4 $ ($\sim\mathbb{R}^4/\mathbb{Z}_1 $), which is in fact type IIb little string theory of type $A_{NM-1}$ which preserves indeed $\mathcal{N}=(2,0)$ supersymmetry. However, it should be noted that the theories labelled by $(N,M)$ and $(\tfrac{NM}{k},k)$ live in different points of the parameter space of little string theories: indeed, as we have already seen in the dual description through M5-brane webs, the parameters $\rho$ and $\tau$ are different in the two cases\footnote{Notice in this regard, while the parameters $(\rho,\tau)$ are simply two K\"ahler parameters from the perspective of the Calabi-Yau manifold, the rather play the role of external parameters from the perspective of the LST. See \cite{Hohenegger:2015btj,Hohenegger:2016eqy} for more explanations.}. Furthermore, the little string partition function is invariant under flop transitions and, therefore, relates the BPS states of the two dual theories. We have generalised an emergent pattern of these examples to the generic case and have made a conjecture how to obtain the full duality map between the parameters of the two theories.

The duality we discussed has consequences for other systems as well which are related to little string theories. The partition function of the $(N,M)$ little strings is related to the elliptic genus of a $(0,2)$ sigma model with target space which is the product of $M$ $U(N)$ instanton moduli spaces of rank (of various charges) \cite{Haghighat:2013tka,Haghighat:2013gba,Hohenegger:2013ala}. The duality between $X_{N,M}$ and $X_{NM/k,k}$ again implies that there are relations between sigma-models with different $(N,M)$. However, since the duality map is rather complex and mixes in a non-trivial fashion the distances of the M5-branes, the duality relation at the level of the sigma-models is non-perturbative and will be explored elsewhere \cite{toappear}.

\section*{Acknowledgement}
A.I. would like to acknowledge the ``2016 Simons Summer Workshop on Mathematics and Physics" for hospitality during this work. A.I. was supported in part by the Higher Education Commission grant HEC-20-2518.

\end{document}